%% file: main.tex
\PassOptionsToPackage{table}{xcolor}

\documentclass[sigconf]{acmart}

\newcommand{\old}[1]{\textcolor{gray}{#1}}

\usepackage{booktabs,tabularx,ragged2e}
\usepackage{array}
\newcolumntype{Y}{>{\RaggedRight\arraybackslash}X}
\usepackage{enumitem}
\usepackage{multicol}

\usepackage{graphicx}
\usepackage{makecell}
\definecolor{okgreen}{RGB}{102,204,102}
\definecolor{lightgray}{RGB}{235,235,235}

\renewcommand{\arraystretch}{1}

\newcommand{\name}{\textit{CAMeleon}}
\AtBeginDocument{%
  }

\usepackage{multirow}

\renewcommand{\old}[1]{}

\renewcommand\footnotetextcopyrightpermission[1]{}
\begin{document}

\title{Comparing Fabrication Workflows in CAD to Support Design Reasoning}


\author{Shuo Feng}
\affiliation{%
  \institution{Cornell Tech}
 \city{New York}
\country{USA}
}
\email{sf522@cornell.edu}

\author{Xuening Wang}
\affiliation{%
  \institution{Cornell Tech}
 \city{New York}
\country{USA}
}
\email{xw672@cornell.edu}

\author{Yifan (Lavenda) Shan}
\affiliation{%
  \institution{Cornell Tech}
 \city{New York}
\country{USA}
}
\email{ys2253@cornell.edu}

\author{Krista U Singh}
\affiliation{%
  \institution{Macaulay Honors College}
\city{New York}
\country{USA}
}
\email{krista.singh@macaulay.cuny.edu}

\author{Bo Liu}
\affiliation{%
  \institution{Cornell Tech}
 \city{New York}
\country{USA}
  }
  \email{bl685@cornell.edu}

\author{Amritansh Kwatra}
\affiliation{%
  \institution{Cornell Tech}
 \city{New York}
\country{USA}
  }
  \email{ak2244@cornell.edu}

  \author{Ritik Batra}
\affiliation{%
  \institution{Cornell Tech}
 \city{New York}
\country{USA}
  }
  \email{rb887@cornell.edu}

  \author{Tobias M Weinberg}
\affiliation{%
  \institution{Cornell Tech}
 \city{New York}
\country{USA}
  }
  \email{tmw88@cornell.edu}

\author{Thijs Roumen}
\affiliation{%
  \institution{Cornell Tech}
  \city{New York}
  \country{USA}}
\email{thijs.roumen@cornell.edu}

\renewcommand{\shortauthors}{Feng et al.}

\begin{abstract}
When novices fabricate, they naturally start by choosing a workflow (e.g., laser cutting, 3D printing, wire bending) and the corresponding software (e.g., Adobe Illustrator, Fusion 360, Rhino) from a narrow set of options they know. As they advance their design, another workflow might better suit their design intent, but their models remain committed to the original workflow. This prohibits exploration, a learning mechanism fostering informed decision-making.

In this paper, we investigate how to develop CAD interfaces to support exploration and comparison of workflows. We focus on how such a comparison can advance users' reasoning regarding design decisions. A prototype interface, \name{}, was developed to enable side-by-side comparison of fabrication workflows in CAD. Users can load their 3D models and preview fabrication outcomes generated by various workflows. We hypothesize that presenting alternative outcomes from multiple workflows supports exploration and scaffolds informed decision-making. Upon workflow confirmation, \name{} allows users to export their designs for fabrication and provides instructions for manual execution of the workflow.

We interviewed seven fabrication educators to understand how such tools can be integrated into teaching and to demonstrate how we adjust our tool based on the resulting insights. In our user evaluation (N = 12), we found that guided comparison helped participants consider a broader range of workflows, reflect on trade-offs, and experiment with new ways of planning fabrication.

\end{abstract}

\begin{teaserfigure}
\centering
  \includegraphics[width=1\textwidth]{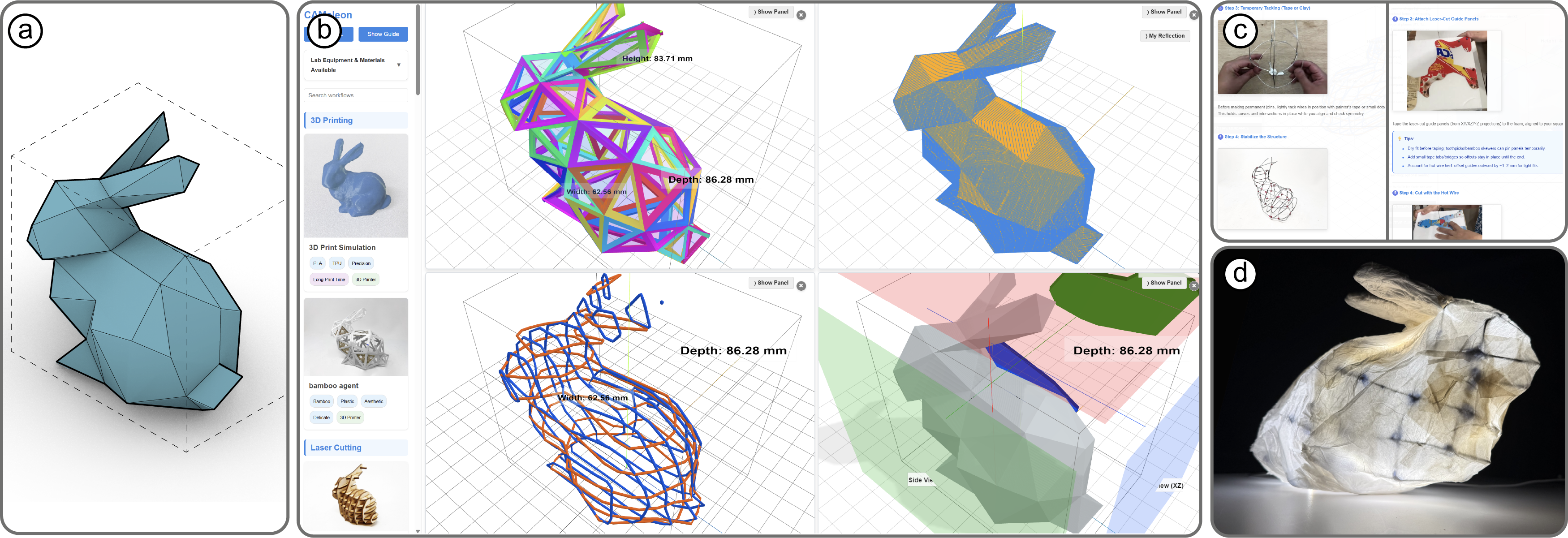}
  \caption{We study the value of comparison in CAD to assist reasoning around fabrication workflows. To this end, we implement an exemplary tool, \name{}, that enables learners to explore and compare fabrication workflows when initial design assumptions no longer align with their goals. 
  (a) A learner imports a Stanford bunny model into \name{}. 
  (b) The model displays with side-by-side previews of alternative workflows (e.g., wire forming), highlighting differences in assembly processes, materials, and machine constraints. 
  (c) Learners examine detailed technical information through the workflow information panel for deeper comparison.
  (d) The final physical outcome (e.g., a bunny lantern) is fabricated using the selected workflow.}
  \label{fig:teaser}
\end{teaserfigure}

\maketitle

\input{src/00_introduction.tex}

\input{src/03_walkthrough}

\input{src/04_relatedwork}

\input{src/05_system}

\input{src/07_expert-eval}
\input{src/09_user-eval}

\input{src/10_discussion}

\input{src/11_conclusion}


\begin{acks}

We thank the fabrication experts who shared their insights and feedback through interviews throughout this research. Their input was invaluable in shaping the direction of this work. We also thank Niti, Sebastian, and the MakerLab at Cornell Tech for their support of this work.

\end{acks}

\bibliographystyle{ACM-Reference-Format}
\bibliography{reference,thijs-papers}
\input{src/12_appendix}


\end{document}

%% file: src/00_introduction.tex
\section{Introduction}

Computer-Aided Design (CAD) and computational fabrication are becoming increasingly accessible to a broader range of users~\cite{baudisch_personal_2017}. This trend has enabled diverse applications, such as customized assistive technology~\cite{hurst_empowering_2011, parry-hill_understanding_2017, higgins_creating_2022,higgins_towards_2023}, DIY robotic devices~\cite{zisimatos_open-source_2014,Taylor_Tanjim_Sack_Hirsch_Cheng_Ching_George_Roumen_Jung_Lee_2025}, medical devices~\cite{Kharat_Dudhani_Kouser_Subramanian_Bhattacharjee_Jhamb_2024, Mamo_Adamiak_Kunwar_2023}, and artistic works~\cite{Mongeon_2016, Hoskins_2018, Fraile-Narváez_Chidean_2025}. As a result, computational fabrication curricula are expanding in educational settings, including middle and high schools~\cite{hartikainen_making_2024,baudisch_kyub_2019}, maker spaces~\cite{blikstein2013digital, fernandez_toward_2021}, and graduate engineering and design programs~\cite{Gershenfeld_2005,qiu_curriculum_2013, buechley_computational_2022}.

Through such curricula, learners are well-equipped to get started with a preset of \textit{specific} workflows. For example, a learner may approach a simple box design with 3D printing without considering whether laser cutting or other workflows might better suit their project needs. However, professionals routinely evaluate multiple workflows before committing to one. They assess trade-offs between aesthetics, functionality, and fabricability to reach their design goals~\cite{Comparing_Novice_and_Expert_Designers’_Approaches_to_Design_Thinking_and_Decision_Making}. As access to diverse computational fabrication workflows matures, fabrication education faces the challenge of helping learners develop this comparative analysis capability, enabling them to make informed decisions to achieve their goals.

Existing research has developed several design tools for “fabrication aware”~\cite{bermano_state_2017} workflows, including \textit{Kyub}~\cite{baudisch_kyub_2019} and \textit{FlatFitFab}~\cite{McCrae_Umetani_Singh_2014} for laser cutting, \textit{Bridging the gap}~\cite{Dumas_Hergel_Lefebvre_2014} and \textit{Stress relief}~\cite{Stava_Vanek_Benes_Carr_Měch_2012} for 3D printing, and \textit{Plushie}~\cite{Mori_Igarashi_2007} for textile design. While each tool excels within its domain, this specialization means learners develop expertise in siloed design environments, making it hard to compare alternative workflows. Similarly, instructional resources on Instructables or YouTube tutorials emphasize step-by-step procedures. These resources boost confidence, but procedural instruction without conceptual grounding may result in ‘inert knowledge’—users may replicate steps but struggle to transfer reasoning across contexts~\cite{DeCaro_2016, Langer_2000, Markovits_Sowder_1994, McNeil_Alibali_2005}. As \citet{Hudson_Alcock_Chilana_2016} observed, siloed workflows may inadvertently limit users by obscuring inter-dependencies across fabrication decisions. Research in design education suggests that comparing alternatives enhances understanding and improves design outcomes~\cite{Tohidi_Buxton_Baecker_Sellen_2006, Dow_Glassco_Kass_Schwarz_Schwartz_Klemmer_2010}, yet fabrication tools, being tied to a machine or process, rarely support such comparison. 

This lack of comparative tools can lead to technical “lock-in”. Although learners acquire multiple fabrication skills over time, they often default to familiar workflows when approaching new design problems, unconsciously adhering to processes they know well~\cite{Brennan_Miney_Simpson_Jablokow_McComb_2023}. Consequently, they tend to frame challenges through the constraints of their most accessible methods rather than critically evaluating which approach best serves their design goals. Once a workflow is selected, its inherent fabrication constraints and modeling strategies further reinforce the initial choice~\cite{Tan_Otto_Wood_2017}. In contrast, exposure to diverse examples has been shown to enhance design outcomes~\cite{Kulkarni_Dow_Klemmer_2014}. Without opportunities to systematically compare different workflows, learners often struggle to cultivate independent, evaluative judgment. They become reliant on instructor-provided solutions rather than building the reflection-in-action that is characteristic of design expertise~\cite{Schön_1983}. This dependency limits their ability to make informed, independent design decisions.

\begin{figure*}[h]
  \centering
  \includegraphics[width=1\linewidth]{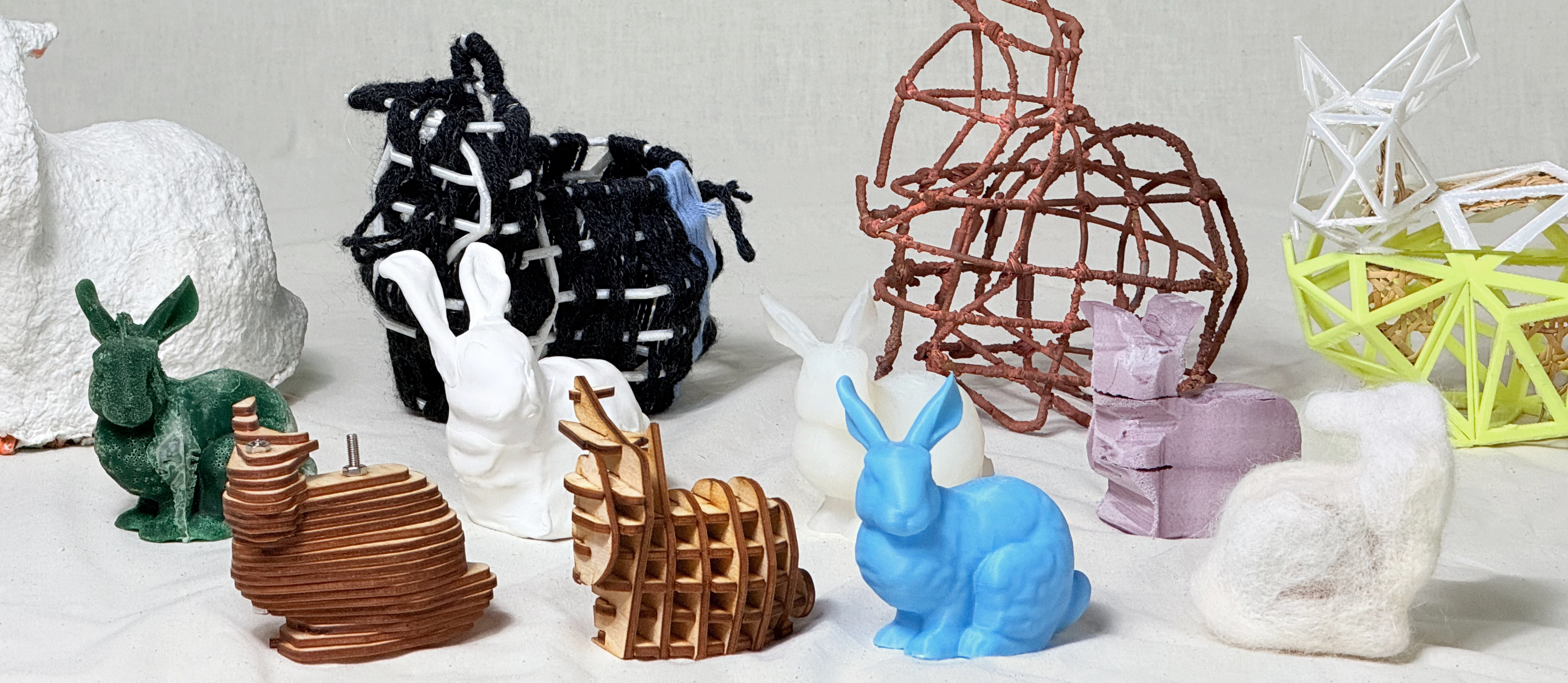}
\caption{Example outputs designed using \name{} and fabricated using a range of machines, showing how the same model behaves across different fabrication workflows, each with distinct material and structural characteristics.}
  \label{fig:workflowoverview}
\end{figure*}

In this paper, we investigate how to support comparison and exploration during CAD design, with the goal of fostering more informed reasoning and decision-making. To study this in practice, we implement an exemplar CAD interface \name{} that explicitly makes users compare fabrication workflows. The key idea is that users import a CAD model and preview how various workflows would fabricate the same geometry. As shown in Figure~\ref{fig:teaser}, users can compare these outcomes side-by-side before deciding which workflow to pursue. The prototype generates previews, provides fabrication metadata (e.g., materials, machines, assembly steps), and highlights workflow-specific considerations (e.g., required post-processing and warnings for possible points of failure). After comparison, users can export models directly from \name{} to fabricate them in a range of machines, as shown in Figure~\ref{fig:workflowoverview}. By making workflows explicitly comparable, we can observe how users explore fabrication workflows by comparing and contrasting them before making decisions, helping us understand how this comparative exploration informs their design reasoning.

To align the notion of workflow exploration within teaching practice, we conducted formative interviews with seven fabrication educators. Their insights into curriculum design provided rich guidance for designing tools like \name{}. We demonstrate how we refined our tool based on these insights. We then conducted a study with twelve students, examining how they explored workflows and made decisions. Our observations suggest that exposure to multiple workflows broadened the range of options learners considered and encouraged a shift from feasibility-driven reasoning (“what can I do?”) toward goal-oriented reasoning (“what best fits my project?”). We see this as a crucial first step towards understanding the impact of such reasoning on long-term learning and quality of design.

\section{Contributions, benefits and limitations}
We make the following contributions:

\begin{itemize}
\item Our main contribution is to refocus CAD tools on exploration and comparison as a vehicle to support reasoning and decision-making. We investigate this idea by implementing an exemplar interface, which we call \name{}, centered on these design principles.

\item We report empirical findings from interviews with expert educators that surface considerations for designing exploration-based fabrication tools, which we demonstrate through our exemplar interface \name{}.

\item We conduct an evaluation with learners to show how participants engaged with cross-workflow comparisons in the short term. Our findings highlight the potential of comparison and exploration in CAD and suggest future research on the long-term effects of comparative reasoning.

\end{itemize}

While studying in the context of our exemplar interface, these contributions provide insights into the design of tools that support exploration and reveal the strategies learners use when exploring fabrication workflows. Long-term deployment in real educational contexts would be required to draw definitive conclusions regarding the impact on learning/education. The findings in this paper serve as starting points for further exploration and design suggestions. Conducting focused studies like ours is essential for understanding how learners engage with comparative tools and for providing evidence-based insights that can inform the design of future tools before wider educational deployment.

We explicitly consider \name{} an exemplar implementation of an interface to compare fabrication workflows/decisions in CAD. While the design is well-considered and based on detailed formative interviews and surveys, we acknowledge that the idea of comparison can be implemented in many ways. The implementation of the interface itself serves as a source of inspiration for others who develop similar tools based on this paper.

No system like \name{} will ever be \textit{complete}; its value is proportional to the number of supported workflows. We implemented 16 workflows that reflect a gamut of workflows found online. We use open source to advance the technical back-end and enable non-technical users to access the interface.

%% file: src/03_walkthrough.tex
\section{Walkthrough: Comparing Fabrication Workflows for a Stool}

Alex, an undergraduate design student enrolled in a computational fabrication course, wants to create a stool as part of a semester project. However, with only limited exposure to 3D printing and little knowledge of other fabrication workflows, Alex is unsure which workflow is best for the project.

\begin{figure}[h]
\centering 
\includegraphics[width=1\linewidth]{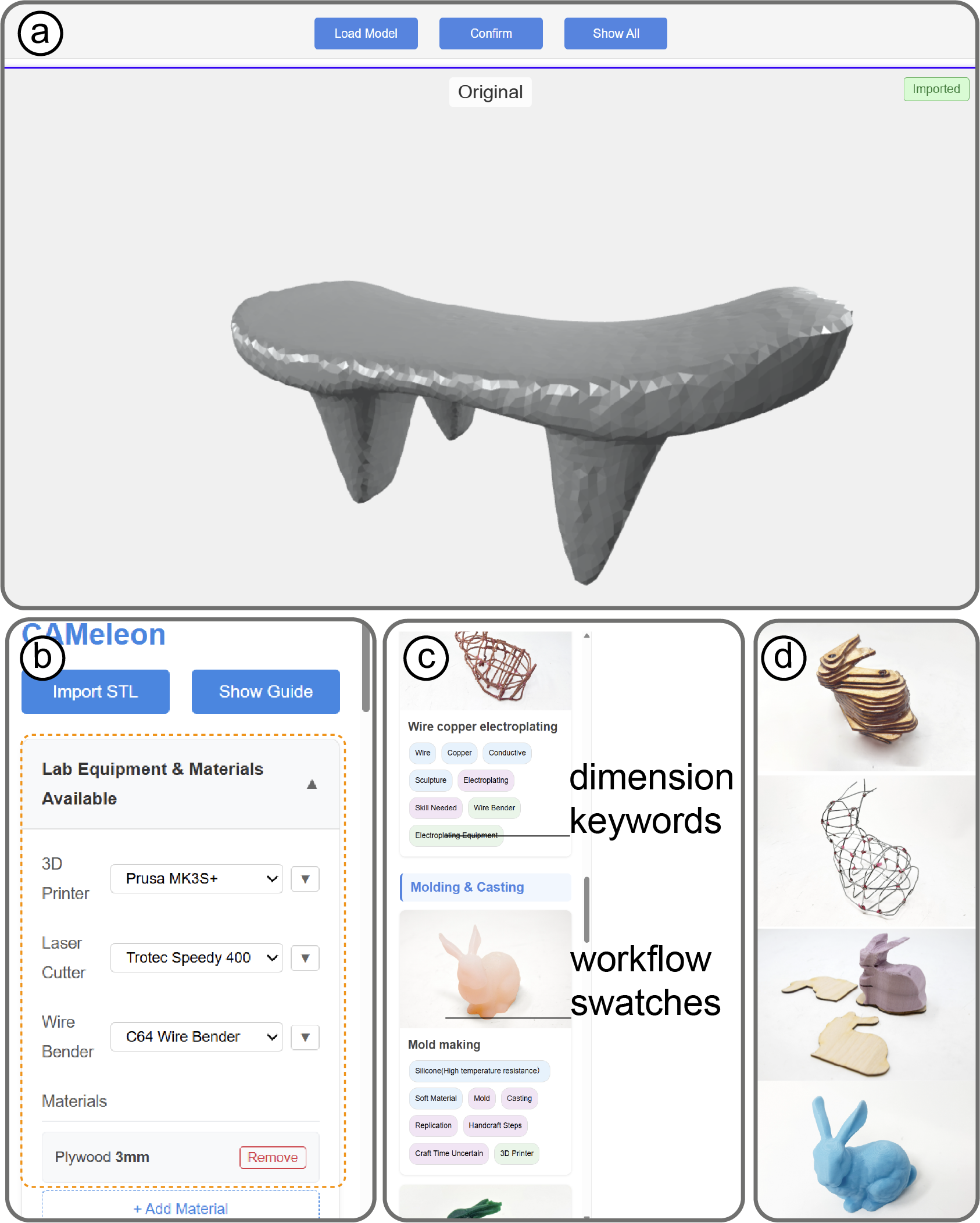}
\caption{Workflow selection: (a) Alex imports the stool in \name{}, (b) workflows are constrained by available tools, (c) sidebar with swatches and keywords for each workflow, (d) four workflows that Alex selects to compare.}
\label{fig:equipment-workflow-selection}
\end{figure}

Alex's instructor recommends using \name{} to explore fabrication workflows beyond 3D printing. Upon entering \name{}, Alex imports a stool model generated using \textit{Hyper3D}\footnote{https://hyper3d.ai/}, which generates 3D models from text prompts. (Figure~\ref{fig:equipment-workflow-selection}a). Alex's school gives access to the equipment shown in (Figure~\ref{fig:equipment-workflow-selection}b). 

Browsing the sidebar, Alex examines the workflow swatches and keywords for each workflow (Figure~\ref{fig:equipment-workflow-selection}c). The swatches help Alex understand potential outcomes, while keywords like “durable” and “lightweight” provide guidance on each workflow's characteristics. Alex initially gravitates toward 3D printing due to familiarity, but decides to explore other options after seeing the diverse possibilities. Alex selects four workflows to compare in detail: epoxy laminating  (promising strength characteristics), wire forming (appealing aesthetic), hot wire foam cutting (attractive low-cost foam material), and 3D printing (familiar technology)(Figure~\ref{fig:equipment-workflow-selection}d).

\begin{figure}[t]
\centering 
\includegraphics[width=1\linewidth]{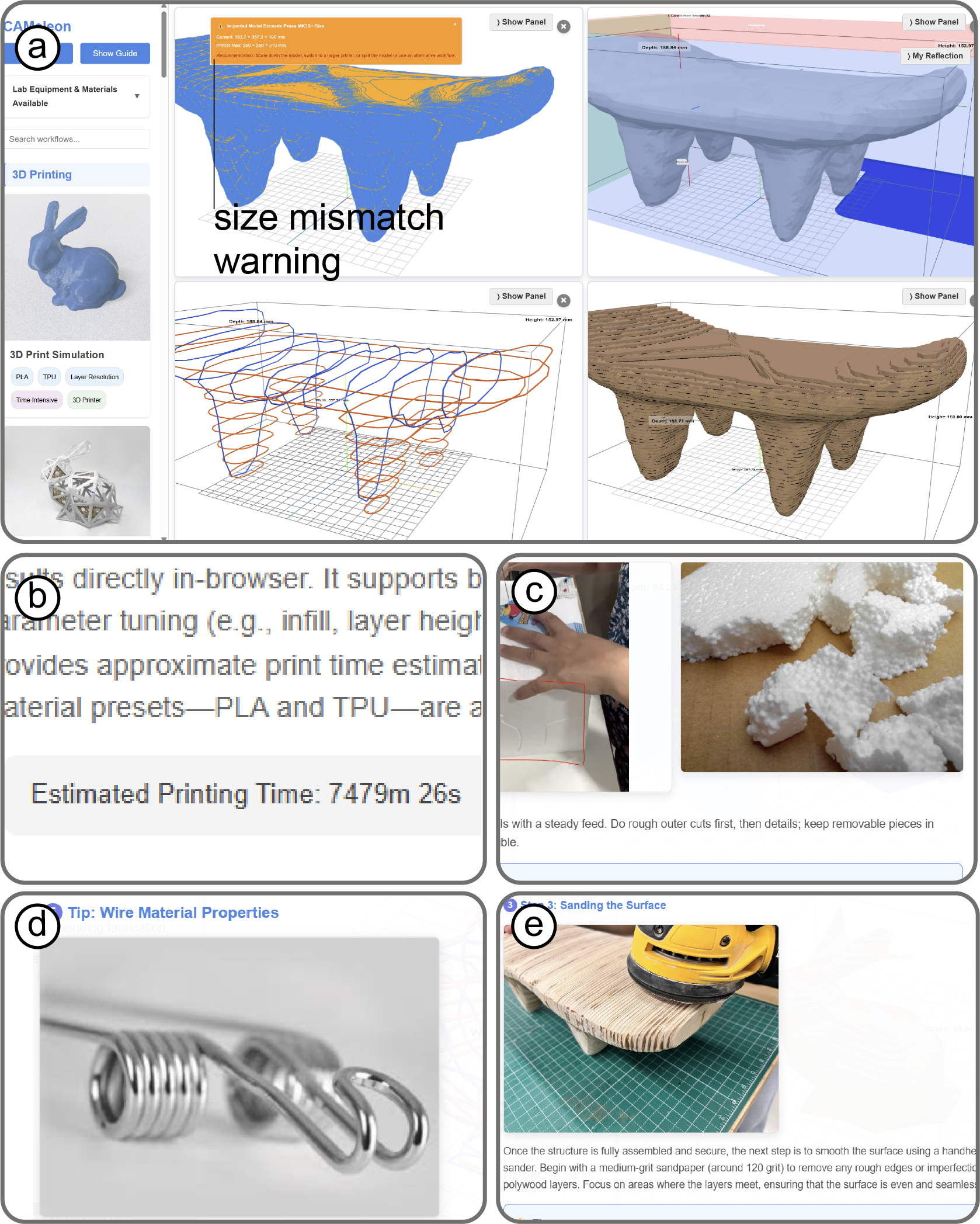} 
\caption{ Comparison interface (a) side-by-side view of four candidate workflows, (b) 3D printing workflow showing print time estimates and size warnings, (c) the hot wire foam cutting workflow and assembly process, with material properties indicating limitations, (d) wire forming showing structural specifications and strength characteristics, (e) epoxy laminating  presenting strength and finish characteristics.} 
\label{fig:workflow-comparison} 
\end{figure}

The side-by-side workflow comparison provides technical information to help Alex decide (Figure~\ref{fig:workflow-comparison}a). This serves as the core of \name{}, enabling Alex to evaluate trade-offs across workflows. For 3D Printing, while Alex is comfortable with the technology, the comparison reveals an estimated print time of 7,479 minutes based on the selected print settings (20\% infill, generic PLA)  (Figure~\ref{fig:workflow-comparison}b). \name{} warns that the model dimensions may require splitting into multiple parts for the available printer bed size (Figure~\ref{fig:workflow-comparison}a). Hot Wire Foam Cutting uses inexpensive foam material, with the following guidance note: \textit{“Foam strength tip: Standard EPS/XPS foam has limited load-bearing capacity.”} Alex considers this note when evaluating the workflow for furniture applications (Figure~\ref{fig:workflow-comparison}c). Wire Forming displays structural specifications showing that wire construction would create a flexible, springy frame that may not provide the stable seating surface needed for the stool, leading Alex to question this workflow. (Figure~\ref{fig:workflow-comparison}d). Finally, Epoxy Laminating provides detailed information about material strength, surface finish quality, and the multi-step process involving laser cutting, assembly, and resin application (Figure~\ref{fig:workflow-comparison}e).

After reviewing the specifications, process complexity, and material properties, Alex chooses Epoxy laminating.

\begin{figure*}[h]
\centering
\includegraphics[width=1\textwidth]{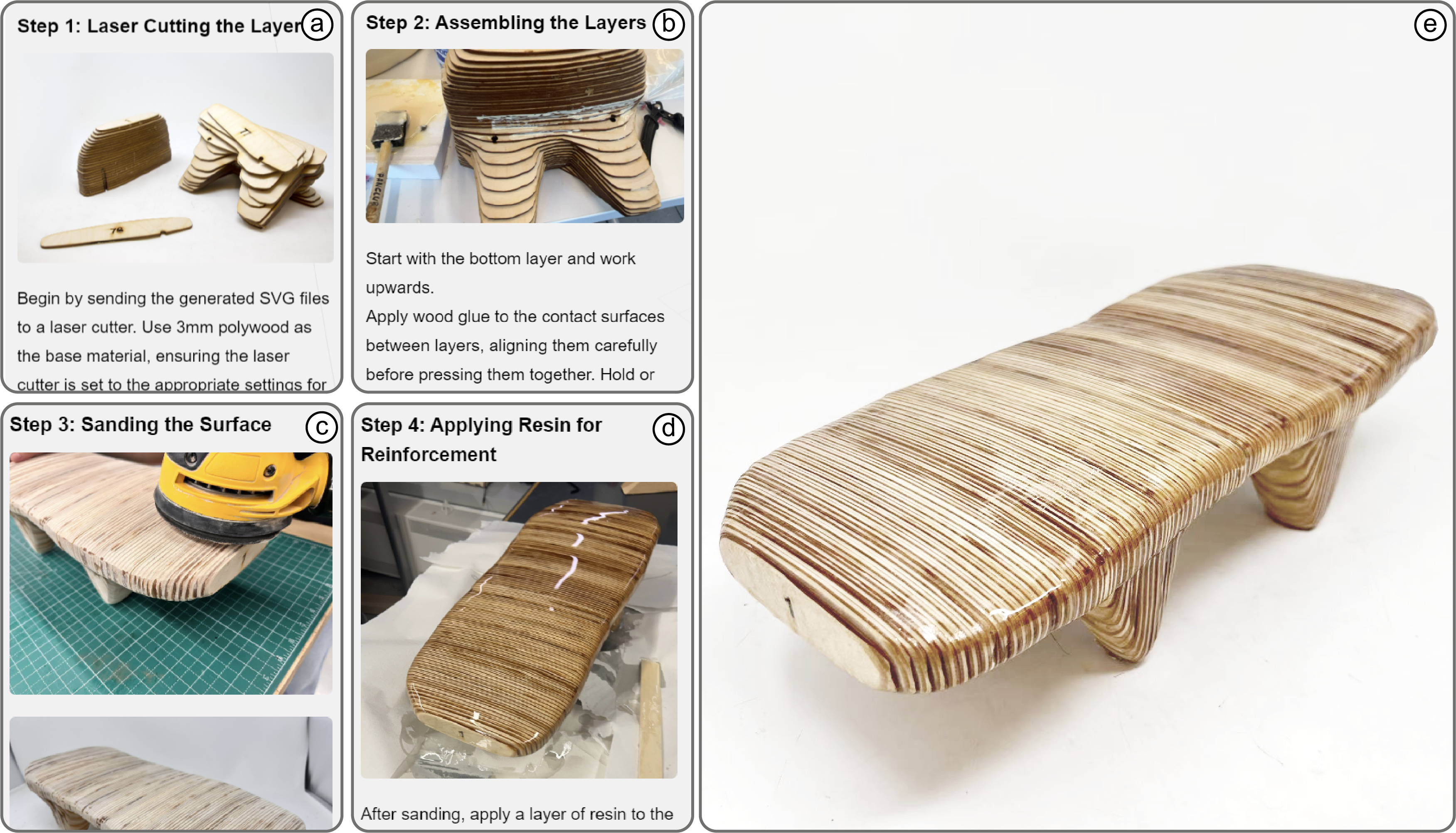}
\caption{Epoxy Laminating workflow execution. (a)~Alex laser-cuts the SVG files, (b)~assembles and glues the pieces, (c)~sands the assembly for surface preparation, (d)~applies a layer of epoxy resin to seal and protect the surface, (e)~completed stool.}
\label{fig:fabrication-process}
\end{figure*}

Alex decides to choose the epoxy laminating workflow for the stool. After exporting the required SVG files for the epoxy-laminating workflow, \name{} assists Alex step by step with the manual steps to build this stool. Although the guidance note serves as a general framework, \name{} provides step-by-step reference instructions and videos to outline the typical steps involved in epoxy laminating. Figure~\ref{fig:fabrication-process} illustrates the epoxy laminating workflow guidance. 

Through \name{}'s comparison, Alex moved beyond 3D printing to adopt a more appropriate fabrication workflow for the project. Besides fabricating the stool using a previously unknown workflow, this exploration experience helps Alex to understand why and when to use such a workflow for future projects.

\begin{figure}[b]
\centering
\includegraphics[width=1\linewidth]{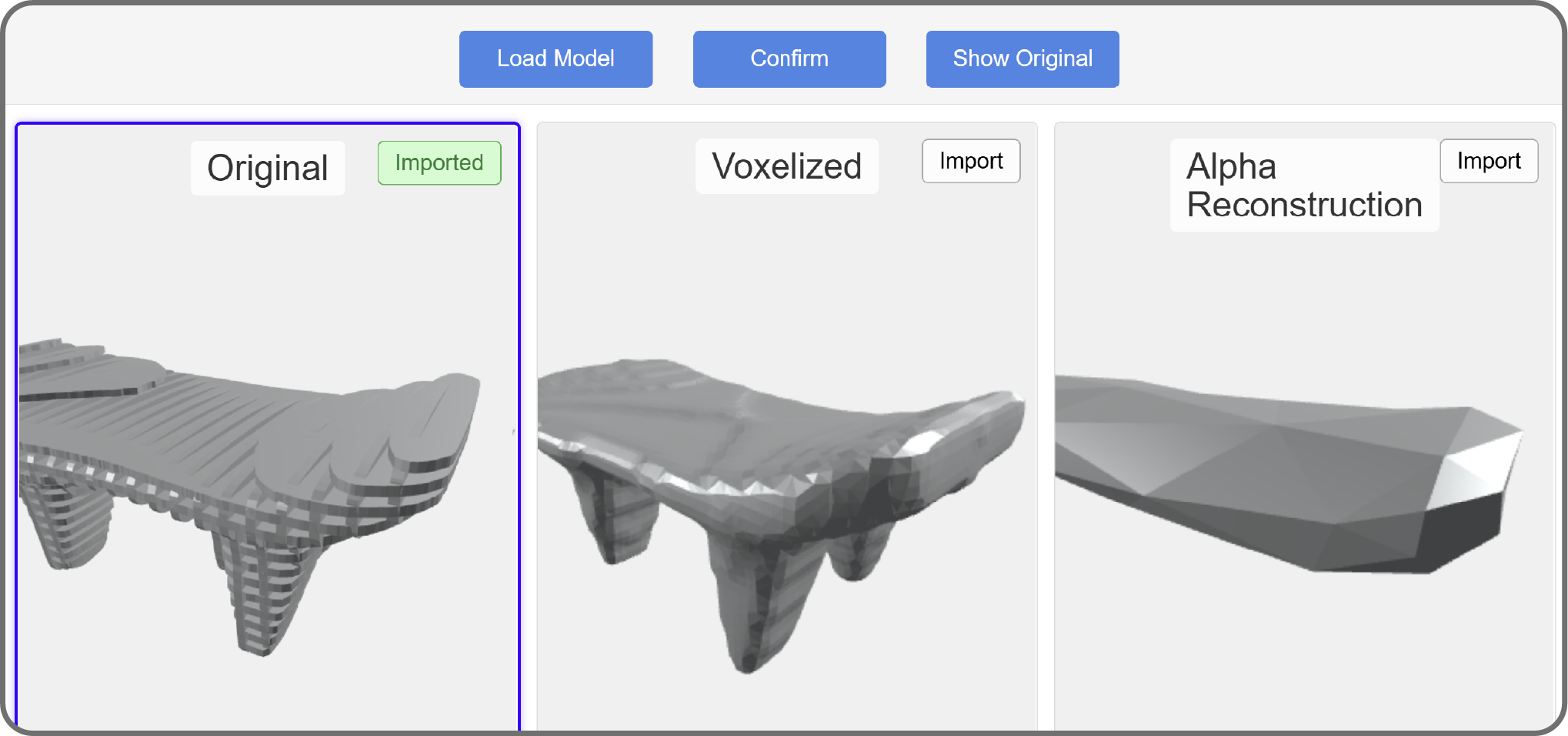}
\caption{Original interlocking model designed for laser cutting by Bailey, which is shown alongside its voxelized reconstruction and alpha-shape reconstruction in \name{}.}
\label{fig:initial-approach}
\end{figure}

Alex's workflow illustrates a not-atypical, but straightforward, situation common in fabrication classrooms in which a ready-made CAD file serves as input for \name{}. In practice, users may create CAD files that make assumptions about specific workflows, for example, Alex's classmate Bailey initially followed YouTube tutorials to design a stool using interlocking laser-cut plates (Figure~\ref{fig:initial-approach}). However, when Bailey brings the design to class, the teaching assistant warns that the interlocking method is unlikely to provide the structural integrity needed for a functional stool.

Confronted with this challenge, Bailey explores alternative workflows that might be better suited for furniture. However, the existing model has been designed exclusively for laser cutting—the SVG files and interlocking geometry cannot be directly adapted for other fabrication equipment, such as wire benders or CNC milling machines. Bailey faces a dilemma: either accept a compromised design with questionable performance, or restart the entire design process from scratch to accommodate a different fabrication workflow.

Bailey instead uses the “reconstruction import” of \name{}. By importing the existing interlocking design as an STL file (either from direct export or through 3D reconstruction \cite{roumen_assembler3_2021, roumen_autoassembler_2021}), \name{} converts the interlocking model into a volumetric representation. \name{} supports different reconstruction methods, and presents previews of the results as shown in Figure~\ref{fig:initial-approach}. Bailey preserves the fundamental form while breaking free from the laser-cutting constraints using the \textit{voxelized reconstruction}. Now featuring the reconstructed geometry, Bailey continues using \name{} as usual, illustrating how the reconstruction import overcomes the lock-in to a specific workflow and allows transitioning between workflows.

%% file: src/04_relatedwork.tex
\section{Related Work}

We build on prior research in computational fabrication education, software for personal fabrication, and interactive tools for comparative design exploration to support design reasoning.

\subsection{Computational Fabrication Education}

Educational research emphasizes the importance of scaffolding: structured guidance that helps novices tackle complex tasks while gradually shifting responsibility from instructor to student as their competence grows~\cite{Vygotskiĭ_Cole_1978, Wood_Bruner_Ross_1976}. In digital domains, \textit{worked examples} and \textit{completion problems} effectively scaffold skill acquisition while supporting transfer to new contexts~\cite{Sweller_Ayres_Kalyuga_2011}. However, the \textit{“blank slate problem”}~\cite{Tabula_rasa_2025} remains a significant barrier for novices~\cite{Burnett_Stumpf_Macbeth_Makri_Beckwith_Kwan_Peters_Jernigan_2016}.

Scaffolding has been applied across creative domains that share challenges with fabrication—namely, complex workflows, multiple possible approaches, and the need to bridge conceptual understanding with practical execution. In photography, systems overlay composition guides~\cite{E_Fried_Lu_Zhang_Mech_Echevarria_Hanrahan_Landay_2020} or provide lighting setup assistance~\cite{E_Fried_Agrawala_2019}. Visual design tools like \textit{Vinci}~\cite{Guo_Jin_Sun_Li_Li_Shi_Cao_2021} apply layout principles for generating posters, while \textit{ProcessGallery}~\cite{Yen_E_Jin_Li_Lin_Pan_Dow_2024} highlights expert-driven workflows. Programming education uses narrative-based systems like \textit{Storytelling Alice}~\cite{Kelleher_Pausch_Kiesler_2007} and more recent tools such as \textit{CodeToon}~\cite{Suh_Zhao_Law_2022}, which map code to stories and comics to support conceptual understanding, as well as debugging-focused approaches like \textit{Gidget}~\cite{Lee_2014} and \textit{NuzzleBug}~\cite{Deiner_Fraser_2024}, which integrate breakpoints, step execution, and interrogative debugging into block-based environments to scaffold problem-solving skills. These systems typically employ feedback, overlays, or reflective prompts to guide skill acquisition.

Fabrication workflows pose unique challenges that remain under-supported. Popular instructional resources like Instructables or YouTube tutorials typically focus on step-by-step procedures through project-based examples. While these resources can boost learner confidence, research in educational psychology demonstrates that purely procedural instruction often leads to "inert knowledge"—learners replicate steps but struggle to transfer reasoning~\cite{DeCaro_2016, Langer_2000, Markovits_Sowder_1994, McNeil_Alibali_2005}. Studies comparing instructional approaches find that learners receiving only procedural instruction show impaired performance when encountering novel problems compared to receiving integrated conceptual and procedural training~\cite{Cheung_Kulasegaram_Woods_Brydges_2019}. In fabrication, learners gravitate toward applying familiar workflows regardless of requirements, preventing the development of reflection-in-action that characterizes design expertise~\cite{Schön_1983} during skill acquisition.

Research in education focuses on lowering the threshold for participation or encouraging reflection during and after making. For example, \textit{LilyPad Arduino}~\cite{Buechley_Eisenberg_Catchen_Crockett_2008} provides classroom curricula and step-by-step tutorials designed for beginners, while \textit{MakerWear}~\cite{Kazemitabaar_McPeak_Jiao_He_Outing_Froehlich_2017} enables students to design programmable garments by assembling pre-built modules, allowing them to explore rather than code. \textit{Mechanix}~\cite{Tseng_Bryant_Blikstein_2011} captures students' tangible constructions and provides visual feedback to support reflection on engineering concepts, and \textit{Process Products}~\cite{Tseng_Tsai_2015} embeds digital traces of the design process into artifacts to support reflection on the process. Among the few fabrication-focused systems providing in-process guidance, \citet{Turakhia_Jiang_Liu_Leake_Mueller_2022} provides in-situ prompts during making, built on expert-authored content. Classroom studies suggest that making—when paired with guidance— enhances engagement and learning in educational settings~\cite{Kulkarni_Magda_2025,Pitkänen_Iwata_Laru_2020}. Yet, despite these contributions, existing work on tool building has not systematically applied scaffolding methods to help novices overcome the blank slate problem or to support workflow exploration during fabrication.

\subsection{Software for Personal Fabrication}

Personal fabrication tools have made fabrication accessible by abstracting technical complexity and supporting novice users~\cite{baudisch_personal_2017}. 

Current personal fabrication software predominantly focuses on a single fabrication workflow. Tools like \textit{Kyub}~\cite{baudisch_kyub_2019} and \textit{FlatFitFab}~\cite{mccrae_flatfitfab_2014} streamline laser-cutting workflows by embedding assembly constraints directly in the design. This focus on one workflow furthermore allows advanced construction to resist loads \cite{abdullah_fastforce_2021}, going off the grid by manipulating individual plates \cite{roumen_structure-preserving_2022}, and strategic placement of joints for assembly~\cite{park_foolproofjoint_2022}. Similarly, \textit{Bridging the gap}~\cite{Dumas_Hergel_Lefebvre_2014} and \textit{Stress relief}~\cite{Stava_Vanek_Benes_Carr_Měch_2012} optimize 3D printing outcomes by generating support structures~\cite{Schmidt_Umetani_2014} or remixing 3D printed machines~\cite{roumen_grafter_2018}. Other domain-specific tools include \textit{Plushie}~\cite{Mori_Igarashi_2007} for textile workflows, \textit{CoilCAM}~\cite{Bourgault_Wiley_Farber_Jacobs_2023} for clay printing, \textit{Imprimer}~\cite{TranO’Leary_Benabdallah_Peek_2023} for CNC milling, and geometric decomposition like \textit{Intersecting Planar Pieces}~\cite{Schwartzburg_Pauly_2013} for laser cutting. While this single-process approach is optimal within workflows, it creates constraints that limit design evolution when users explore alternative workflows.

Recent work has begun to address workflow switching across cross-process platforms and programming frameworks. \textit{Vespidae}~\cite{fossdal2023VespidaeProgrammingFrameworkb} provides a framework for developing workflows, and \textit{Tandem}~\cite{tran_oleary_tandem_2024} enables reproducible fabrication through computational notebooks. This serves as a backbone to support interfaces like \name{}. Other researchers have explored combining fabrication workflows. \textit{Platener}~\cite{Beyer_Gurevich_Mueller_Chen_Baudisch_2015} combines 3D prints with laser-cut plates for enhanced speed, \textit{E-Joint}~\cite{Li__2024} combines 3D printed conductive parts with copper-plated mortise-and-tenon joints, while \citet{Magrisso_Mizrahi_Zoran_2018} uses 3D printing to create generative connectors for hybrid carpentry, \textit{faBrickation}~\cite{Mueller_Mohr_Guenther_Frohnhofen_Baudisch_2014} substitutes parts with LEGO bricks, and \textit{Patching Physical Objects}~\cite{Teibrich_Patching_Physical_Objects} combines CNC milling and 3D printing to modify existing objects. More complex systems, such as \textit{LaserFactory}~\cite{Nisser_Liao_Chai_Adhikari_Hodges_Mueller_2021}, integrate fabrication processes, such as pick-and-place and laser cutting. To date, no tools appear to specifically provide scaffolding for novice users to compare fabrication workflows.

\subsection{Interactive Tools for Comparative Design Exploration}

\begin{figure*}[t]
  \centering
  \includegraphics[width=1\textwidth]{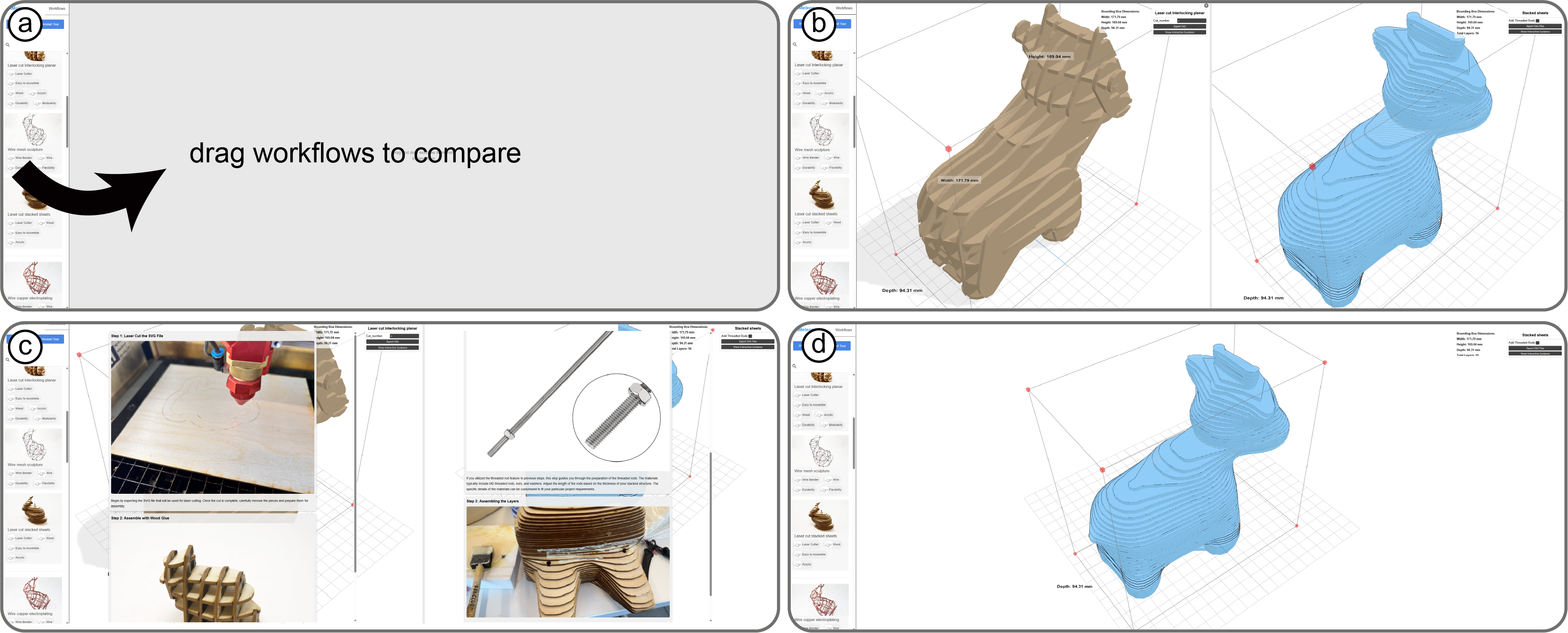}
  \caption{The comparison tool. (a)~users drag workflows into the scene (b)~each window shows the same 3D model but previewed for that specific workflow, up to four windows can be displayed simultaneously, (c)~manual steps can also be displayed, and (d)~when users converge on their workflow, they close the alternative windows to free up screen estate.}
  \label{fig:comparison-tool}
\end{figure*}

Design research has emphasized that exploring alternatives 
can lead to better design outcomes.~\cite{Dow_Fortuna_Schwartz_Altringer_Schwartz_Klemmer_2011}. For example, parallel prototyping~\cite{Dow_Glassco_Kass_Schwarz_Schwartz_Klemmer_2010} has been shown to produce diverse ideas and improved final results. Presenting options also elicits richer feedback, allowing users to compare trade-offs and refine decisions. \citet{Buxton_2007} advocates for sketching multiple concepts in parallel to avoid design fixation and encourage reflection.

This principle of comparative visualization is widely adopted in commercial creative tools. For instance, image editing software such as \textit{Adobe Lightroom} and \textit{Photoshop} provide side-by-side previews for filter effects. Similarly, video editing platforms like \textit{DaVinci Resolve} allow users to contrast different timeline versions or grading effects in parallel. These tools demonstrate how side-by-side comparison helps users reflect on detailed differences and make deliberate adjustments. In HCI research, interactive systems have been developed to bring this comparison capability into design toolkits. For instance, \textit{Juxtapose}~\cite{Hartmann_Yu_Allison_Yang_Klemmer_2008} allows interface designers to develop multiple UI alternatives in parallel. \textit{SUPPLE}~\cite{Gajos_Weld_2004} automatically generates and optimizes UI layouts based on user abilities and preferences.  \textit{Scout}~\cite{Swearngin_Wang_Oleson_Fogarty_Ko_2020} supports exploration of UI layouts by generating variations based on design constraints. Similarly, \textit{Parallel Paths}~\cite{Terry_Mynatt_Nakakoji_Yamamoto_2004} and \textit{Subjunctive Interfaces}~\cite{Lunzer_Hornbæk_2008} allow users to manipulate and evaluate multiple, alternative interactive states, encouraging side-by-side reasoning rather than linear iteration.

This principle has been extended to physical products and generative design contexts. \textit{CAMBRIA}~\cite{Kolarić_Woodbury_Erhan_2014} manages 2D graphic layout variants in parallel, while \textit{f3.js}~\cite{Kato_Goto_2017} is a web-based parametric tool that enables designers to create and compare variations for physical computing. In circuit design, \citet{Lin_Ramesh_Pandhare_Tay_Dutta_Hartmann_Mehta_2024} introduces an interactive tool to explore component-level alternatives. \textit{Dream Lens}~\cite{Matejka_Glueck_Bradner_Hashemi_Grossman_Fitzmaurice_2018} allows users to explore generative design spaces using multi-dimensional performance criteria. 

While these systems streamline design tasks or generate optimized outputs, our focus is different: we aim to support deliberate decision-making when choosing among fabrication workflows. Rather than optimizing a single workflow, we encourage systematic comparison and reasoning, particularly for novice designers.

%% file: src/05_system.tex
\section{\name{}: a CAD interface based on guided comparison}

In this section, we describe the exemplar interface \name{}, which we designed to study the value of guided comparison in CAD. Our interface consists of a side-by-side comparison view, supported by fabrication workflows. We specify design considerations as well as the selection of supported workflows.

\subsection{Interface Design}
\name{} is a browser tool. Users compare workflows using the comparison tool side-by-side. Once a workflow is selected, the interface provides interactive guides for fabrication. 

\subsubsection{Comparison Windows}
The comparison tool (Figure~\ref{fig:comparison-tool}) lets users drag up to four workflows into the viewport. All workflows load the same 3D model, enabling side-by-side comparison of how the same design would be fabricated across different workflows. In addition to visual comparison, the comparison panel also includes information about each workflow.

Users preview manual steps and required machines. This feature allows workflows to be compared not only by their results but also by the resources they require, supporting informed decisions.

\begin{figure}[h]
  \centering
  \includegraphics[width=1\linewidth]{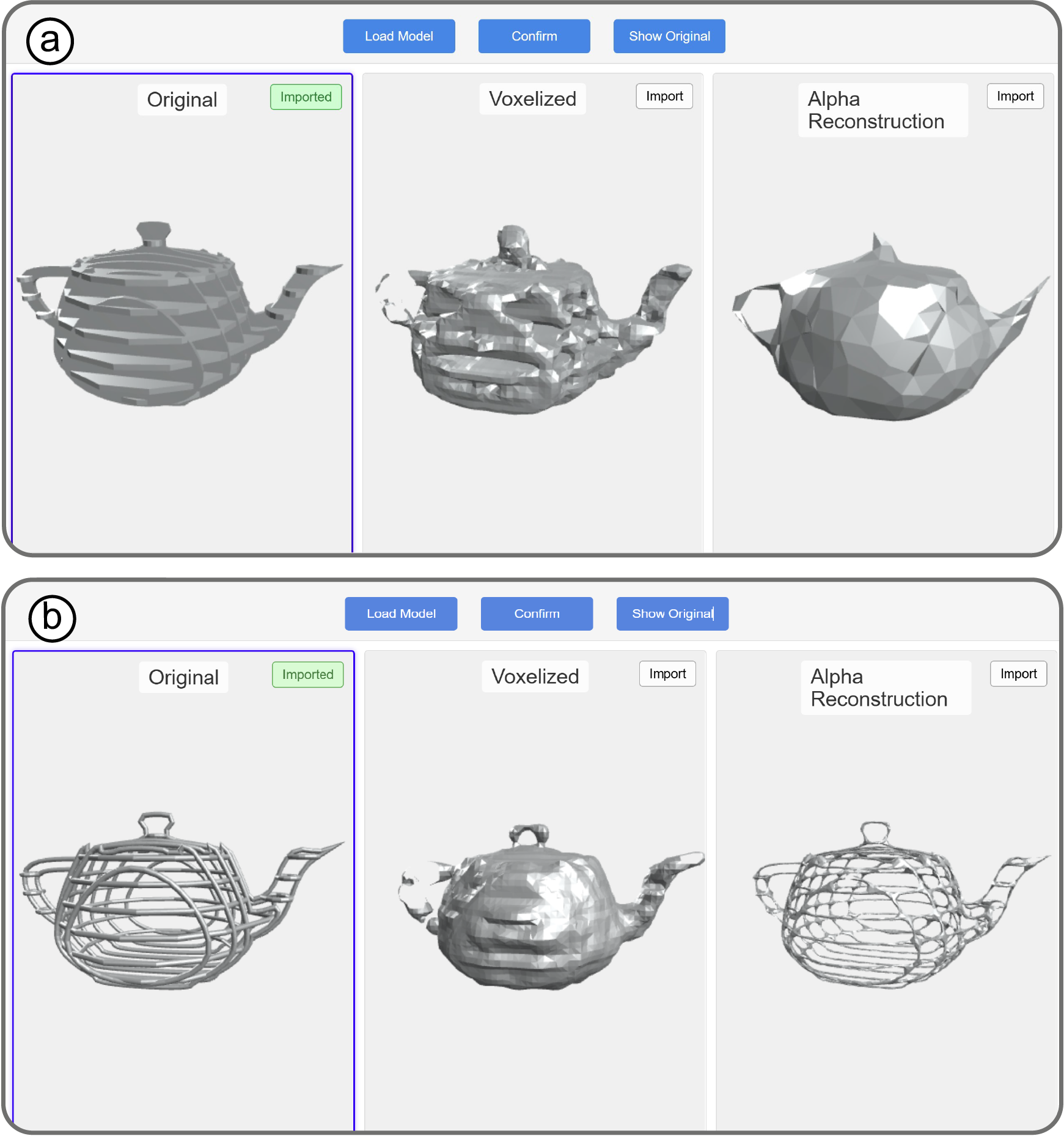}
  \caption{Importing machine-specific models: (a)~Laser-cut interlocking planar structure (Alpha shape reconstruction); (b)~Wire-bending structure surrounding the external form (Voxelization with Poisson surface reconstruction).}
  \label{fig:reconstruction-comparison}
\end{figure}

\subsubsection{Loading Tool}

As illustrated in the walkthrough (Figure~\ref{fig:initial-approach}), the loading tool accepts STL models. When users have designed a model for a specific fabrication workflow outside \name{} (such as a structure of laser-cut plates), \name{}'s import recovery reconstructs a volumetric representation. Two reconstruction methods are supported: Voxelization with Poisson surface and Alpha shape reconstruction, each addressing different structural characteristics.

Figure~\ref{fig:reconstruction-comparison} illustrates the application of these two methods using the Teapot. For a laser-cut interlocking planar structure (Figure~\ref{fig:reconstruction-comparison}~a), Alpha shape reconstruction accurately recovered the inner geometry. For this wire-bending structure (Figure~\ref{fig:reconstruction-comparison}~b), Voxelization with Poisson surface reconstruction produces a smooth recovery of the exterior. The preview of the different imports gives users a sense of what their model will look like under different algorithms.

\subsubsection{Interactive Guide}

\begin{figure}[t]
  \centering
  \includegraphics[width=1\linewidth]{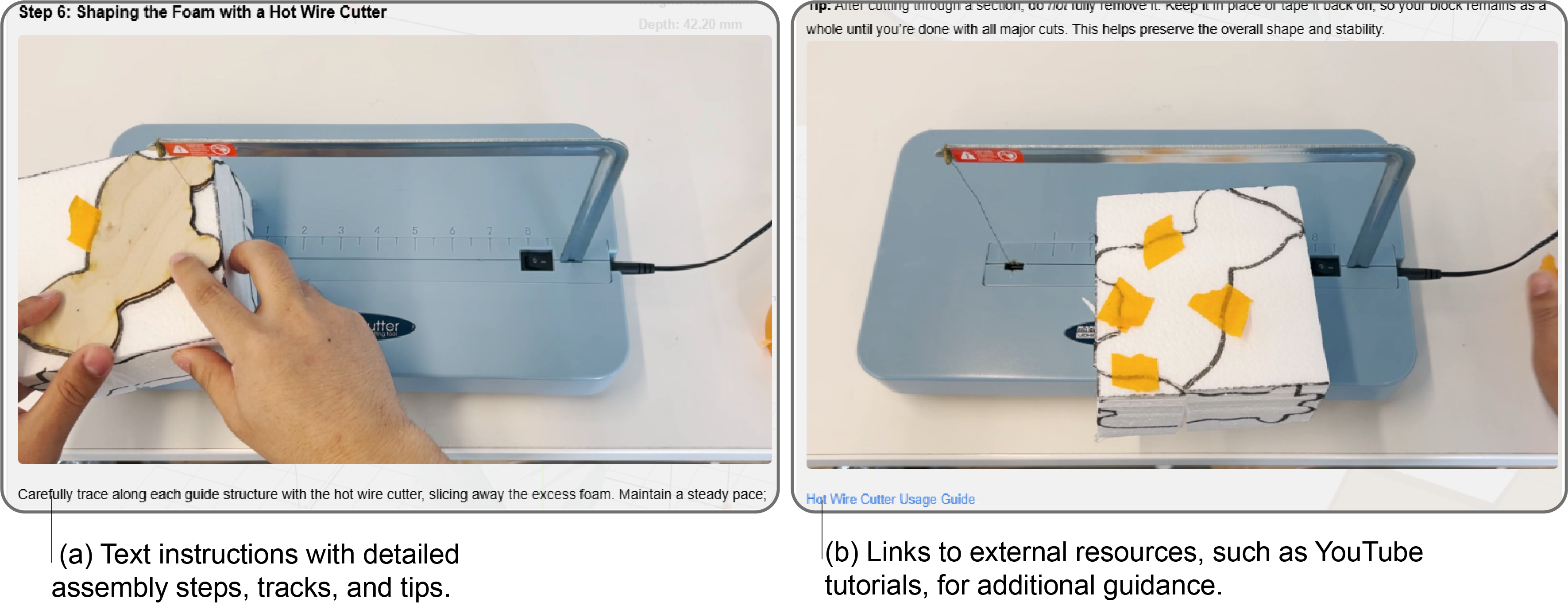}
  \caption{The interactive guide in \name{} showing a step-by-step assembly process. (a)~Text instructions with detailed assembly steps. (b)~And links to external resources.}
  \label{fig:Interactive Guide}
\end{figure}

Figure~\ref{fig:Interactive Guide} shows such guidance; users click through the steps one at a time. (a)~This guide includes textual information, detailing how to assemble each component, with tips highlighted to ensure clarity. (b)~Additionally, the guide includes links to external resources, such as YouTube tutorials, for in-depth guidance on the specific technique or examples of similar processes. These guides are currently not specific to the user's design; they serve as a general explanation of the workflow. Future work could explore how to leverage dynamic rendering techniques to contextualize these instructions for the model at hand.

\subsection{Surveying and Integrating Fabrication Workflows}

Any exploration tool is proportionally valuable to the range and amount of workflows it supports. To understand the space of existing workflows, we explored workflows shared in online communities. We reviewed 150 winning projects from the 'Workshop' category on \textit{Instructables}\footnote{\url{https://www.instructables.com}} (sorted by 'I Made It'). Workflows centered on microprocessor-based products, purely 3D printed items without additional processes, and those focused solely on polishing were excluded to maintain an emphasis on \textit{fabrication workflows}.

\begin{figure}[h]
\centering
\includegraphics[width=1\linewidth]{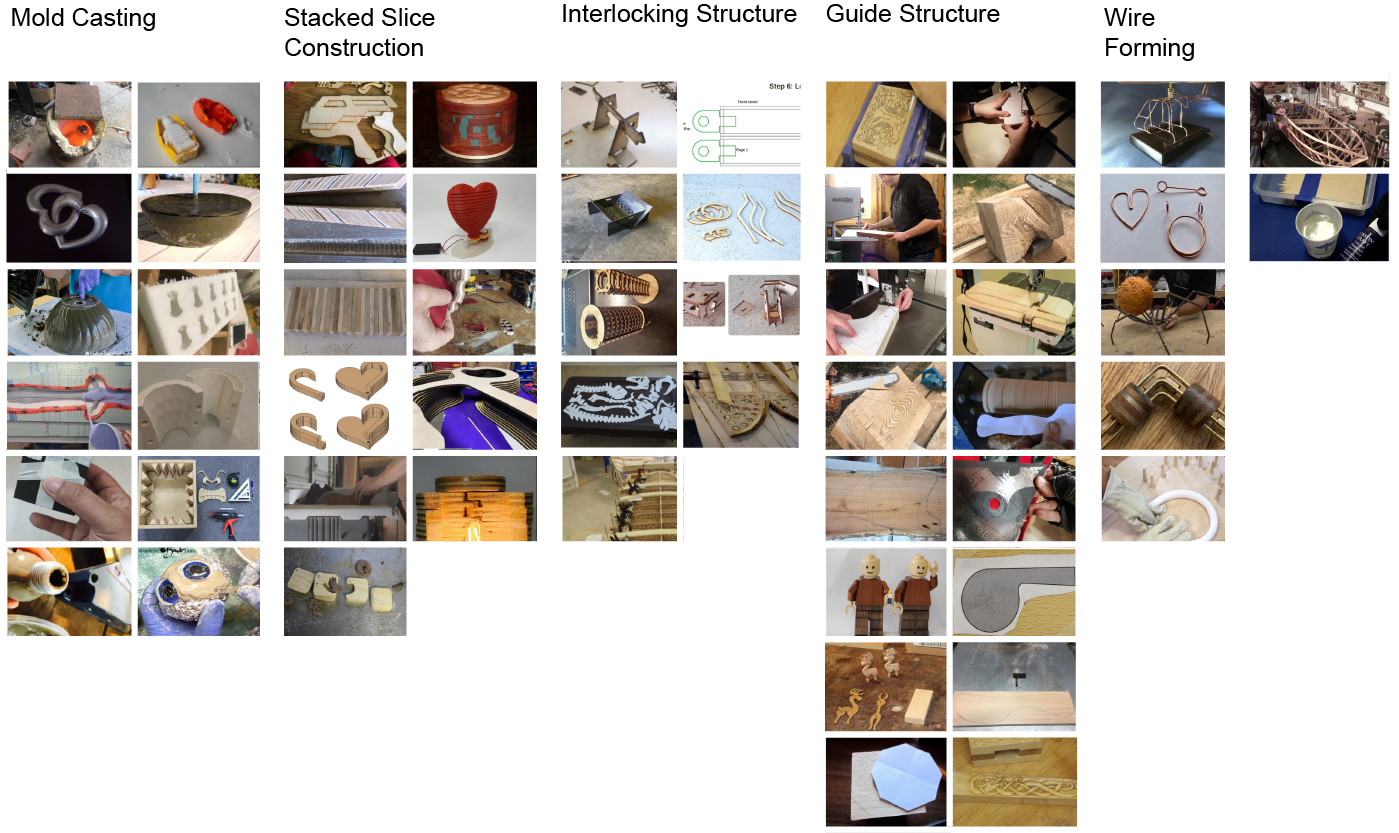}
\caption{An overview of 55 fabrication workflows, organized into five primary categories, with two additional workflows that do not fit any specific category.}
\label{fig:craft-workflows}
\end{figure}

This left us with 55 projects. The appendix contains a list of all workflows used for this process with links to their instruction pages. We clustered these into categories based on their fabrication characteristics to lay out the design space. In Figure~\ref{fig:craft-workflows}, the different processes are shown and divided into the following categories: Mold Casting, Stacked Slice Construction, Interlocking Structure, Guide Structure, and Wire Forming. Two workflows did not fit this categorization; one of them was creating shapes with epoxy resin, and the other was building boats from wooden spars and frames.

\begin{figure}[b]
\centering
\includegraphics[width=1\linewidth]{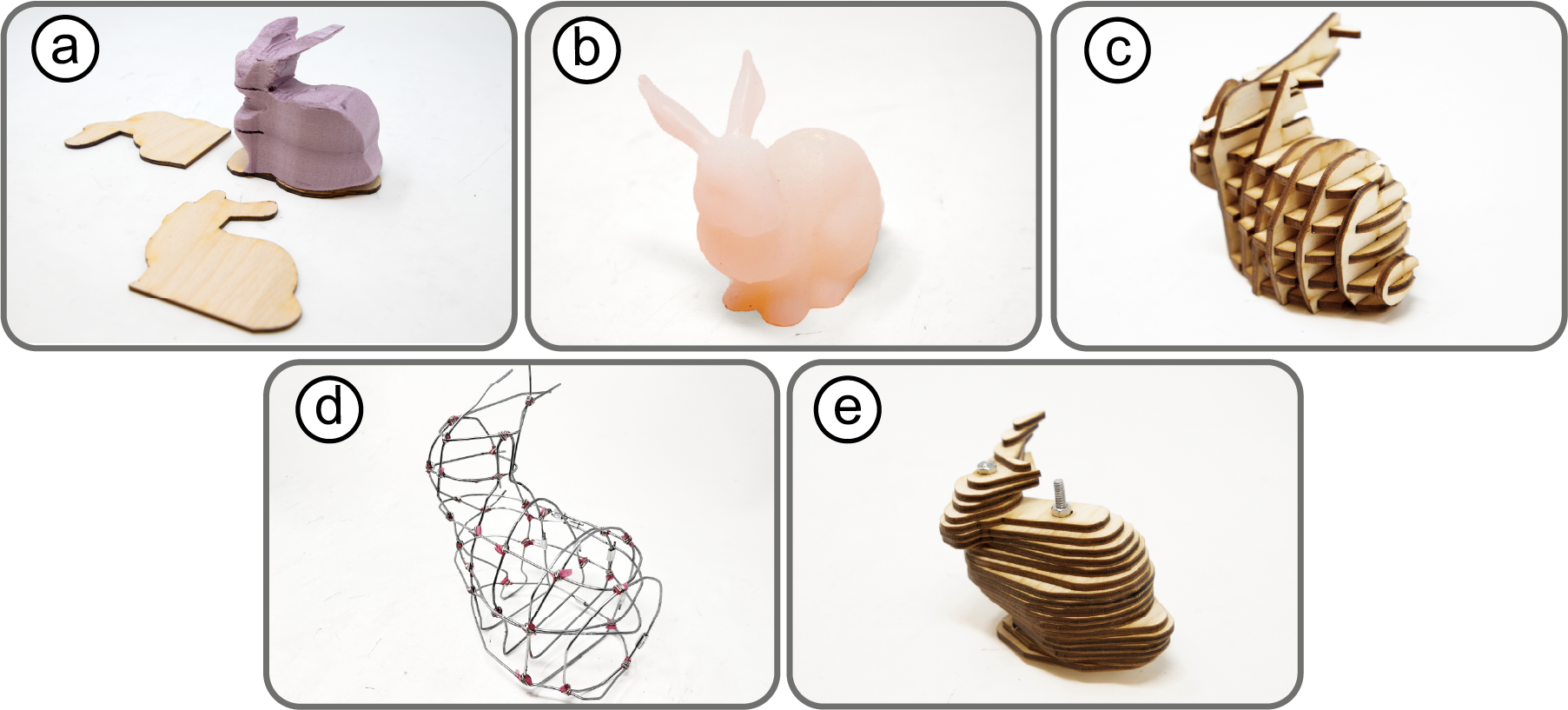}
\caption{Examples of the \textit{representative workflows}. }
\label{fig:foundational-workflows}
\end{figure}

We selected one representative workflow from each category to implement in \name{}, providing learners with diverse fabrication workflows across the spectrum of methods to compare. This selection provides learners with meaningful comparison opportunities between fundamentally different fabrication paradigms (Figure~\ref{fig:foundational-workflows} shows an overview of these workflows applied to the Stanford Bunny). Under the hood, each workflow follows a standardized template comprising four key components: scene initialization to ensure consistent 3D environments, fabrication-specific CAD logic for geometry processing, integrated tutorial guidance, and export functionality for machine-appropriate file formats.

\begin{figure}[h]
\centering
\includegraphics[width=1\linewidth]{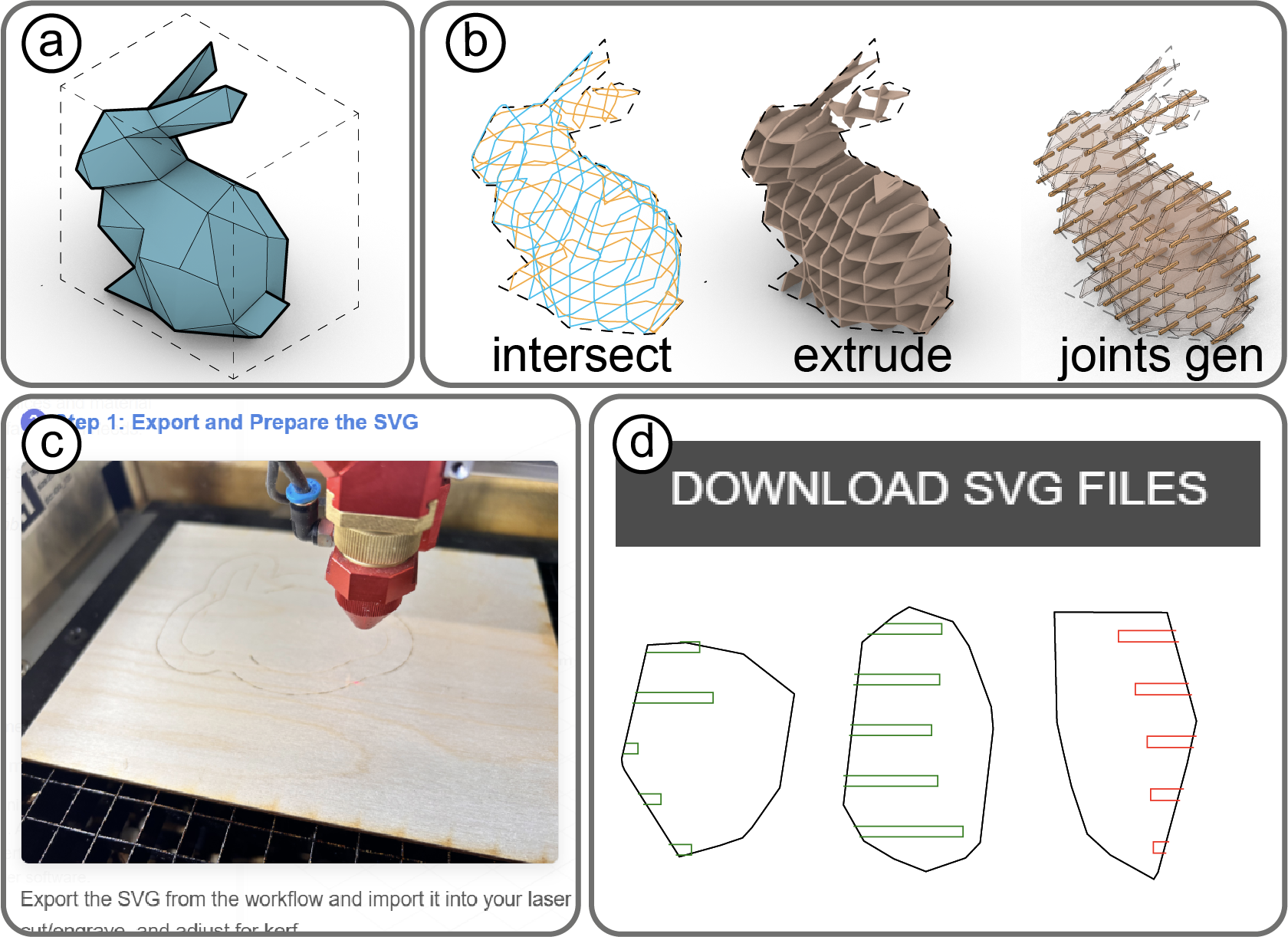}
\caption{Template implementation breakdown using the Laser Cut Interlocking workflow as an example. The four template components are shown with their corresponding code modules: (a) Scene initialization with camera setup and visual helpers, (b) CAD logic modules for slicing and joint generation, (c) Tutorial system with step-by-step guidance, and (d) SVG export functionality.}
\label{fig:template-breakdown}
\end{figure}

Figure~\ref{fig:template-breakdown} illustrates how these template components are instantiated using the Laser Cut Interlocking workflow as an example. The scene initialization layer handles standard 3D setup—configuring the camera, lighting, synchronized viewports, and basic visual helpers such as axes and bounding boxes. The fabrication-specific CAD logic performs a sequence of geometric operations: slicing the mesh into horizontal and vertical contours, generating 3D geometry from these contours, computing intersection points where joints should be placed, and producing the interlocking features accordingly (overview in Table~\ref{tab:core_modules}). The tutorial component presents structured guidance for this workflow, including assembly steps, troubleshooting tips, and safety notes. Finally, the export module produces laser-cutter-ready SVG files optimized for fabrication.

\section{Implementation}
\name{} employs a multi-layer architecture where React manages user interface components and application state, Three.js (using WebGL) renders 3D model visualization, Node.js provides the development runtime environment, and Python handles computationally intensive mesh processing operations.

\begin{figure}[h]
    \centering
    \includegraphics[width=1\linewidth]{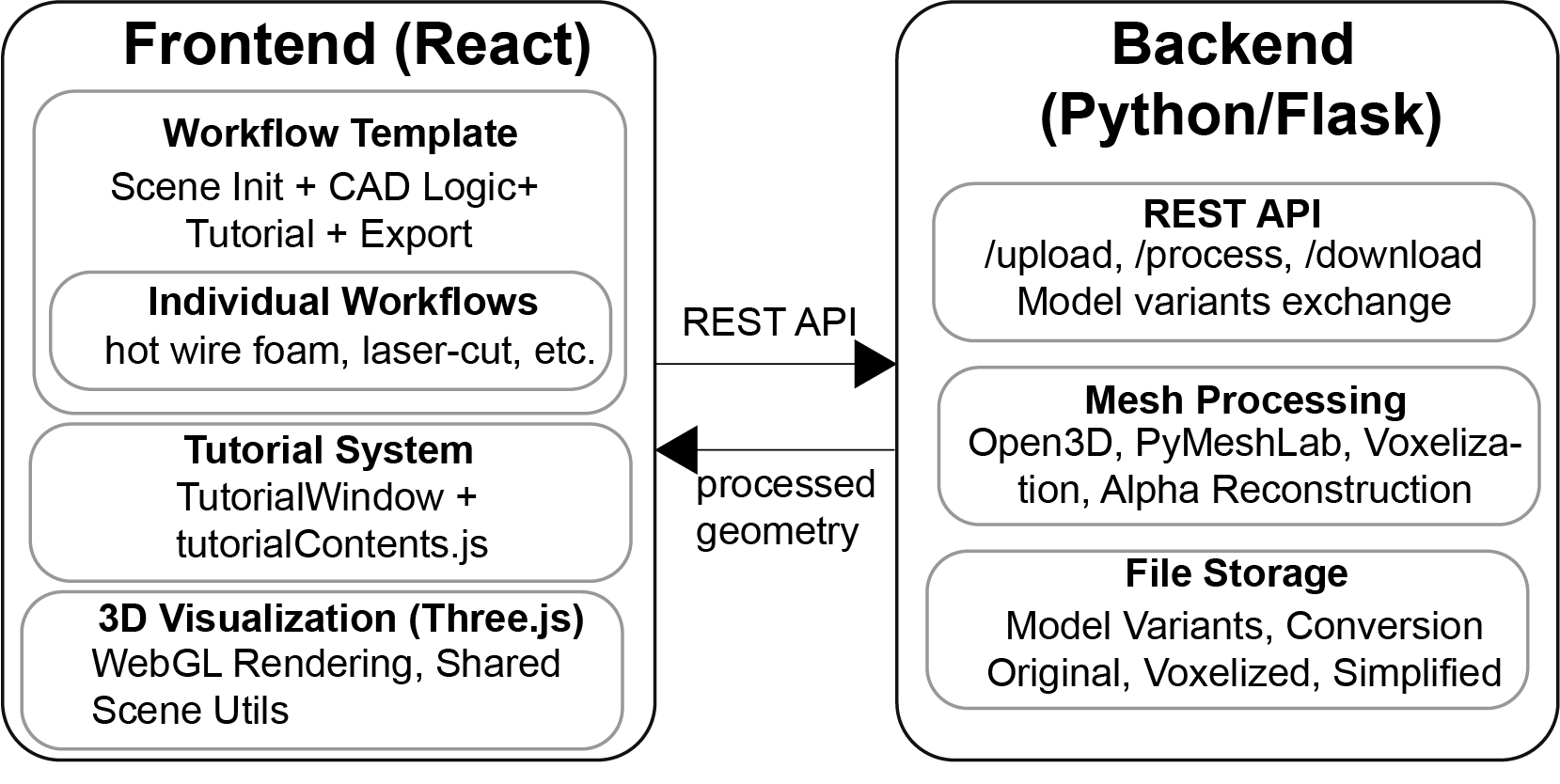}
    \caption{System architecture showing workflow integration in \name{}. Individual workflows inherit from the workflow template and integrate with tutorial components, while communicating with the backend for intensive mesh processing operations.}
    \label{fig:system-architecture}
\end{figure}

While implementing these workflows, we identified recurring computational patterns and interface requirements. For instance, both laser-cut stacked sheets and laser-cut interlocking workflows use \texttt{Intersect} for planar mesh slicing to generate 2D profiles. Mold making and 3D printing both require \texttt{Calc Draft Angles} for fabricability assessment. Based on these findings, we developed a standardized template architecture that offers visual and computational support for fabrication workflows. 

As shown in Figure~\ref{fig:system-architecture}, workflows inherit from this template and interact with the backend for intensive mesh processing tasks.

\subsubsection{Workflow Modules}

\name{} modules are organized into four categories and are listed in Table~\ref{tab:core_modules}, covering the complete end-to-end workflow for 3D visualization and fabrication:

\begin{itemize}
    \item \textbf{Model Loading.} Handles the import and initial processing of 3D models, including loading and geometry adjustments.

    \item \textbf{Geometry Processing.} Performs core computational tasks on 3D geometry, such as intersections, contour simplifications, and projections for specific workflows.

    \item \textbf{Visualization.} Enhances understanding of 3D models through tools like draft angle analysis, dimension labeling, and color-coded feedback for manufacturing validation.

    \item \textbf{Export.} Converts processed models or data into formats such as STL, SVG, or CSV for fabrication workflows.
\end{itemize}

These modular components can be extended and customized by developers to create tailored fabrication solutions.

The step-wise tutorial component represents a critical part of workflow implementation, as it bridges the gap between computational design and physical fabrication. Each workflow’s tutorial content is structured as an array of steps in \texttt{tutorialContents.js}, where each step contains a title, instructional content, optional images or videos, safety tips, and troubleshooting guidance. The \texttt{TutorialWindow} component renders this predefined content, providing users with consistent guidance throughout the fabrication process. This embedded guidance mechanism ensures that users can successfully transition from digital design to physical artifacts.

For computationally intensive operations, such as mesh reconstruction with \textit{voxelization} and \textit{alpha-shape reconstruction}, the front-end communicates with a Python back-end through REST APIs built on Flask. As shown in Figure~\ref{fig:system-architecture}, when a user loads a model, the WorkflowController handles loading and pre-processing before delegating tasks like simplification and reconstruction to the back-end, which uses \textit{Open3D} and \textit{PyMeshLab} for efficient processing.

\subsubsection{Distribution}
The source code for \name{} is now publicly available, providing full access to all implementation details and enabling the community to further research and develop.

%% file: src/07_expert-eval.tex
\section{Formative study to understand the integration of comparison tools in existing curricula}

We conducted a formative study with seven expert fabrication educators to gather feedback on comparison features and how one would integrate this into existing curricula. Our experts have 5-14 years of teaching experience in a range of educational contexts: K-12 schools (n=3), maker spaces (n=2), and universities (n=2).

Each educator participated in a short session in which we demonstrated the current \name{} prototype and asked them to explore a sample model using the comparison interface. After this hands-on exploration, we conducted a semi-structured interview focusing on the clarity of the comparison features and how such capabilities might support or fit within their instructional practice.

\subsection{Refining \name{} Based on Experts' Insights}
Based on experts' insights, we iteratively refined \name{}’s workflow comparison interface. The following sections summarize insights from the interviews and the subsequent refinement of the prototype to better align with educational practice.

\subsubsection{Guided On-boarding}

\begin{figure}[b]
\centering
\includegraphics[width=1\linewidth]{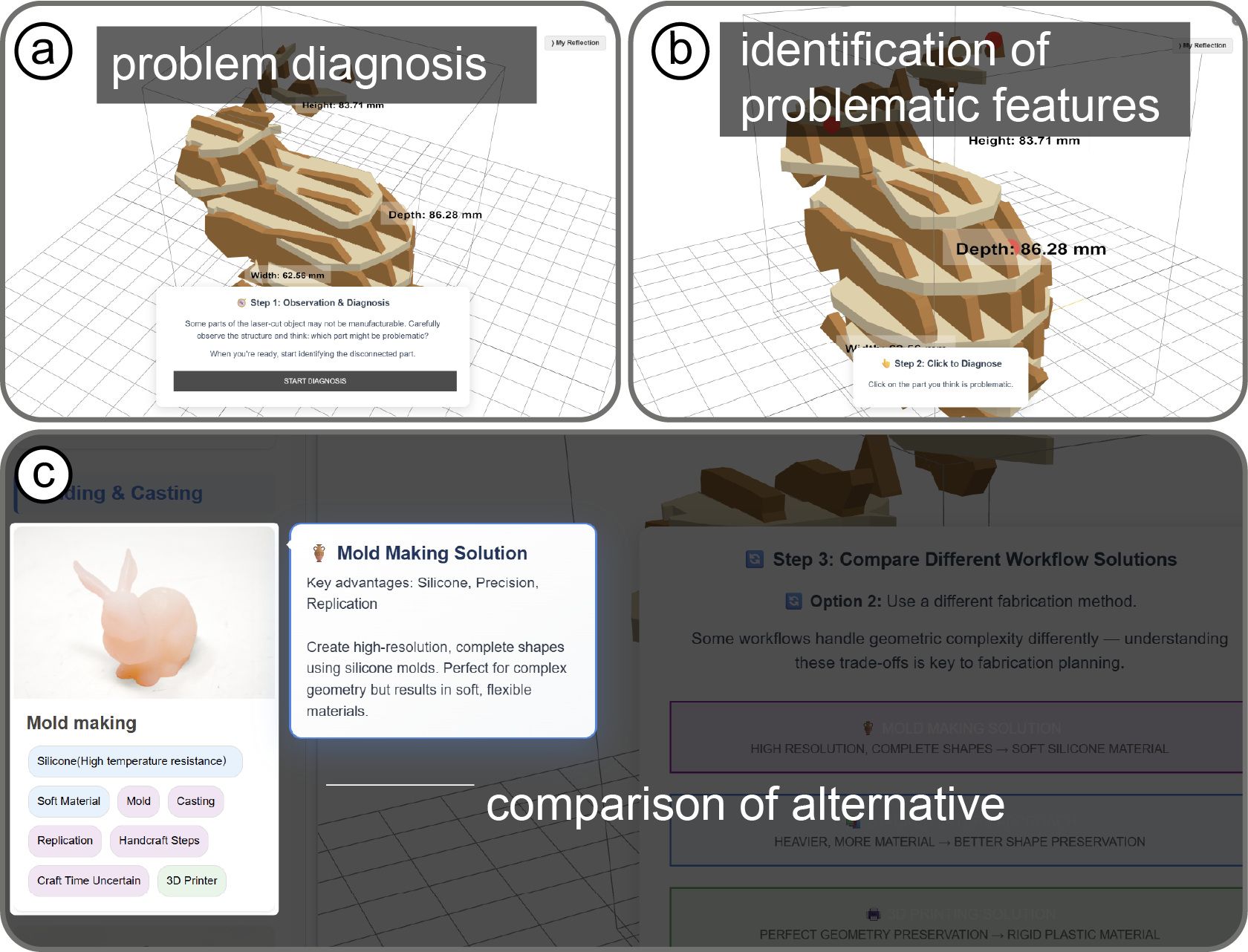}
\caption{Guided on-boarding task in \name{}. A laser-cut interlocking structure with floating parts prompts learners to compare how different workflows handle disconnected geometry through interactive workflow comparison.}
\label{fig:guidedtasks}
\end{figure}

Expert feedback emphasized that first-time users, especially those unfamiliar with computational fabrication, often need structured support to begin comparing workflows meaningfully. E4 described the value of using such tools as self-guided resources (\textit{“a standalone link that allows them to go through it at any point.”}) that students can revisit flexibly. 

To address this, we designed guided onboarding tasks that help learners recognize fabrication challenges and explore alternative solutions through workflow comparison. These onboarding tasks appear when \name{} is first launched (though they can be skipped), presenting interactive challenges that require learners to diagnose problems and compare workflow alternatives. Each task progresses through three stages: observation and diagnosis, identification of problematic features, and comparison of alternative workflows to find suitable solutions. For example, the task shown in Figure~\ref{fig:guidedtasks} presents a laser-cut interlocking structure with floating, disconnected parts. The interface highlights the problematic geometry and prompts:\textit{ “Why does this part fail in laser cutting?”} Upon identification, \name{} suggests alternative workflows such as mold making or wire mesh and presents them in the comparison view.

\subsubsection{Machine and Material Parameter Integration}  

\begin{figure}[t]
\centering
\includegraphics[width=1\linewidth]{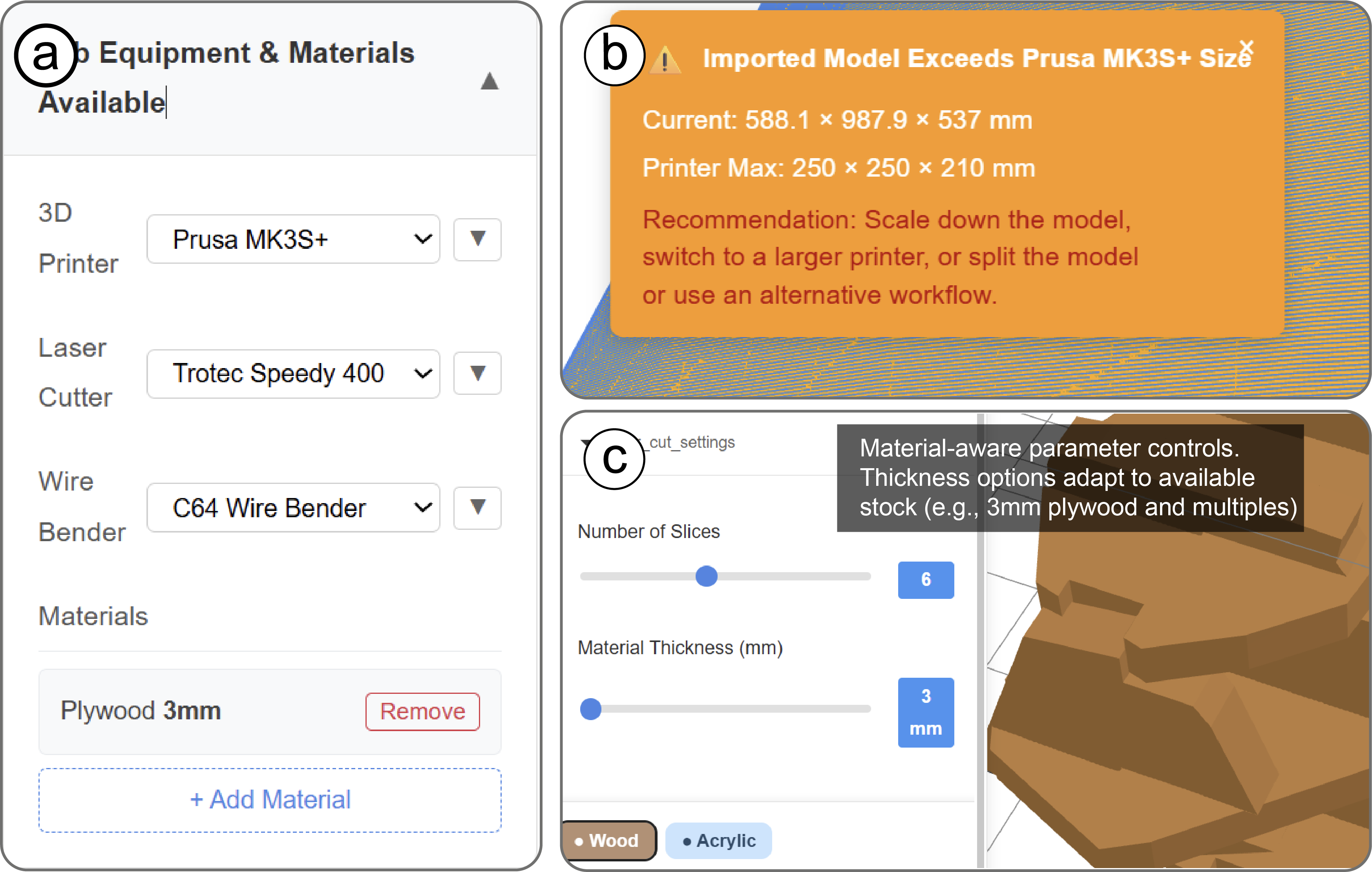}
\caption{Machine and material parameter configuration interface. (a) Dropdown menus for selecting common machines such as \textit{Prusa} or \textit{Ender} printers, each with pre-loaded working dimensions. (b) Warning indicators when workflow outputs exceed selected machine settings. (c) Material presets, aligning generated geometries to realistic stock sizes.}
\label{fig:constraints}
\end{figure}

Educators stressed that workflow comparisons should reflect actual fabrication constraints. E6 noted the need to align workflows with the machines learners have access to: \textit{“You give it like the machines you have access to... then it generates workflows based on that.”}

To help learners compare workflows under realistic fabrication settings, we integrated machine and material parameters into the interface. As shown in Figure~\ref{fig:constraints}a, users can select from a list of common machines such as \textit{Prusa} or \textit{Ender} 3D printers, or standard laser cutters and wire benders. Each machine type comes with its typical working dimensions pre-loaded. When a workflow is dragged into the viewport, \name{} automatically checks these parameters: if a component exceeds the selected machine’s capacity, \name{} displays visual warnings (Figure~\ref{fig:constraints}b) to highlight fabrication limits.  

For materials, we added presets based on available stock in typical fabrication environments. For instance, if a workshop only stocks 3\, mm plywood, the related workflow can be configured to reflect this. When students adjust design parameters for workflows requiring plywood (such as laser-cut interlocking structures), \name{} automatically limits material thickness values, ensuring the models align with the actual material stock available. This prevents learners from designing with unrealistic material specifications and ensures workflow comparisons reflect the real fabrication limits of their specific environment (Figure~\ref{fig:constraints}c). For further integration in common curricula, \name{} lets teachers define a “template” for their course, so all students have access to the same materials, workflows, and machines (and machine settings).

\subsubsection{Feasibility Analysis}

Educators expressed the need for comparison tools to help learners recognize not only what a workflow can produce but also where a design may fail before fabrication begins. E7 emphasized that previews should make such issues visible, noting that tools should \textit{“show what is not working.” }

\begin{figure}[h]
\centering
\includegraphics[width=1\linewidth]{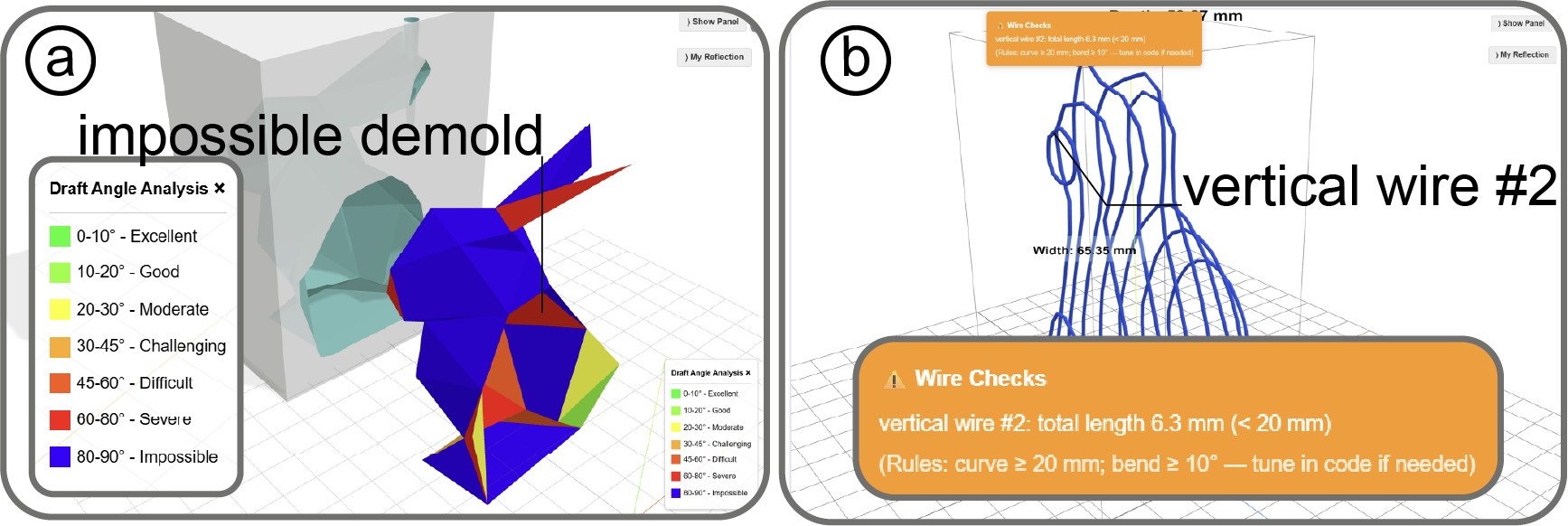}
\caption{Feasibility analysis for different workflows. (a) Mold making draft angle visualization using a color legend. (b) Wire form analysis calculating minimum wire length and bend angle, with warnings for impossible bends.}
\label{fig:feasibility}
\end{figure}

We implemented basic feasibility analysis to help learners see whether their designs might successfully fabricate across workflows, using simple checks tailored to each workflow. Instead of applying a single generic rule, \name{} highlights different feasibility issues depending on the chosen workflow. For example, in the \textit{mold making} workflow (Figure~\ref{fig:feasibility}a), \name{} visualizes draft angles directly on the model with a color legend in the interface. Similarly, for the \textit{wire forming} workflow (Figure~\ref{fig:feasibility}b), \name{} calculates the minimum feasible segment length and bend angle for each wire. If a segment is too short to bend with a standard wire bender, or if a bend angle is sharper than what can realistically be achieved, \name{} warns users.  

The full matrix summarizing the specific feasibility checks implemented across all sixteen fabrication workflows is provided in Figure \ref{fig:feasibility_matrix} in the Appendix for reference.

\subsubsection{Reflective Comparison Support}

Educators highlighted that workflow comparison can serve as an opportunity for reflection rather than only a technical choice. E6 emphasized embedding prompts that foreground goals and criteria: \textit{“Maybe your first thing is like, what are we making today?... when you teach CAD to CAM, it's always like, what's your goal? What are you making?”} E4 similarly suggested assignments where learners revisit the same workflows later in the semester to reflect on how their reasoning has changed with experience. E6 additionally described how reflection occurs across multiple fabrication cycles, noting that learners often “sketch first, then model and put in [the tool], then reflect on laser cutting results, modify design, and 3D print.”

\begin{figure}[h]
\centering
\includegraphics[width=1\linewidth]{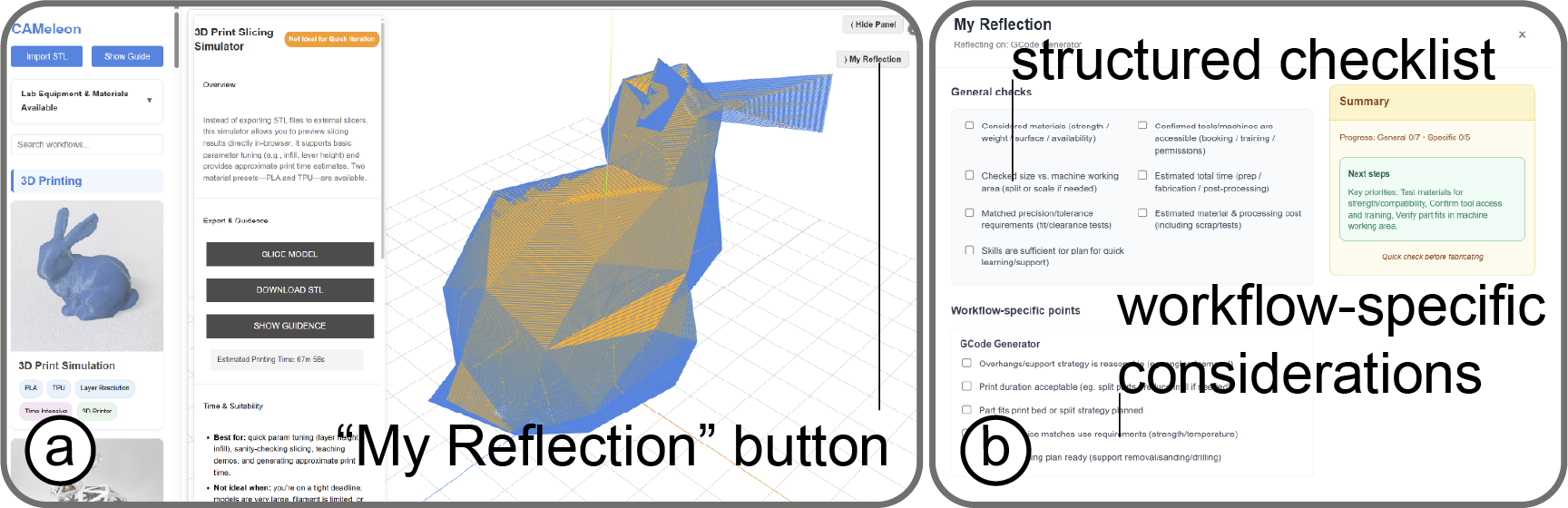}
\caption{"My Reflection" checklist feature. (a) A button on each workflow page opens the reflection tool. (b) The structured checklist shows general considerations (materials, tools, size, time, precision, cost, skills) and workflow-specific questions (e.g., overhang strategies and post-processing for 3D printing workflows, kerf compensation and nesting for laser cutting workflows).}
\label{fig:reflection}
\end{figure}

To help learners clarify their goals when exploring different fabrication workflows, we integrated a lightweight reflection tool into \name{}. Each workflow page includes a “My Reflection” button (Figure~\ref{fig:reflection}a), which opens a structured checklist. The checklist combines general considerations (such as whether time, materials, and machine availability have been considered) with workflow-specific questions tailored to each workflow (Figure~\ref{fig:reflection}b).

For example, in the case of 3D printing workflows, learners are asked to check whether print duration is acceptable, whether the part fits the print bed, whether the material choice matches use requirements, and what post-processing they plan to use.

In line with feature requests from our expert evaluation, \name{} thus supports more deliberate, comparative reasoning when choosing among different fabrication workflows. Learners can revisit these reflections to compare their reasoning across workflows and see how their thinking changes as they gain more experience.

\subsubsection{Expanding the Number of Workflows}

\begin{figure}[h]
\centering
\includegraphics[width=1\linewidth]{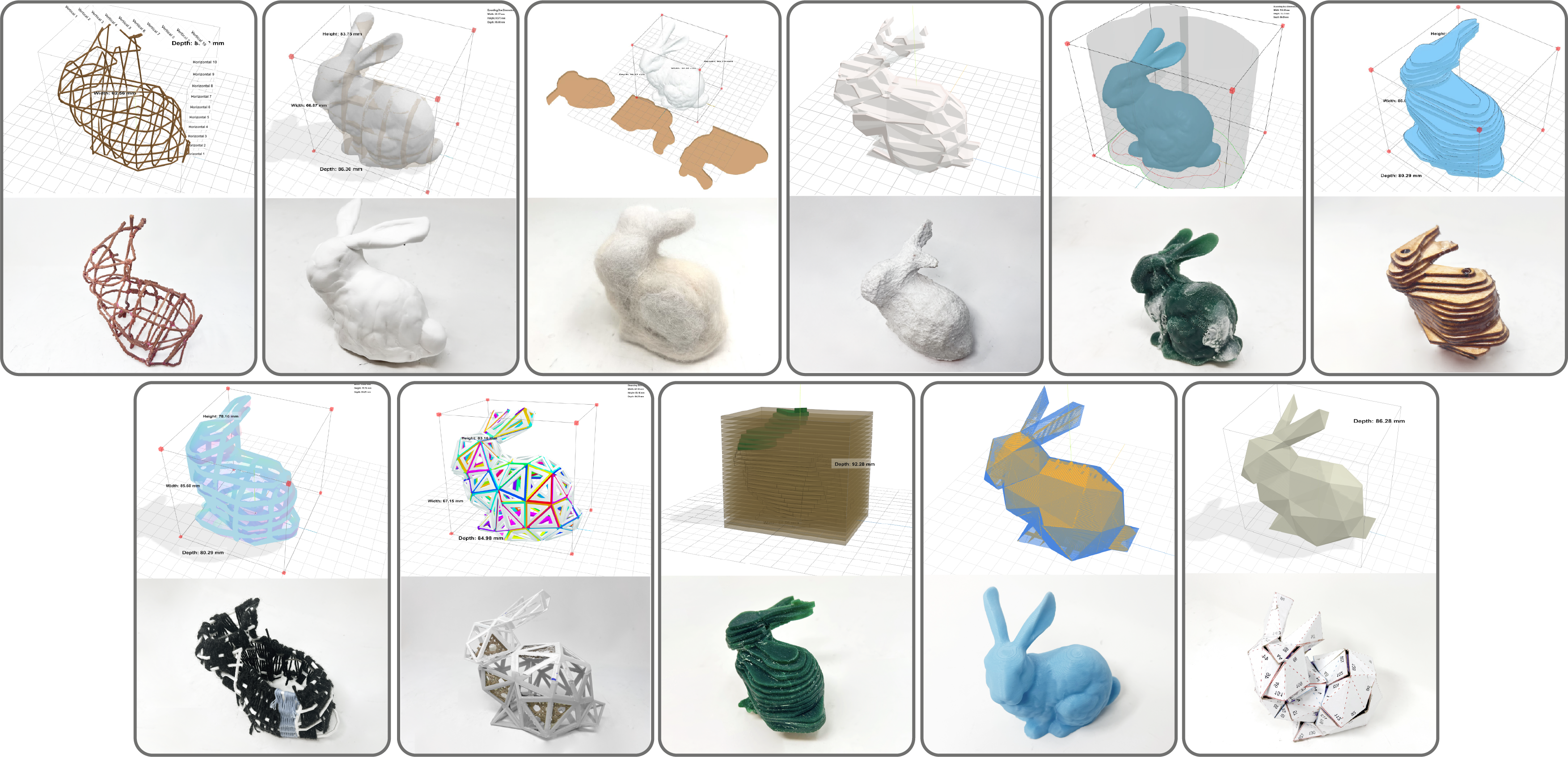}
\caption{Eleven additional workflows added to \name{} based on expert recommendations. These workflows include classroom teaching methods (3D printing, laser-cut clay sculpture, paper mache), low-cost approaches (paper folding), post-processing techniques (wire copper electroplating, epoxy laminating , needle felt sculpture, silicone molding), and workflows from prior research projects (bamboo agent, escape loom, stackmold).}
\label{fig:expanded-workflows}
\end{figure}

Four experts (E4, E5, E6, E7) identified the need for broader workflow coverage to enable meaningful fabrication comparison. We expanded \name{} from five foundational workflows: laser-cut interlocking, mold making, wire forming, stacked layers, and hot wire foam cutting—to sixteen workflows, responding to these expert recommendations (Figure~\ref{fig:expanded-workflows}).

We added workflows based on expert teaching practices and recommendations. E4 suggested incorporating traditional tool fabrication workflows, which led us to identify and implement laser-cut clay sculpture and papier-mache that experts use in arts education. E5 and E6 emphasized the importance of low-cost approaches and specifically recommended paper folding for resource-constrained environments. We included 3D printing since experts noted students often default to this method and requested it as a comparison baseline. Based on expert observations that fabrication workflows extend beyond the core fabrication process, we implemented post-processing workflows, including wire-copper electroplating, epoxy laminating, and needle-felt sculpture. We implemented three workflows from prior research projects: EscapeLoom~\cite{Deshpande_Takahashi_Kim_2021}, StackMold~\cite{Valkeneers_Leen_Ashbrook_Ramakers_2019}, and Bamboo Agents~\cite{Gao_Gao_Yang_Liu_Shi_Li_2023}. These workflows integrate traditional crafts with computational fabrication tools and expand the range of comparison opportunities.

This expansion increases the range of fabrication approaches available for comparison, supporting more comprehensive workflow decision-making in educational settings. It also demonstrates the modularity of \name{}'s architecture to allow it to be extended to a broad range of workflows in future iterations.

The comprehensive implementation details, including all tunable parameters, machine requirements, and specific notes for all sixteen workflows, are summarized in Appendix Table \ref{tab:final-summary}.

%% file: src/09_user-eval.tex
\section{User Study: The Value of Guided Comparison in CAD}

\begin{table*}[h]
\caption{Participant demographics and self-reported fabrication experience.}
\label{tab:participants}
\begin{small}
\begin{tabular}{llllll}
\toprule
\textbf{ID} & \textbf{Gender} & \textbf{Age} & \textbf{Background} & \textbf{Machines Known} & \textbf{Fabrication Level} \\
\midrule
P1  & Female & 24 & Architecture (UG)     & Laser Cutter, 3D Printer   & Novice \\
P2  & Female & 25 & Architecture (Grad) & Laser Cutter, 3D Printer, CNC, Vinyl Cutter & Advanced \\
P3  & Male   & 18 & Mech. Engineering (UG) & Laser Cutter, 3D Printer   & Advanced \\
P4  & Male   & 29 & Product Mgmt  & None                        & Novice \\
P5  & Female & 27 & CS (Grad)            & 3D Printer                 & Intermediate \\
P6  & Male   & 22 & Architecture (UG)      & Laser Cutter, 3D Printer, CNC & Intermediate \\
P7  & Female & 26 & Marketing (PhD)        & None                        & Novice \\
P8  & Male   & 24 & Computer Vision (PhD)  & 3D Printer                 & Novice \\
P9  & Male   & 26 & Biomedical Eng. (PhD)  & None                        & Novice \\
P10 & Male   & 24 & Materials Sci. (PhD)   & Minimal (3D Printer / Laser Cutter) & Intermediate \\
P11 & Female & 28 & Landscape Arch. (Grad)  & Laser Cutter               & Novice \\
P12 & Female & 27 & Urban Tech (Grad)      & 3D Printer (workshop only) & Novice \\
\bottomrule
\end{tabular}
\end{small}
\end{table*}

To develop an initial understanding of the role of comparing fabrication outcomes in CAD workflows, we conduct a lab-based user study using our interface. The objective of this study is to ensure the utility and value of the idea of comparison. Resulting findings, therefore, serve as suggestions for design, observations around the usage and role of comparison through our prototype, and initial qualitative reflection on decision-making in CAD. They do not evaluate the specific implementation of \name{} or its impact on learning, which would require a longer-term deployment study.

In this study, 12 students planned fabrication workflows for given CAD models with and without guided comparison of \name{}, in a controlled setting. We measure three aspects of participants’ planning behavior: (1) the workflows they considered and selected before and after using guided comparison, (2) their exploration behaviors captured through interaction logs, and (3) their self-reported confidence and reasoning preferences from survey responses.

\subsection{Participants}
We recruited 12 students (6f/6m; mean age=25.0, SD=3.0) from our university. Table~\ref{tab:participants} summarizes participant demographics and experience levels. None had used \name{} before. We assigned each participant an experience level (novice, intermediate, or advanced) based on self-reported experience with digital fabrication machines.

\subsection{Task Materials}

\begin{figure}[b]
    \centering
    \includegraphics[width=1\linewidth]{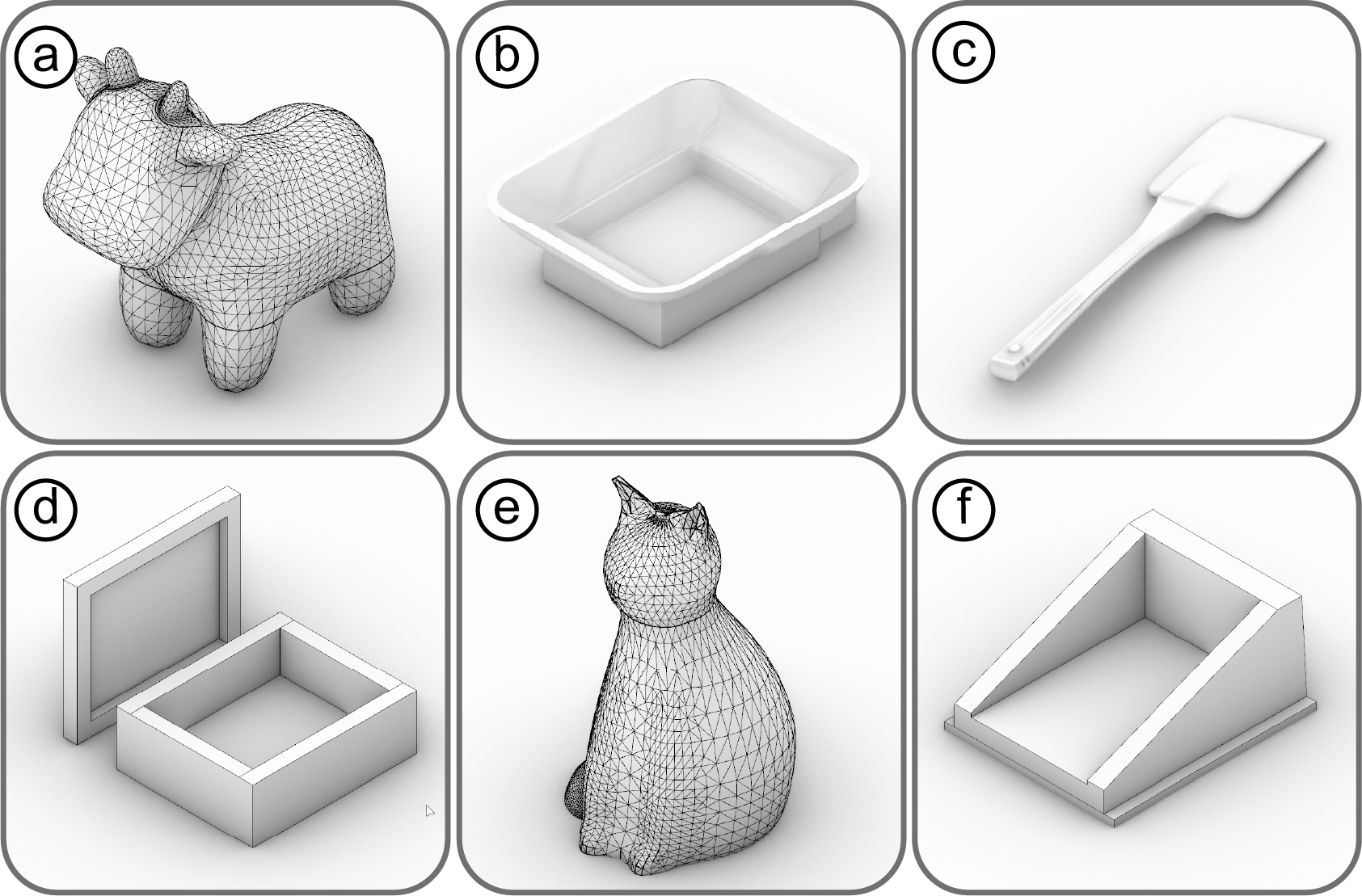}
    \caption{Models used in the study: (a) Decorative cow sculpture, (b) Jewelry tray, (c) Heat-resistant spatula, (d) Transparent display box, (e) Large cat sculpture, (f) shelf-box.}
    \label{fig:study_models}
\end{figure}

Participants completed the same fabrication planning task twice: once before using \name{} to establish a baseline (unsupported condition), and once again with \name{} (\name{} condition). To evaluate both aesthetic and technical considerations, we provided users with the CAD models depicted in Figure~\ref{fig:study_models}.

\subsubsection{Procedure}
Each session lasted approximately 1 hour and followed a four-step protocol. During the study session, an experimenter was present to resolve technical issues and prompt participants to think aloud throughout the session.

\paragraph{Phase 1: Pre-Study Assessment (10 min). }

Participants described their fabrication workflow choices for the models and explained their rationale. To reduce ordering and learning effects, the assignment of models (cow or tray) was counterbalanced across subjects.

\paragraph{Phase 2: Learning with \name{} (30 min). }

Participants explored fabrication workflows using \name{}. At the beginning of this phase, the researcher briefly introduced the tool’s interface and features. Participants then completed the following tasks:

\begin{itemize}
    \item \textbf{Completion problems.} Participants used \name{} to explore workflows to fabricate the following items, each with a specific technical requirement (using Figure~\ref{fig:study_models} models c-e):
    
    (1) A heat-resistant spatula that can withstand high temperatures during cooking.  
    
    (2) A transparent display box that has clear surfaces.  
    
    (3) A large cat sculpture with a built volume larger than that of typical 3D printers in the market.  

    \item \textbf{Open-ended exploration.} Participants were asked to invent their own intended use case for a shelf-box model and then explore which workflows would be appropriate for that purpose while thinking aloud.
\end{itemize}

During the open-ended task, we logged participants’ interaction behaviors, capturing how many workflows they accessed, the sequence and patterns of their selections, their use of filtering features, and the total time they spent on individual workflows.

\paragraph{Phase 3: Post-Study Assessment (10 min). }

Participants completed the same task again, this time using \name{}, allowing us to compare the kinds of workflows they described with and without support in comparison in their CAD interface.

\paragraph{Phase 4: Semi-Structured Interview (10 min).}

Participants filled out a survey. The survey consisted of six 5-point Likert-scale questions to assess how confident participants felt about their reasoning. It also contained four categorical and multiple-choice questions: whether the number of considered workflows increased, and which types of information were most helpful, such as time, equipment, difficulty, and material properties. 

Following the survey, we conducted a semi-structured interview to dig deeper into the reasoning and experience of using an interface focused on comparison. In the interview, we examined the following: 1) how participants decided which workflows to explore; 2) what surprised them during exploration; 3) how guided comparison changed their fabrication planning or thinking; and 4) where they could have used a tool like this in the past.

\subsection{Data Analysis}
We analyzed three data sources to characterize participants’ reasoning and exploration behaviors.

\textbf{Survey Responses.} We summarized the surveys using descriptive statistics to provide an overview of participants’ responses and to highlight patterns observed in the collected data. 

\textbf{Qualitative Data.} We used inductive thematic analysis of the pre-/post-rationales, think-aloud, and interview transcripts. Following a semantic approach, one researcher conducted open coding and created initial codes that captured distinct observations about participants' reasoning patterns and decision-making behaviors (e.g., "defaults to familiar method," "mentions material properties," "expresses uncertainty about durability," "compares time vs. quality trade-offs," "rejects workflow due to assembly complexity"). These codes were iteratively refined, merged, and reorganized through two major rounds of analysis: the first round consolidated semantically similar codes (e.g., merging "checks if transparent," "wants clear surfaces," and "needs light transmission" into "transparency considerations"), and the second round organized these into broader thematic categories (e.g., "workflow exploration breadth," "decision criteria evolution"). Throughout this process, we revisited earlier codes to check internal consistency and ensure that themes captured patterns visible across participants. The resulting themes aim to capture shifts in fabrication planning from the unsupported condition to the \name{} condition.

\textbf{Interaction Logs.} We examined the workflow exploration logs to characterize participants’ strategies. We measured breadth using the number of distinct workflows accessed, depth as repeated visits to the same workflow, and comparison patterns as the sequence of switching between similar or contrasting workflows.  

\begin{table*}[h]
\renewcommand{\arraystretch}{1.1}
\centering
\footnotesize
\setlength{\tabcolsep}{2pt}

\caption[Pre $\rightarrow$ Post workflow changes]{Pre $\rightarrow$ Post workflow changes with separated columns for Cow/Tray, counts, and chosen workflows. Participant IDs include experience level indicators: (N)=Novice, (I)=Intermediate, (A)=Advanced.}
\label{tab:experience_changes}
\begin{tabularx}{0.7\linewidth}{c 
 >{\centering\arraybackslash\bfseries}p{1.6cm} 
 >{\arraybackslash}p{1.5cm} 
 >{\arraybackslash}p{1.7cm} 
 >{\arraybackslash}p{1.5 cm} 
 >{\arraybackslash}X}
\toprule
\textbf{ID} & \textbf{Cow/Tray} & \textbf{unsupported condition: \#~workflows} & \textbf{unsupported condition: chosen workflow} &  \textbf{\name{} condition: \#~workflows} & \textbf{\name{} condition: chosen workflow} \\

\midrule
\rowcolor{gray!8}
P1(N)  & Cow  & 2 & 3D printing                    & 2 & Wire Mesh \\
\rowcolor{gray!8}
   & Tray & 1 & 3D printing                   & 1 & Casting \\

   P4(N)  & Cow  & 1 & Clay molding                   & 1 & Wire mesh \\
   & Tray & 1 & Mold pressing                 & 2 & Silicone mold \\

   \rowcolor{gray!8}
P7(N)  & Cow  & 1 & Paper folding                  & 3 & Wire mesh \\
\rowcolor{gray!8}
   & Tray & 1 & Paper folding                 & 3 & 3D printing \\

   P8(N)  & Cow  & 1 & Multi-part 3D~print + glue     & 2 & Hot-wire foam cutting \\
   & Tray & 1 & 3D~printing                   & 1 & Epoxy laminating  \\

\rowcolor{gray!8}
P9(N)  & Cow  & 2 & Wood carving                   & 3 & Stacked wood sections/slices \\
\rowcolor{gray!8}
   & Tray & 2 & Wood carving                  & 4 & Hybrid (stacked + clay coat) \\

P11(N) & Cow  & 1 & 3D printing                    & 3 & Hybrid (lasercut \& papermache) \\
   & Tray & 1 & 3D printing                   & 3 & Silicone molding \\

   \rowcolor{gray!8}
P12(N) & Cow  & 2 & Laser cutting                  & 3 & Molding \\
\rowcolor{gray!8}
   & Tray & 1 & 3D printing                   & 3 & Stack mold \\

P5(I)  & Cow  & 2 & 3D~printing                    & 3 & Felt \\
   & Tray & 3 & 3D~printing                   & 2 & 3D printing \\

\rowcolor{gray!8}
P6(I)  & Cow  & 1 & 3D~printing                    & 3 & Hybrid (laser-cut inside + fur) \\
\rowcolor{gray!8}
   & Tray & 2 & Hybrid (lasercut + 3D~print)  & 3 & Hybrid (lasercut + molded) \\

   P10(I) & Cow  & 2 & 3D~printing                    & 3 & Wire mesh \\
   & Tray & 2 & Laser cutting                & 4 & Silicone mold \\

\rowcolor{gray!8}
P2(A)  & Cow  & 3 & 3D printing                    & 3 & 3D printing \\
\rowcolor{gray!8}
   & Tray & 4 & 3D printing                   & 6 & Epoxy/Resin/Silicone \\

P3(A)  & Cow  & 2 & 3D printing                    & 3 & Wire mesh \\
   & Tray & 2 & 3D printing                   & 1 & 3D printing \\

\bottomrule
\end{tabularx}
\end{table*}

\subsection{Results}
We divide our results into three categories: how participants described their decision-making with and without guided comparison; exploration strategies based on interaction logs; and, finally, the subjective experience of participants based on interviews.

\subsubsection{Shifts in Workflow Considerations and Reasoning}

We analyzed participants’ responses to identify changes in their fabrication decisions before and after using \name{}. Table \ref{tab:experience_changes} presents the workflows they considered and selected.

\textbf{Expanded Exploration.} Participants considered more fabrication workflows when using guided comparison. In the unsupported condition, participants considered an average of 1.67 workflows for the cow (SD = 0.65) and 1.83 for the tray (SD = 0.94), totaling 3.50 workflows (SD = 1.45). In the \name{} condition, this increased to 2.67 for the cow (SD = 0.65) and 2.75 for the tray (SD = 1.48), totaling 5.42 workflows (SD = 1.98). A paired-samples t-test confirmed this increase was statistically significant, t(11) = 4.60, p < .001, Cohen's d = 1.33. Nine participants increased their workflow considerations. Eleven participants changed their final selection for at least one object, with eight participants changing selections for both objects. 

Notably, the few cases where participants retained their original choices were concentrated among intermediate (P5) and advanced users (P2, P3), with more nuanced reasoning. For instance, P2 (advanced) maintained 3D printing for the decorative cow, but shifted from convenience-based reasoning (\textit{“super easy”}) to material-property considerations involving transparency, durability, and environmental context. P3 (advanced) retained 3D printing for the jewelry tray, but evolved from a technical specification focus (\textit{“ABS, stronger for daily use”}) to comparative evaluation that explicitly weighed durability against alternative materials (\textit{“a lot of these materials, it wouldn't last. I think 3D printing is my safest option”}). 

Participants listed more workflows when using guided comparison. In the unsupported condition, participants tended to concentrate on familiar methods: 3D printing accounted for 10/24 total selections, while laser cutting (2/24) and various manual methods filled in the remainder—highlighting participants’ reliance on workflows that were most familiar and readily accessible. In the \name{} condition, participants explored a broader range of workflows: wire-based fabrication emerged as the most striking example—absent from initial considerations yet selected by four participants (P1, P3, P4, P7, P10). Similarly, molding and casting, as well as hybrid workflows, expanded from 1 to 4 selections each.

\textbf{Novice users} (P1, P4, P7, P8, P9, P11, P12) showed the largest shifts, often replacing initial choices entirely. P4 adopted new approaches for both objects, while P8 expanded from shape-based reasoning to considering weight and safety: \textit{“I still think it’s a good toy for kids, so it being lightweight is very important.”} \textbf{Intermediate users} (P5, P6, P10) moved beyond familiar defaults. P5 experimented with felt to balance aesthetics and feasibility, while P10 combined wire bending with silicone molding to improve durability. \textbf{Advanced users} (P2, P3) layered hybrid fabrication methods on top of their original option, with P2 describing \name{} as a \textit{“dictionary”} that expanded their design vocabulary.

A recurring pattern, especially among novices like P9, was \textit{assumption → contradiction → revision}. P9 initially assumed that laser-cut stacked wood would be a good option for food-related objects, but revised their plan upon realizing it involved adhesives: \textit{“I thought of wood, but layer-by-layer glue might not be food safe.”} They expressed preference for carving from a solid block—\textit{“Because it contains no glue… but I don’t see that option here”}—before acknowledging that this option was not available. They ultimately selected silicone molding for its durability and heat resistance. 

\begin{figure*}[h]
\centering
\includegraphics[width=1\textwidth]{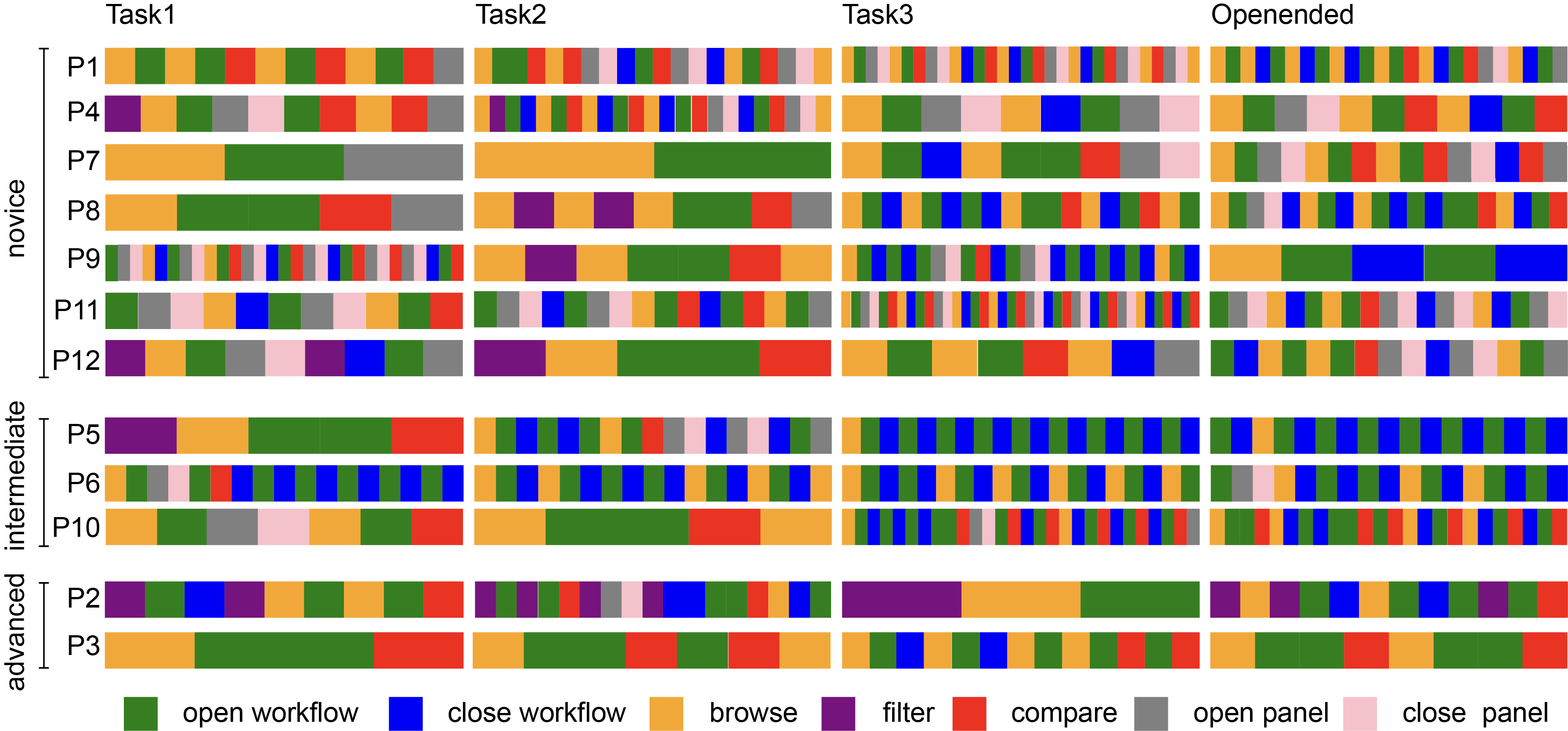}
\caption{Raw interaction data for all participants across tasks. Colors indicate different interaction types (open workflow, browse, filter, compare, open panel, close panel). Participants are grouped by experience level: novices (P1, P4, P7, P8, P9, P11, P12), intermediates (P5, P6, P10), and advanced users (P2, P3).}
\label{fig:task_strategies}
\end{figure*}

\textbf{Differences Between Feasibility-Focused and Goal-Oriented Considerations.} After using \name{}, participants’ explanations broadened from focusing on feasibility or familiarity to multiple criteria and goal-aligned evaluations. Post-use accounts explicitly weighed time/effort, material properties, assembly complexity, equipment/size constraints, durability, and aesthetics. 

In participants’ post-study interviews, more references to goal-related criteria appeared alongside feasibility considerations. (Table~\ref{tab:experience_changes}). P1 (novice), who initially defaulted to 3D printing, ultimately chose wire-bending for the decorative cow and casting for the tray to access more material choices. P3 (advanced) described \name{} as \textit{“eye-opening”}: for the cow, they moved from 3D print to wire mesh as a more striking \textit{“talking piece,”} while still preferring 3D printing for the jewelry tray due to everyday durability. 

\textbf{Use of Conditional and Context-Dependent Considerations.} Several participants used conditional \textit{"if…then…"} reasoning in their post-study explanations. For example, P2 demonstrated this sophisticated approach: \textit{“If I need transparency, I'll use epoxy/resin; if I want softness, silicone.”} P8 showed similar conditional thinking, prioritizing weight and the ability to produce complex geometry: \textit{“I'd prefer hot-wire foam cutting...it's very lightweight and still handles complex shapes”}—over their initial plan to 3D print in parts.

In their post-study interview, some participants referenced object function when explaining their choices. For the cow model, selections emphasized visual appeal: P3 chose wire mesh as \textit{“the coolest one”}, while P4 noted, \textit{“I'm seeing it more as an artwork. So I didn't bother about durability.”} For the tray, decisions focused on practical requirements like durability and material properties.

\subsubsection{Exploration Strategies by Experience Level}

Figure~\ref{fig:task_strategies} demonstrates how exploration patterns varied widely. Advanced participants explored different workflows in unique ways: P2 relied heavily on filtering to narrow options, whereas P3 emphasized open workflow inspections and multi-workflow comparisons with minimal filtering. Novice participants also varied in their approaches, ranging from minimal interactions (e.g., P7 in Task 2: 2 interactions; Task 1: 3 interactions) to extended exploration sequences (e.g., P9 in Task 1: 29 interactions; P11 in Task 3: 39 interactions). While Tasks 1-3 showed extreme participant differences, the open-ended task showed more balanced sequence lengths, suggesting that participants' exploration patterns became more consistent. Across later tasks, several novices (P1, P7, P8, P11) used comparisons more frequently. This pattern may reflect that some novices relied more heavily on comparison features during their exploration.

Despite this diversity, several consistent patterns emerged across all participants. A \textbf{two-phase progression} was evident throughout: participants typically began with \textit{filtering and browsing} actions during initial exploration, then transitioned to \textit{multi-workflow comparisons} in later stages. Advanced participants like P3 frequently eliminated workflows after brief inspection phases. For example, P3 ruled out options by aligning keywords with requirements (\textit{"3D printing would deform under heat... clay sculpture wouldn't withstand impact"}), illustrating how P3 referenced material constraints when excluding certain options. In contrast, novices often spent more time inspecting multiple workflows and panels. Longer exploration sequences consistently included more \textit{open panel} actions, indicating participants' increasing reliance on detailed workflow information as exploration deepened, with these panels serving primarily for in-depth understanding rather than initial exploration.

Intermediate participants (P5 and P6) used strategies that resembled both novice and advanced approaches. Notably, in Task 3 and the open-ended task, they rarely conducted large-scale comparisons. Instead, they adopted sequential workflow inspection before making final decisions. This pattern suggests a \textbf{transition from parallel to sequential exploration} as participants gained familiarity with available workflows. Sequential exploration often displayed a layered structure. Participants inspected several workflows consecutively before concentrating on the panels, confirming information panels served as summary steps rather than continuous reference points. Several participants (P9, P11) displayed clear mid-sequence transitions, beginning with open workflow and browsing actions before shifting to dense clusters of comparisons and panel reading, showing a shift from initial exploration actions to more comparison- and panel-oriented actions later in the sequence.

\subsubsection{User Perceptions and Survey Responses}
\begin{figure}[h]
    \centering
    \includegraphics[width=1\linewidth]{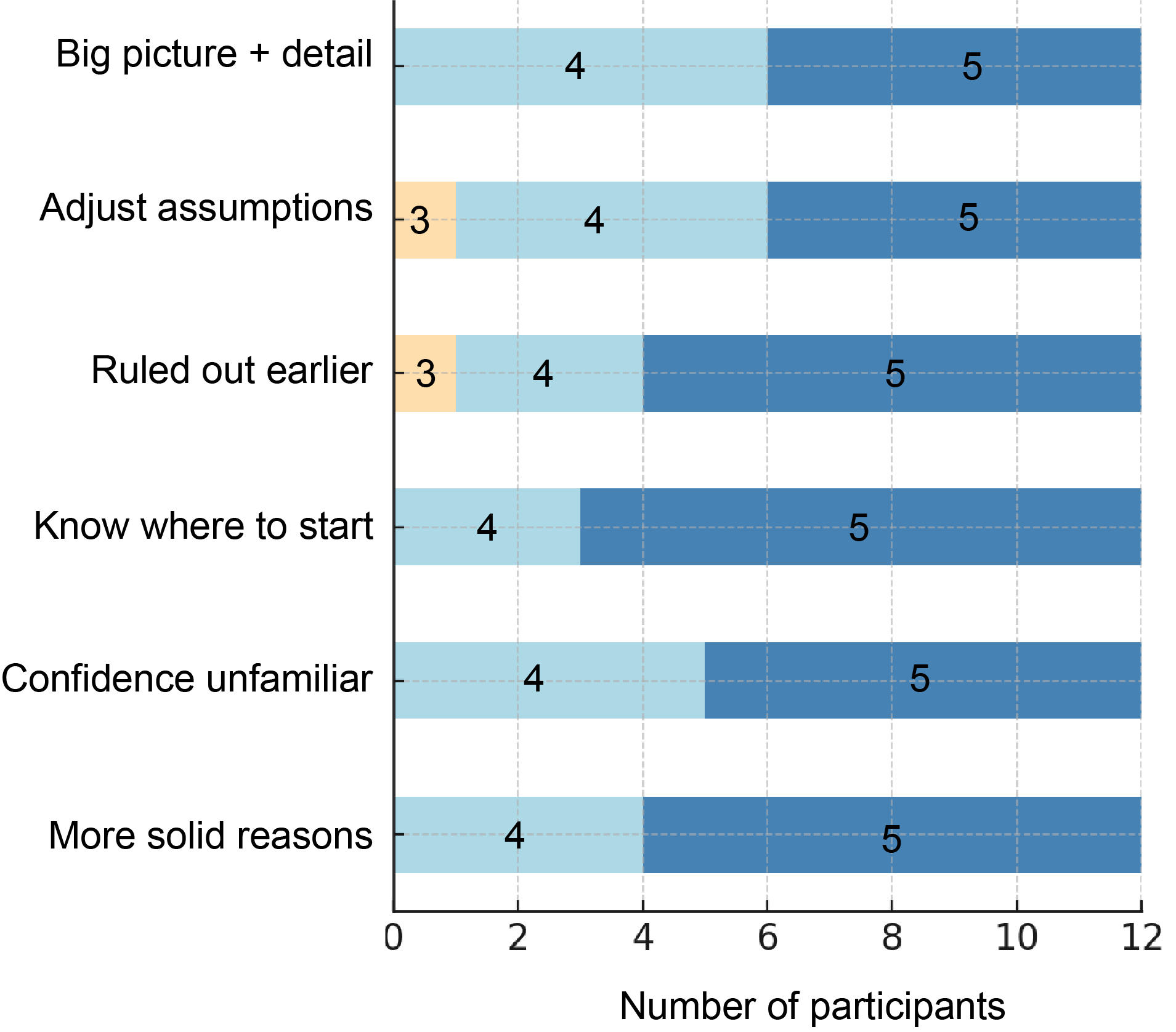}
    \caption{Distribution of post-study survey responses (N=12). 
    Each bar shows the number of participants selecting ratings from 1 (low) to 5 (high). 
   }
    \label{fig:survey_likert}
\end{figure}

As shown in Figure~\ref{fig:survey_likert}, Likert-scale ratings were consistently high across items. Participants most frequently reported that \name{} resulted in solid reasons for their choices (M=4.67, 8/12 rated 5), increased confidence in trying unfamiliar methods (M=4.67, 8/12 rated 5), and helped them adjust design assumptions after exploring alternatives (M=4.67, 9/12 rated 5). When asked which types of information were most helpful in making choices, participants highlighted equipment requirements (9/12) and final quality (8/12), followed by difficulty level (6/12) and time required (4/12). 

Participants from design disciplines highlighted how \name{} addressed persistent gaps in fabrication education. Two participants with architecture backgrounds (P1, P11) highlighted a persistent lack of fabrication education in their curricula. P1 described their only formal fabrication training as a first-year wood shop safety course covering three basic tools. Beyond this, most model-making decisions relied on ad hoc support from staff or online tutorials. As they explained: \textit{“With those tutorials—say I go on YouTube—it’s like the person says, ‘Today we’ll teach you how to make this 3D printed file’ or ‘how to do this casting.’ Then they take out their own file and go step-by-step. What I can’t do there is take my own file, drag it in, and see, for a specific method, exactly what process it would go through.”} P1 described this difference as meaningful for their own learning process: \textit{“Reading a panel like in \name{} is a process in itself—I’m thinking while looking. Without that direct visual experience, I don’t learn as fast.”} They reflected that conventional tutorials often left them imitating without understanding: \textit{“Schools don’t really teach you these things—it’s normal. You have to learn a lot yourself.”} In contrast, P1 described \name{} as fitting into what they viewed as a ‘learning through making’ loop: \textit{“Making something is really about a ‘learning through making’ process. You can’t just watch once and remember—you have to make and reflect at the same time.”}  

\textbf{P11} similarly noted that their fellow students often have limited prior experience in fabrication, leaving many without a good way to evaluate if a design is fabricable. They described that even when fabrication appeared in the studio, it was project-specific and not generalizable: \textit{“It was only one class… she showed us, and everybody just followed that. It wasn’t like a basic knowledge kind of thing... That knowledge is not applicable to other models.”} Like P1, they expressed regret about past missteps: \textit{“I have chosen in my past studio projects some wrong methods. If I had this kind of, like a software or like a guideline, it would have helped me a lot to decide… less time consuming, less hectic.”} They praised \name{} as both a practical guide and a way to visualize alternatives, for informed decision-making: \textit{“Even though I don’t have any knowledge about the other methods, it helped me a lot… it really helped me to visualize how it gonna build my model if I choose that specific method”}.

\subsubsection{Critiques and Feature Requests}

While participants broadly valued \name{}, interviews surfaced several limitations and feature requests that suggest directions for improvement.

\textbf{Information density vs. user experience level.} Participants differed in how much detail they wanted at different stages and experience levels. For example, P6 (intermediate) reported rarely opening the long information panel, whereas novices (e.g., P1, P11) relied on the panel when encountering unfamiliar workflows. To address this, \name{} could adopt progressive disclosure tailored to user needs: minimal keywords and thumbnails for experienced users; expandable bullets, step-by-step panels, or optional \textit{“Compare Against”} views for novices seeking more guidance.

\textbf{Cost awareness and repeat production.} P7 requested cost estimates and guidance for batch production:\textit{ “If I want to repeatedly produce, can we find methods or materials that lower cost?”} To address this, \name{} could integrate cost calculators and \textit{“batch mode”} recommendations (e.g.,  reusable jigs, vendor suggestions) to inform production planning and better align with their goals.

\textbf{Faster shortlisting via intent prompts.} P4 and P5 suggested combining AI-driven prompts with manual exploration:\textit{ “If I give one line—food-safe, transparent, under 3 hours—the system could shortlist two or three workflows with reasons”}. In the future, \name{} could add an intent-based filtering mode that ranks workflows by user-specified requirements before users conduct deeper side-by-side comparisons, as an optional layer.

%% file: src/10_discussion.tex
\section{Discussion}

Our findings illustrate how participants used guided comparison to make decisions on their CAD workflows, pointing to potential considerations for educational contexts. We organize the discussion into five themes: comparison as a lens for reasoning, learning from limits, designing for a spectrum of expertise, educational integration, and a speculation about alternative use cases of \name{}.

\subsection{Comparison as a Lens for Reasoning}
Comparison appeared to shape how participants framed their decisions, particularly during planning, a finding consistent with cognitive psychology research showing that side-by-side comparison facilitates abstract thinking and pattern recognition by making similarities and differences explicit~\cite{Gentner_Markman_1997, Loewenstein_Thompson_Gentner_1999}. Prior research in education has shown that combining procedural knowledge (“how to”) with conceptual understanding (“why to”) leads to deeper and more transferable learning outcomes~\cite{Rittle-Johnson_Schneider_Star_2015,Cheung_Kulasegaram_Woods_Brydges_2019}. 

The shift in reasoning observed in the study from \textit{“Can I make this?”} to \textit{“What approach best fits my goals?”} reflects a more conceptual and goal-oriented perspective. These observations indicate that participants described their choices in more detailed and comparative terms when using guided comparison in CAD. This finding highlights how tools like \name{} may support both the “how” and the “why” of fabrication decisions, and points to the importance of a curriculum that facilitates this dual perspective.

It supports users in developing conditional rules: \textit{“if I need this property, then use this method”}, which is valuable beyond that specific workflow but will help them in the future when confronted with similar questions. We speculate that this will, over time, help them cement a \textit{decision-making framework}, and further long-term deployment studies should shed more light on this.

While we do not have a long-term account of the learning of our participants, we did observe reflection patterns during comparison—such as revisiting initial assumptions after encountering workflow constraints—that align with mechanisms described in prior learning sciences work~\cite{Kapur_2016, Bjork_1994}.

\subsection{Learning from Limits}

Conventional fabrication tools often aim to let non-experts achieve professional-looking results by bypassing complexity. While this lowers entry barriers, it removes opportunities for learners to understand underlying principles and explore the alternatives. Our approach was different: we explicitly showed how certain workflows might fail for specific designs.

Participants valued seeing limitations and potential failure points, in line with prior work~\cite{Rosner_Ryokai_2009,Kapur_2014,Song_2024} showing how encounters with breakdowns and unmaking processes can inform design reasoning. The fabricability warnings that surfaced geometric limits, model sizes, and constraints made the underlying fabrication considerations more explicit than they typically are in tutorials.

Showing the limits of workflows made more experienced users aware that a multi-part object might require different workflows for different parts. Instead of treating workflows as mutually exclusive, they started assembling hybrid solutions or using the workflow cards like a \textit{“dictionary”} to generate new possibilities. This suggests that educational tools should embrace complexity rather than hide it. When people see what can go wrong and why, they can become better at making informed decisions. The interface allowed people to consider multiple options long enough to better understand the trade-offs in each decision.

\subsection{Design for a Spectrum of Expertise}

Novices and advanced users used \name{} differently. Advanced users quickly scanned visual previews and eliminated options through rapid pattern matching. Novices needed detailed explanations and systematic back-and-forth comparison to build understanding.

The same interface served both groups; panels that advanced users ignored were essential scaffolding for novices, while previews that novices initially skipped, became powerful tools for advanced users. For interfaces to serve a spectrum of expertise, they should provide multiple pathways and information layers, echoing what \citet{Resnick_Silverman_2005} describes as ‘wide walls’—designs that allow varied points of entry and diverse ways of engaging with a system. By focusing on novice-first design, we made sure all features were in place to support them. We then hierarchically laid out the features so that the ones that are \textit{only} serving novices are deepest in the hierarchy and thus don't get in the way of more experienced users. An exception may be the filtering feature, while it helped novices find starting points, some became too dependent on it and expected \name{} to give them answers similar to how text-based AI tools would respond with ready solutions.

Advanced users used filtering more like browsing a catalog, a quick way to survey options before doing their own evaluation. We argue that educational tools should only sparingly present open text boxes to better scaffold novice learning decision-making. Especially early learners benefit from seeing the different options (and learning to choose) instead of typing away in an open text box.

\subsection{Educational Integration}

We identify several gaps in fabrication education preventing current comparative reasoning. Even in fabrication-oriented disciplines such as architecture, formal instruction was typically confined to safety training or tool-specific workshops, leaving students to rely heavily on staff support or online tutorials. This narrow exposure encouraged imitation of isolated workflows rather than the development of transferable reasoning skills about alternatives.

Fabrication tutorials are commonly organized around individual machines or workflows rather than cross-method decision-making. This machine-centric structure reinforces siloed knowledge: students learn how to operate a laser cutter or 3D printer, but rarely practice weighing trade-offs across methods for a given design goal. As a result, they develop islands of competence but lack integrative strategies for hybrid or comparative reasoning skills increasingly important for complex design tasks. 

Comparison interfaces like \name{} should thus not be positioned as \textit{replacements} for hands-on instruction or operational safety training, but as one way to structure guided comparison—a process in which learners examine how different workflows align with their intentions before committing to fabrication. Educators we interviewed emphasized that such comparison is most meaningful early in the process, when students are forming expectations about what various workflows can produce. In our study, participants used guided comparison to articulate these expectations as they explored alternatives in a low-stakes environment. Our findings suggest directions for further work on how such comparative activities might support learners’ preparation for fabrication. This perspective resonates with flipped-classroom models in applied education~\cite{Bishop_Verleger_2013, Lo_Hew_2017}, where conceptual exploration precedes physical practice. Guided comparison such as the kind with \name{}, however, addresses only one dimension of fabrication learning. Material behavior, tool handling, and tacit knowledge remain grounded in hands-on engagement and cannot be substituted by digital exploration. Rather, guided comparison complements these embodied practices by making visible the criteria learners use when navigating among fabrication options. Several novices defaulted to familiar workflows despite knowing alternatives existed; comparison may not always change this preference but would make their reasons for choosing a familiar method more explicit and deliberate.

\subsection{Alternative use of \name{}}

In this paper, we have focused on the most impactful use case of comparison and exploration of fabrication workflows: supporting comparative reasoning. However, throughout our interviews and personal experience with \name{} in the lab, we identified two other use cases: transitioning a design throughout stages of the design process and replicating a model found online using slightly different tools at hand. While \name{} may not be optimized for these use cases, we see value of more in-depth future exploration.

When transitioning from a mock-up to a functional prototype (or even a model for manufacturing), it is not always clear what it would take to use different tools and fabrication workflows. \name{} inherently demonstrates what the same CAD model would look like when executed on a different workflow, side by side. We recognize that often you would still need to adjust the CAD model in this transition, but seeing how, say a cardboard chair prototype would come out when injection molded, either confirms that the design will consistently work or points to required redesign of the CAD model. Or it might reveal other potential manufacturing workflows for the final design to less experienced designers.

The replication workflow is relatively common and will continue to get more common as sharing and building on the work of others increases over time. You download a model from the web and want to replicate it, but your machines are inherently slightly different from the original~\cite{Roumen_2020}, or maybe you have a 3D printed model you'd like to laser cut (or vice-versa). These transitions are hard and require deep technical skill. \name{} inherently facilitates this, making the shared models much more valuable as they can continue to support a range of workflows and end-users.

\section{Future work}
Besides running a long-term deployment study, we see several other fruitful avenues of future work based on our investigation into guided comparison for CAD.

Much of fabrication knowledge emerges through engagement with tools and materials, where learners adjust plans, respond to machine behavior, and refine techniques through physical action. The comparative reasoning supported in digital environments engages a different moment in the process—when learners are articulating intentions and evaluating possibilities prior to making. Connecting these conceptual explorations with the situated learning that unfolds during fabrication remains an important direction for expanding this line of work.

We acknowledge that fabrication expertise often relies on situated learning—knowledge gained through reacting to material behaviors and machine states in real-time. Our current implementation focuses on pre-fabrication reasoning, which risks missing dynamic, context-specific information (e.g., environmental fluctuations or material inconsistencies). Future iterations could integrate sensor-based feedback to bridge this gap. By surfacing real-time fabrication data, the tool could evolve from a static planning interface into a dynamic support system that connects abstract decision-making with the situated realities of physical production.

%% file: src/11_conclusion.tex
\section{Conclusion}

This paper investigates how comparing fabrication workflows can support reasoning during CAD-based planning. Through the design of \name{}, we explored comparison as a lens that makes trade-offs, constraints, and alternative fabrication paths more visible to learners. Our lab study provides initial evidence that guided comparison broadens the range of workflows people consider, surfaces conditional reasoning, and helps participants articulate more goal-aligned decisions. These findings point to guided comparison as a promising direction for future CAD tools and fabrication education. Future research should examine how such tools integrate into curricula, support long-term learning, and generalize to other fabrication contexts. By foregrounding the reasoning behind fabrication choices, we hope this work contributes to CAD environments that better support reflective, fabrication-aware decision-making.

%% file: src/12_appendix.tex
\appendix
\section{Workflow Sources}

We compiled a diverse set of fabrication workflows from publicly available instructional resources. The list below includes examples spanning different materials, machines, and construction techniques, with links to their original instruction pages.

\begin{itemize}
    \item \url{https://www.instructables.com/Creating-a-University-Monument-and-Replica-Bronzes/}
    \item \url{https://www.instructables.com/Huge-DIY-Concrete-Face-Garden-Sculpture/}
    \item \url{https://www.instructables.com/Helicopter-BatterySolar-Wire-Bending/}
    \item \url{https://www.instructables.com/Twisty-Toy/}
    \item \url{https://www.instructables.com/Spider-Pumpkins-Arachnophobia-Jack-O-Lanterns/}
    \item \url{https://www.instructables.com/Free-the-Heart-II-Wire-Puzzle-Level-2-Wire-Bending/}
    \item \url{https://www.instructables.com/Build-a-6-0-tall-Wooden-T-Rex-Model/}
    \item \url{https://www.instructables.com/Build-Your-Own-Portable-or-Stationary-Firepit/}
    \item \url{https://www.instructables.com/Hollow-Wooden-Surfboard-my-Magic-Carpet/}
    \item \url{https://www.instructables.com/Laser-Cut-Cryptex/}
    \item \url{https://www.instructables.com/Fairytale-Story-Proposal-With-Puzzles/}
    \item \url{https://www.instructables.com/Lasercut-RC-Comet-Tank/}
    \item \url{https://www.instructables.com/Wooden-Silhouette-Test-Tube-Vases/}
    \item \url{https://www.instructables.com/Reclaimed-timber-Hollow-Wood-Surfboard-92/}
    \item \url{https://www.instructables.com/Lasercut-RC-PakWagen/}
    \item \url{https://www.instructables.com/Carved-Plywood-Rocking-Elephant/}
    \item \url{https://www.instructables.com/Screwdriver-Handle/}
    \item \url{https://www.instructables.com/Sculptural-Customizable-Plywood-Lamp/}
    \item \url{https://www.instructables.com/Laser-Cut-M-6-Carnifex-Rubber-Band-Gun-from-Mass-E/}
    \item \url{https://www.instructables.com/A-3D-Printed-Animated-Valentine-Heart-for-My-Valen/}
    \item \url{https://www.instructables.com/Secret-Heart-Box/}
    \item \url{https://www.instructables.com/Wooden-Fidget-Cube/}
    \item \url{https://www.instructables.com/Codex-Puzzle-Box/}
    \item \url{https://www.instructables.com/561-Segment-Pentagon-Pen-or-Feather-Pen-You-Decide/}
    \item \url{https://www.instructables.com/Sculpted-Reclaimed-Wood-Bench-With-Steel-Base/}
    \item \url{https://www.instructables.com/Wooden-Knuckle-Dusters/}
    \item \url{https://www.instructables.com/Wooden-Reindeer-Figures/}
    \item \url{https://www.instructables.com/Easy-Toy-Sword/}
    \item \url{https://www.instructables.com/Impossible-Marble-in-Truncated-Cube/}
    \item \url{https://www.instructables.com/A-Modern-Bandsaw-Reindeer/}
    \item \url{https://www.instructables.com/Make-a-Bird-Whistle/}
    \item \url{https://www.instructables.com/Wooden-Cream-Cheese-Knife/}
    \item \url{https://www.instructables.com/Epic-Thors-Carving-Mallet/}
    \item \url{https://www.instructables.com/T-Rex-Coin-Chomper-Dinosaur-Piggy-Bank/}
    \item \url{https://www.instructables.com/How-to-Build-a-Tahitian-Ukulele/}
    \item \url{https://www.instructables.com/Batarang-From-a-Saw-Blade/}
    \item \url{https://www.instructables.com/How-to-make-a-pair-of-bellows/}
    \item \url{https://www.instructables.com/Giant-wooden-Lego-men/}
    \item \url{https://www.instructables.com/Twisted-Stool-End-Table/}
    \item \url{https://www.instructables.com/Carvers-Mallet/}
    \item \url{https://www.instructables.com/How-to-Bend-PVC-Make-Incredible-Shapes/}
    \item \url{https://www.instructables.com/How-to-Make-a-Concrete-Base-for-Many-Applications/}
    \item \url{https://www.instructables.com/3D-Printing-a-Mold-for-a-Mold/}
    \item \url{https://www.instructables.com/Concrete-Light-Bulb-Coat-Rack-W-Reclaimed-Wood/}
    \item \url{https://www.instructables.com/Real-Life-Super-Mario-Concrete-Thwomp/}
    \item \url{https://www.instructables.com/Casting-Objects-in-Sugru-Make-your-own-candle-wax-/}
    \item \url{https://www.instructables.com/Entwined-Hearts-Cast-in-Aluminum/}
    \item \url{https://www.instructables.com/SuperTiny-Skulls-Silicone-Concrete-Mold/}
    \item \url{https://www.instructables.com/Cute-Concrete-Planter/}
    \item \url{https://www.instructables.com/Aluminium-Casting-and-why-you-need-to-be-careful/}
    \item \url{https://www.instructables.com/Skyrim-Chillrend-Prop-Sword-Made-of-Epoxy-Resin/}
    \item \url{https://www.instructables.com/3d-Printed-Molds/}
    \item \url{https://www.instructables.com/Jeepers-Creepers-Rock-Eye-Peepers/}
    \item \url{https://www.instructables.com/A-Boat-From-a-Single-2x4/}
    \item \url{https://www.instructables.com/Secret-Wood-Rings-DIY/}
\end{itemize}

\clearpage
\subsection{Feasibility Check Matrix Details} 

\begin{figure*}[h]
    \centering
    \includegraphics[width=\textwidth]{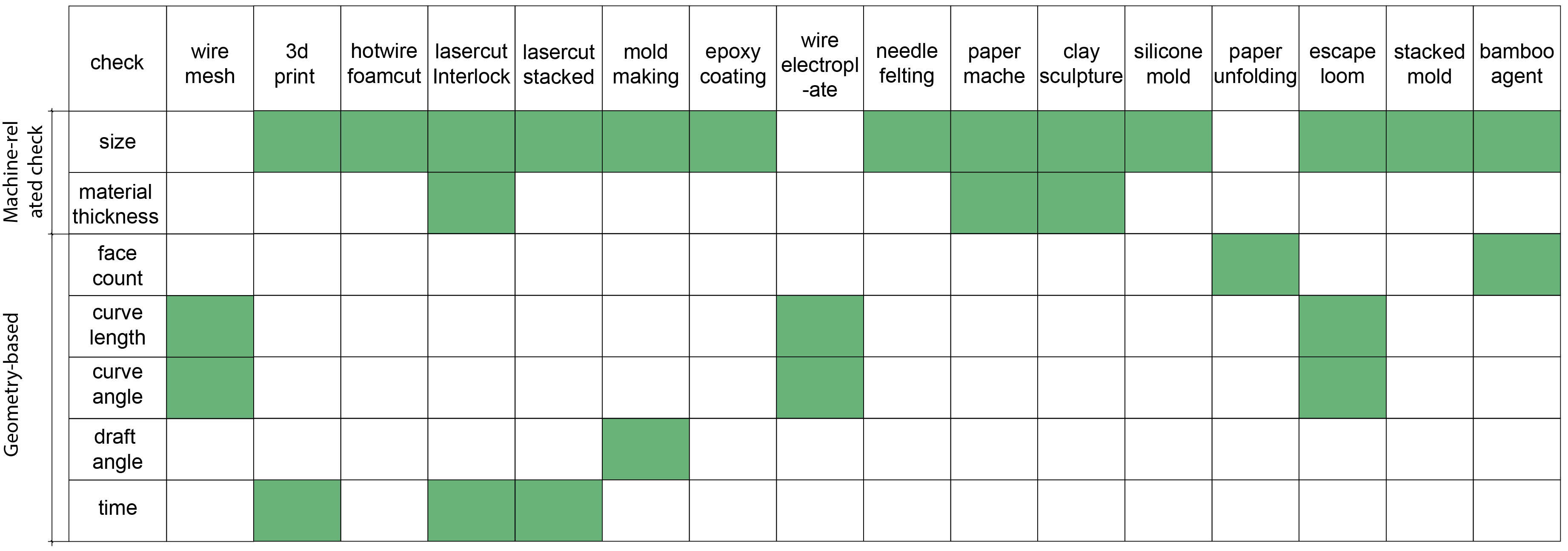}
    \caption{Feasibility checks implemented across all fabrication workflows.}
    \label{fig:feasibility_matrix}
\end{figure*}

The matrix below (Fig~\ref{fig:feasibility_matrix}) summarizes all feasibility checks implemented across all 16 fabrication workflows.

\subsection{Core Module Overview}
\label{app:core_modules_overview}

The table below (Table~\ref{tab:core_modules}) summarizes the core software modules implemented in \name{}. Modules are grouped by functional category, including model loading, geometry processing, visualization, and export. For each module, we provide a brief description of its primary role within the tool.

\begin{table*}[h]
\centering
\caption{Core Modules in \name{}}
\label{tab:core_modules}
\begin{tabular}{ p{1.5cm}|p{2.8cm}|p{8cm} }
\textbf{Category} & \textbf{Module} & \textbf{Description} \\ \hline
Model Loading     & Model Loading & Load and process STL files, with options for centering, scaling, mesh rotation (x/y/z), and bounding box handling. \\ \hline
\multirow{2}{1.5cm}{Geometry Processing} 
                  & Hollow Triangle Gen & Generate hollow triangles from  vertices, faces, and normals. \\
                  & Tube Gen      & Generate tubes with specified color, opacity, and radius. \\
                  & Interlock Joints Gen & Create interlocking joints based on points and dimensions. \\
                  & Split Volume           & Create positive and negative molds with air vents. \\
                  & Intersect  & Generate intersections and contours using meshes and planes. \\                  
                  & Extrude              & Generate 3D extruded shapes from lines. \\
                  & Loft                 & Create lofted surfaces from inner and outer contours. \\
                  & Calc Draft Angles & Analyze draft angles with color mapping. \\         
                  & Project Mesh           & Project meshes into 2D lines. \\              
                  & Calc Line Inters & Calculate intersection points for lines in specific planes. \\ \hline
Visualization     & Model Info Vis & Display mesh details and properties. \\
                  & Visualize Mold       & Visualize positive and negative mold parts. \\
                  & TutorialWindow     & Provide text, images, and links for user guidance. \\
                  & Color Legend         & Show visual legend for angle ranges and other properties. \\
                  & 3D Visualization     & Render meshes in 3D. \\
                  & Geom Visualization & Visualize points and lines with legends. \\ \hline
Export            & Export SVG           & Save lines as SVG files. \\
                  & Export CSV           & Save lines as CSV files. \\
                  & Export STL           & Save meshes as STL files. \\ \hline
\end{tabular}%
\end{table*}

\subsection{Workflow Implementation Details}
\label{sec:details_revised}
The following table (Table~\ref{tab:final-summary}) details the implementation specifics for the sixteen fabrication workflows supported by \name{}, including machine requirements, fabrication characteristics, and critical implementation notes.

\begin{table*}[h]
\centering
\footnotesize
\setlength{\tabcolsep}{3pt}
\begin{tabular}{|p{2.7cm}|p{2cm}|p{2 cm}|p{2.5cm}|p{6.8cm}|}
\hline
\textbf{Workflow} & \textbf{Adjustable Parameters} & \textbf{Machine Requirements} & \textbf{Fabrication Characteristics} & \textbf{Notes} \\
\hline

\textbf{Wire Mesh Structure} & Slice count (uniform XY) & CNC Wire Bender & Flexible and lightweight for large-scale fabrication & Output can be directly fabricated. Mesh curvature is approximated through segmentation. The workflow automatically detects unbendable segments caused by geometric constraints and flags them for the user. \\
\hline

\textbf{3D Printing} & Layer height, infill density, preset PLA/TPU material & 3D Printer & Time intensive & Visualization-only. The slicing preview does not generate G-code and requires further toolpath generation in the printer's slicer. Basic printability checks ensure the model fits within the build volume. Fabrication time is estimated based on preset profiles. \\
\hline

\textbf{Hot Wire Foam Cutting} & None (template thickness fixed at 3mm) & Hot Wire Cutter, Laser Cutter & Low-resolution rapid volume prototyping, soft, lightweight & Final precision depends on manual skill and trimming. Templates are laser-cut from simplified multi-view contours (front/side/top) and used to guide the hot wire cutter. Manufacturability checks ensure compatibility with the laser cutter’s working envelope. \\
\hline

\textbf{Laser Cut Interlocking Planar} & Slice count (uniform XY), material thickness, rendering material (wood/acrylic) & Laser Cutter & Modular planar assembly, durability & Fine features not intersecting slices are lost. Manufacturability is constrained by material thickness and the laser cutter’s working area. Fabrication time is estimated based on cutting complexity. \\
\hline

\textbf{Laser Cut Stacked Layers} & Layer count, rendering material (wood/acrylic) & Laser Cutter & Bulky due to solid material stacking, durability & Each layer is laser-cut and manually aligned with adhesive. Exterior contour extraction may smooth fine geometry. Manufacturability is constrained by material thickness and the laser cutter’s working area. Fabrication time is estimated based on the total number of layers and cutting path. \\
\hline

\textbf{Mold Making} & Draft angle analysis, split-line placement, air hole size & 3D Printer & Repeatable silicone casting parts, craft time uncertain & Mold is 3D-printed. This workflow requires manual casting and silicone curing. Draft angle analysis and printability checks ensure manufacturability within the printer’s build volume. Final accuracy depends on print resolution and alignment. \\
\hline

\textbf{Epoxy Coating} & Layer count & Laser Cutter Manual Tools & Smoothed and reinforced surface finishing & Wooden core is laser-cut, then manually assembled and sanded before applying epoxy resin. The underlying wooden core must comply with size constraints (working envelope of the laser cutter). Craftsmanship and curing determine final durability. Surface detail is constrained by the underlying stacked-layer geometry. \\
\hline

\textbf{Wire Electroplating} & Slice count (uniform XY) & Electroplating Setup, Wire Bender & Conductive and durable copper reinforcement & Built on the wire mesh structure. Requires chemical handling, immersion control, and manual monitoring of current and time. Coating thickness and uniformity depend on electrical parameters and user skill. \\
\hline

\textbf{Needle Felt} & None (foam thickness fixed at 3mm) & Needle Felting Tools, Hot wire foam cutter, laser cutter & Soft tactile skin over foam core & Built on the foam-cut base shape. Requires iterative manual felting to build volume and stiffness. Final texture and hardness depend on wool density and craftsmanship. Surface detail is constrained by the underlying foam resolution. \\
\hline

\textbf{Paper Mache} & Slice count (uniform XY) & Manual Tools, laser cutter & Paper skin over interlocking core & Built on the planar interlocking skeleton. Requires layered pulp wrapping and drying time. Final strength and smoothness depend heavily on craftsmanship. Surface detail is constrained by the underlying interlocking geometry. The workflow ensures the core and paper layers comply with material thickness constraints. \\
\hline

\textbf{Clay Sculpture} & Slice count (uniform XY) & Manual Tools, laser cutter & Clay skin over interlocking core & Built on the planar interlocking skeleton. Requires clay application and mesh reinforcement to maintain moisture and adhesion. Drying and smoothing determine accuracy. Manufacturability is constrained by the underlying interlocking geometry and the core's Material Thickness limitations. \\
\hline

\textbf{Silicone Mold} & Draft angle & Laser Cutter + 3D Printer & Food-safe reusable mold for casting & Base fixture is laser-cut to anchor a 3D-printed model. Silicone is poured and cured to form a reusable mold. Final demolding success depends on smoothness and draft angles. \\
\hline

\textbf{Paper Unfolding} & None & Paper Cutter & Lightweight and fast-iterating DIY forms & Based on a Python unfolding algorithm (\textbf{repo: felixfeliz/paperfoldmodels}). Limited to fewer than 100 triangulated faces for foldability. Small or dense features must be simplified. Alignment accuracy depends on manual folding and gluing. \\
\hline

\textbf{Escape Loom} & Strut thickness & 3D Printer, Manual Tools for weaving & Lightweight frame for hand weaving & Based on the “EscapeLoom” project. Strut thickness affects strength and tension tolerance. Only the frame is fabricated—final textiles depend entirely on weaving skill. \\
\hline

\textbf{Stack Mold} & Layer thickness & Laser Cutter, Manual Tools & Modular mold for multi-material casting & layer thickness defines contour count for laser cutting. Accuracy and demolding success depend on alignment, draft conditions, and manual assembly skill. \\
\hline

\textbf{Bamboo Agent} & None & 3D Printer & Strut geometry supporting bamboo weaving & Based on the “Bamboo Agent” workflow. Mesh faces limited to less than 100 to ensure attachable supports. Size constrained by printer build volume. \\
\hline
\end{tabular}
\caption{Implementation summary of fabrication workflows supported and extended by \name{}.}
\label{tab:final-summary}
\end{table*}

%% file: reference.bib
@inproceedings{McCrae_Umetani_Singh_2014, address={Honolulu Hawaii USA}, title={FlatFitFab: interactive modeling with planar sections}, ISBN={9781450330695}, url={https://dl.acm.org/doi/10.1145/2642918.2647388}, DOI={10.1145/2642918.2647388}, booktitle={Proceedings of the 27th annual ACM symposium on User interface software and technology}, publisher={ACM}, author={McCrae, James and Umetani, Nobuyuki and Singh, Karan}, year={2014}, month=oct, pages={13–22}, language={en} }

@article{Dumas_Hergel_Lefebvre_2014, title={Bridging the gap: automated steady scaffoldings for 3D printing}, volume={33}, ISSN={0730-0301, 1557-7368}, url={https://dl.acm.org/doi/10.1145/2601097.2601153}, DOI={10.1145/2601097.2601153}, number={4}, journal={ACM Transactions on Graphics}, author={Dumas, Jérémie and Hergel, Jean and Lefebvre, Sylvain}, year={2014}, month=jul, pages={1–10}, language={en} }

@article{Mori_Igarashi_2007, title={Plushie: an interactive design system for plush toys}, volume={26}, ISSN={0730-0301, 1557-7368}, url={https://dl.acm.org/doi/10.1145/1276377.1276433}, DOI={10.1145/1276377.1276433}, number={3}, journal={ACM Transactions on Graphics}, author={Mori, Yuki and Igarashi, Takeo}, year={2007}, month=jul, pages={45}, language={en} }

@inproceedings{Bourgault_Wiley_Farber_Jacobs_2023, address={Hamburg Germany}, title={CoilCAM: Enabling Parametric Design for Clay 3D Printing Through an Action-Oriented Toolpath Programming System}, ISBN={9781450394215}, url={https://dl.acm.org/doi/10.1145/3544548.3580745}, DOI={10.1145/3544548.3580745}, booktitle={Proceedings of the 2023 CHI Conference on Human Factors in Computing Systems}, publisher={ACM}, author={Bourgault, Samuelle and Wiley, Pilar and Farber, Avi and Jacobs, Jennifer}, year={2023}, month=apr, pages={1–16}, language={en} }

@inproceedings{TranO’Leary_Benabdallah_Peek_2023, address={Hamburg Germany}, title={Imprimer: Computational Notebooks for CNC Milling}, ISBN={9781450394215}, url={https://dl.acm.org/doi/10.1145/3544548.3581334}, DOI={10.1145/3544548.3581334}, booktitle={Proceedings of the 2023 CHI Conference on Human Factors in Computing Systems}, publisher={ACM}, author={Tran O’Leary, Jasper and Benabdallah, Gabrielle and Peek, Nadya}, year={2023}, month=apr, pages={1–15}, language={en} }

@inproceedings{Valkeneers_Leen_Ashbrook_Ramakers_2019, address={New Orleans LA USA}, title={StackMold: Rapid Prototyping of Functional Multi-Material Objects with Selective Levels of Surface Details}, ISBN={9781450368162}, url={https://dl.acm.org/doi/10.1145/3332165.3347915}, DOI={10.1145/3332165.3347915}, booktitle={Proceedings of the 32nd Annual ACM Symposium on User Interface Software and Technology}, publisher={ACM}, author={Valkeneers, Tom and Leen, Danny and Ashbrook, Daniel and Ramakers, Raf}, year={2019}, month=oct, pages={687–699}, language={en} }

@article{Schwartzburg_Pauly_2013, title={Fabrication‐aware Design with Intersecting Planar Pieces}, volume={32}, ISSN={0167-7055, 1467-8659}, url={https://onlinelibrary.wiley.com/doi/10.1111/cgf.12051}, DOI={10.1111/cgf.12051}, number={2pt3}, journal={Computer Graphics Forum}, author={Schwartzburg, Yuliy and Pauly, Mark}, year={2013}, month=may, pages={317–326}, language={en} }

@inproceedings{Schmidt_Umetani_2014, address={Vancouver Canada}, title={Branching support structures for 3D printing}, ISBN={9781450329774}, url={https://dl.acm.org/doi/10.1145/2619195.2656293}, DOI={10.1145/2619195.2656293}, booktitle={ACM SIGGRAPH 2014 Studio}, publisher={ACM}, author={Schmidt, Ryan and Umetani, Nobuyuki}, year={2014}, month=jul, pages={1–1}, language={en} }

@inproceedings{Tseng_Tsai_2015, address={Stanford California USA}, title={Process Products: Capturing Design Iteration with Digital Fabrication}, ISBN={9781450333054}, url={https://dl.acm.org/doi/10.1145/2677199.2687891}, DOI={10.1145/2677199.2687891}, booktitle={Proceedings of the Ninth International Conference on Tangible, Embedded, and Embodied Interaction}, publisher={ACM}, author={Tseng, Tiffany and Tsai, Geoff}, year={2015}, month=jan, pages={631–636}, language={en} }

@inproceedings{Deshpande_Takahashi_Kim_2021, address={Yokohama Japan}, title={EscapeLoom: Fabricating New Affordances for Hand Weaving}, ISBN={9781450380966}, url={https://dl.acm.org/doi/10.1145/3411764.3445600}, DOI={10.1145/3411764.3445600}, booktitle={Proceedings of the 2021 CHI Conference on Human Factors in Computing Systems}, publisher={ACM}, author={Deshpande, Himani and Takahashi, Haruki and Kim, Jeeeun}, year={2021}, month=may, pages={1–13}, language={en} }

@inproceedings{Gao_Gao_Yang_Liu_Shi_Li_2023, address={Warsaw Poland}, title={Bamboo Agents: Exploring the Potentiality of Digital Craft by Decoding and Recoding Process}, ISBN={9781450399777}, url={https://dl.acm.org/doi/10.1145/3569009.3572746}, DOI={10.1145/3569009.3572746}, booktitle={Proceedings of the Seventeenth International Conference on Tangible, Embedded, and Embodied Interaction}, publisher={ACM}, author={Gao, Peizhong and Gao, Tanhao and Yang, Yanbin and Liu, Zhenyuan and Shi, Jianyu and Li, Jin}, year={2023}, month=feb, pages={1–13}, language={en} }

@article{Stava_Vanek_Benes_Carr_Měch_2012, title={Stress relief: improving structural strength of 3D printable objects}, volume={31}, ISSN={0730-0301, 1557-7368}, url={https://dl.acm.org/doi/10.1145/2185520.2185544}, DOI={10.1145/2185520.2185544}, number={4}, journal={ACM Transactions on Graphics}, author={Stava, Ondrej and Vanek, Juraj and Benes, Bedrich and Carr, Nathan and Měch, Radomír}, year={2012}, month=aug, pages={1–11}, language={en} }

@book{Schön_1983, address={New York}, title={The reflective practitioner: how professionals think in action}, ISBN={9780465068784}, publisher={Basic Books}, author={Schön, Donald A.}, year={1983}, language={eng} }

@book{Buxton_2007, address={Amsterdam Boston}, title={Sketching user experiences: getting the design right and the right design}, ISBN={9780123740373}, callNumber={658.575 2},  publisher={Elsevier/Morgan Kaufmann}, author={Buxton, William}, year={2007}, language={eng} }

@article{Dow_Glassco_Kass_Schwarz_Schwartz_Klemmer_2010, title={Parallel prototyping leads to better design results, more divergence, and increased self-efficacy}, volume={17}, ISSN={1073-0516, 1557-7325}, url={https://dl.acm.org/doi/10.1145/1879831.1879836}, DOI={10.1145/1879831.1879836}, number={4}, journal={ACM Transactions on Computer-Human Interaction}, author={Dow, Steven P. and Glassco, Alana and Kass, Jonathan and Schwarz, Melissa and Schwartz, Daniel L. and Klemmer, Scott R.}, year={2010}, month=dec, pages={1–24}, language={en} }

@article{Brennan_Miney_Simpson_Jablokow_McComb_2023, title={Manufacturing Fixation in Design: Exploring the Effects of Manufacturing Fixation During Idea Generation}, volume={145}, ISSN={1050-0472, 1528-9001}, url={https://asmedigitalcollection.asme.org/mechanicaldesign/article/145/1/012005/1150237/Manufacturing-Fixation-in-Design-Exploring-the}, DOI={10.1115/1.4056222},  number={1}, journal={Journal of Mechanical Design}, author={Brennan, Jennifer Bracken and Miney, William B. and Simpson, Timothy W. and Jablokow, Kathryn W. and McComb, Christopher}, year={2023}, month=jan, pages={012005}, language={en} }

@article{blikstein2013digital,
  title={Digital fabrication and ‘making’in education},
  author={Blikstein, Paulo},
  journal={FabLab: Of machines, makers, and inventors},
  pages={203--222},
  year={2013},
  publisher={Transcript}
}

@inproceedings{Turakhia_Jiang_Liu_Leake_Mueller_2022, address={Bend OR USA}, title={The Reflective Maker: Using Reflection to Support Skill-learning in Makerspaces}, ISBN={9781450393218}, url={https://dl.acm.org/doi/10.1145/3526114.3558716}, DOI={10.1145/3526114.3558716}, booktitle={Adjunct Proceedings of the 35th Annual ACM Symposium on User Interface Software and Technology}, publisher={ACM}, author={Turakhia, Dishita and Jiang, Peiling and Liu, Brent and Leake, Mackenzie and Mueller, Stefanie}, year={2022}, month=oct, pages={1–4}, language={en} }

@article{Pitkänen_Iwata_Laru_2020, title={Exploring technology-oriented Fab Lab facilitators’ role as educators in K-12 education: Focus on scaffolding novice students’ learning in digital fabrication activities}, volume={26}, ISSN={22128689}, url={https://linkinghub.elsevier.com/retrieve/pii/S2212868920300313}, DOI={10.1016/j.ijcci.2020.100207}, journal={International Journal of Child-Computer Interaction}, author={Pitkänen, Kati and Iwata, Megumi and Laru, Jari}, year={2020}, month=dec, pages={100207}, language={en} }

@inproceedings{Hartmann_Yu_Allison_Yang_Klemmer_2008, address={Monterey CA USA}, title={Design as exploration: creating interface alternatives through parallel authoring and runtime tuning}, ISBN={9781595939753}, url={https://dl.acm.org/doi/10.1145/1449715.1449732}, DOI={10.1145/1449715.1449732}, booktitle={Proceedings of the 21st annual ACM symposium on User interface software and technology}, publisher={ACM}, author={Hartmann, Björn and Yu, Loren and Allison, Abel and Yang, Yeonsoo and Klemmer, Scott R.}, year={2008}, month=oct, pages={91–100}, language={en} }

@inproceedings{Swearngin_Wang_Oleson_Fogarty_Ko_2020, address={Honolulu HI USA}, title={Scout: Rapid Exploration of Interface Layout Alternatives through High-Level Design Constraints}, ISBN={9781450367080}, url={https://dl.acm.org/doi/10.1145/3313831.3376593}, DOI={10.1145/3313831.3376593}, booktitle={Proceedings of the 2020 CHI Conference on Human Factors in Computing Systems}, publisher={ACM}, author={Swearngin, Amanda and Wang, Chenglong and Oleson, Alannah and Fogarty, James and Ko, Amy J.}, year={2020}, month=apr, pages={1–13}, language={en} }

@inproceedings{Kolarić_Woodbury_Erhan_2014, address={Vancouver BC Canada}, title={CAMBRIA: a tool for managing multiple design alternatives}, ISBN={9781450329033}, url={https://dl.acm.org/doi/10.1145/2598784.2602788}, DOI={10.1145/2598784.2602788}, booktitle={Proceedings of the 2014 companion publication on Designing interactive systems}, publisher={ACM}, author={Kolarić, Siniša and Woodbury, Robert and Erhan, Halil}, year={2014}, month=jun, pages={81–84}, language={en} }

@inproceedings{Terry_Mynatt_Nakakoji_Yamamoto_2004, address={Vienna Austria}, title={Variation in element and action: supporting simultaneous development of alternative solutions}, ISBN={9781581137026}, url={https://dl.acm.org/doi/10.1145/985692.985782}, DOI={10.1145/985692.985782}, booktitle={Proceedings of the SIGCHI Conference on Human Factors in Computing Systems}, publisher={ACM}, author={Terry, Michael and Mynatt, Elizabeth D. and Nakakoji, Kumiyo and Yamamoto, Yasuhiro}, year={2004}, month=apr, pages={711–718}, language={en} }

@article{Lunzer_Hornbæk_2008, title={Subjunctive interfaces: Extending applications to support parallel setup, viewing and control of alternative scenarios}, volume={14}, ISSN={1073-0516, 1557-7325}, url={https://dl.acm.org/doi/10.1145/1314683.1314685}, DOI={10.1145/1314683.1314685}, number={4}, journal={ACM Transactions on Computer-Human Interaction}, author={Lunzer, Aran and Hornbæk, Kasper}, year={2008}, month=jan, pages={1–44}, language={en} }

@inproceedings{Kato_Goto_2017, address={Edinburgh United Kingdom}, title={f3.js: A Parametric Design Tool for Physical Computing Devices for Both Interaction Designers and End-users}, ISBN={9781450349222}, url={https://dl.acm.org/doi/10.1145/3064663.3064681}, DOI={10.1145/3064663.3064681}, booktitle={Proceedings of the 2017 Conference on Designing Interactive Systems}, publisher={ACM}, author={Kato, Jun and Goto, Masataka}, year={2017}, month=jun, pages={1099–1110}, language={en} }

@inproceedings{Lin_Ramesh_Pandhare_Tay_Dutta_Hartmann_Mehta_2024, address={Honolulu HI USA}, title={Design Space Exploration for Board-level Circuits: Exploring Alternatives in Component-based Design}, ISBN={9798400703300}, url={https://dl.acm.org/doi/10.1145/3613904.3642009}, DOI={10.1145/3613904.3642009}, booktitle={Proceedings of the CHI Conference on Human Factors in Computing Systems}, publisher={ACM}, author={Lin, Richard and Ramesh, Rohit and Pandhare, Parth Nitin and Tay, Kai Jun and Dutta, Prabal and Hartmann, Bjoern and Mehta, Ankur}, year={2024}, month=may, pages={1–14}, language={en} }

@inproceedings{Matejka_Glueck_Bradner_Hashemi_Grossman_Fitzmaurice_2018, address={Montreal QC Canada}, title={Dream Lens: Exploration and Visualization of Large-Scale Generative Design Datasets}, ISBN={9781450356206}, url={https://dl.acm.org/doi/10.1145/3173574.3173943}, DOI={10.1145/3173574.3173943}, booktitle={Proceedings of the 2018 CHI Conference on Human Factors in Computing Systems}, publisher={ACM}, author={Matejka, Justin and Glueck, Michael and Bradner, Erin and Hashemi, Ali and Grossman, Tovi and Fitzmaurice, George}, year={2018}, month=apr, pages={1–12}, language={en} }

@inproceedings{Dow_Fortuna_Schwartz_Altringer_Schwartz_Klemmer_2011, address={Vancouver BC Canada}, title={Prototyping dynamics: sharing multiple designs improves exploration, group rapport, and results}, ISBN={9781450302289}, url={https://dl.acm.org/doi/10.1145/1978942.1979359}, DOI={10.1145/1978942.1979359}, booktitle={Proceedings of the SIGCHI Conference on Human Factors in Computing Systems}, publisher={ACM}, author={Dow, Steven and Fortuna, Julie and Schwartz, Dan and Altringer, Beth and Schwartz, Daniel and Klemmer, Scott}, year={2011}, month=may, pages={2807–2816}, language={en} }

@inproceedings{Hudson_Alcock_Chilana_2016, address={San Jose California USA}, title={Understanding Newcomers to 3D Printing: Motivations, Workflows, and Barriers of Casual Makers}, ISBN={9781450333627}, url={https://dl.acm.org/doi/10.1145/2858036.2858266}, DOI={10.1145/2858036.2858266}, booktitle={Proceedings of the 2016 CHI Conference on Human Factors in Computing Systems}, publisher={ACM}, author={Hudson, Nathaniel and Alcock, Celena and Chilana, Parmit K.}, year={2016}, month=may, pages={384–396}, language={en} }

@inproceedings{E_Fried_Lu_Zhang_Mech_Echevarria_Hanrahan_Landay_2020, address={Honolulu HI USA}, title={Adaptive Photographic Composition Guidance}, ISBN={9781450367080}, url={https://dl.acm.org/doi/10.1145/3313831.3376635}, DOI={10.1145/3313831.3376635}, booktitle={Proceedings of the 2020 CHI Conference on Human Factors in Computing Systems}, publisher={ACM}, author={E, Jane L. and Fried, Ohad and Lu, Jingwan and Zhang, Jianming and Mech, Radomír and Echevarria, Jose and Hanrahan, Pat and Landay, James A.}, year={2020}, month=apr, pages={1–13}, language={en} }

@inproceedings{E_Fried_Agrawala_2019, address={New Orleans LA USA}, title={Optimizing Portrait Lighting at Capture-Time Using a 360 Camera as a Light Probe}, ISBN={9781450368162}, url={https://dl.acm.org/doi/10.1145/3332165.3347893}, DOI={10.1145/3332165.3347893}, booktitle={Proceedings of the 32nd Annual ACM Symposium on User Interface Software and Technology}, publisher={ACM}, author={E, Jane L. and Fried, Ohad and Agrawala, Maneesh}, year={2019}, month=oct, pages={221–232}, language={en} }

@article{Yen_E_Jin_Li_Lin_Pan_Dow_2024, title={ProcessGallery: Contrasting Early and Late Iterations for Design Principle Learning}, volume={8}, ISSN={2573-0142}, url={https://dl.acm.org/doi/10.1145/3637389}, DOI={10.1145/3637389},number={CSCW1}, journal={Proceedings of the ACM on Human-Computer Interaction}, author={Yen, Yu-Chun Grace and E, Jane L. and Jin, Hyoungwook and Li, Mingyi and Lin, Grace and Pan, Isabelle Yan and Dow, Steven P.}, year={2024}, month=apr, pages={1–35}, language={en} }

@inproceedings{Kelleher_Pausch_Kiesler_2007, address={San Jose California USA}, title={Storytelling alice motivates middle school girls to learn computer programming}, ISBN={9781595935939}, url={https://dl.acm.org/doi/10.1145/1240624.1240844}, DOI={10.1145/1240624.1240844}, booktitle={Proceedings of the SIGCHI Conference on Human Factors in Computing Systems}, publisher={ACM}, author={Kelleher, Caitlin and Pausch, Randy and Kiesler, Sara}, year={2007}, month=apr, pages={1455–1464}, language={en} }

@inproceedings{Lee_2014, address={Melbourne, Australia}, title={Gidget: An online debugging game for learning and engagement in computing education}, ISBN={9781479940356}, url={http://ieeexplore.ieee.org/document/6883051/}, DOI={10.1109/VLHCC.2014.6883051}, booktitle={2014 IEEE Symposium on Visual Languages and Human-Centric Computing (VL/HCC)}, publisher={IEEE}, author={Lee, Michael J.}, year={2014}, month=jul, pages={193–194} }

@inproceedings{Buechley_Eisenberg_Catchen_Crockett_2008, address={Florence Italy}, title={The LilyPad Arduino: using computational textiles to investigate engagement, aesthetics, and diversity in computer science education}, ISBN={9781605580111}, url={https://dl.acm.org/doi/10.1145/1357054.1357123}, DOI={10.1145/1357054.1357123}, booktitle={Proceedings of the SIGCHI Conference on Human Factors in Computing Systems}, publisher={ACM}, author={Buechley, Leah and Eisenberg, Mike and Catchen, Jaime and Crockett, Ali}, year={2008}, month=apr, pages={423–432}, language={en} }

@inproceedings{Tseng_Bryant_Blikstein_2011, address={Ann Arbor Michigan}, title={Collaboration through documentation: automated capturing of tangible constructions to support engineering design}, ISBN={9781450307512}, url={https://dl.acm.org/doi/10.1145/1999030.1999044}, DOI={10.1145/1999030.1999044}, booktitle={Proceedings of the 10th International Conference on Interaction Design and Children}, publisher={ACM}, author={Tseng, Tiffany and Bryant, Coram and Blikstein, Paulo}, year={2011}, month=jun, pages={118–126}, language={en} }

@inproceedings{Kazemitabaar_McPeak_Jiao_He_Outing_Froehlich_2017, address={Denver Colorado USA}, title={MakerWear: A Tangible Approach to Interactive Wearable Creation for Children}, ISBN={9781450346559}, url={https://dl.acm.org/doi/10.1145/3025453.3025887}, DOI={10.1145/3025453.3025887}, booktitle={Proceedings of the 2017 CHI Conference on Human Factors in Computing Systems}, publisher={ACM}, author={Kazemitabaar, Majeed and McPeak, Jason and Jiao, Alexander and He, Liang and Outing, Thomas and Froehlich, Jon E.}, year={2017}, month=may, pages={133–145}, language={en} }

@inproceedings{Gajos_Weld_2004, address={Funchal, Madeira Portugal}, title={SUPPLE: automatically generating user interfaces}, ISBN={9781581138153}, url={https://dl.acm.org/doi/10.1145/964442.964461}, DOI={10.1145/964442.964461}, booktitle={Proceedings of the 9th international conference on Intelligent user interfaces}, publisher={ACM}, author={Gajos, Krzysztof and Weld, Daniel S.}, year={2004}, month=jan, pages={93–100}, language={en} }

@book{Gershenfeld_2005, address={New York, NY}, edition={1. ed}, title={Fab: the coming revolution on your desktop - from personal computers to personal fabrication}, ISBN={9780465027453}, abstractNote={How to make -- Almost anything -- The past -- Hard ware -- The present -- Birds and bikes -- Subtraction -- Growing inventors -- Addition -- Building models -- Description -- Playing at work -- Computation -- Making sense -- Instrumentation -- Network -- Communication -- Art and artillery -- Interaction -- The future -- Joy -- The details}, publisher={Basic Books}, author={Gershenfeld, Neil A.}, year={2005}, language={eng} }

@article{Taylor_Tanjim_Sack_Hirsch_Cheng_Ching_George_Roumen_Jung_Lee_2025, title={Rapidly Built Medical Crash Cart! Lessons Learned and Impacts on High-Stakes Team Collaboration in the Emergency Room}, url={http://arxiv.org/abs/2502.18688}, DOI={10.48550/arXiv.2502.18688},  note={arXiv:2502.18688}, number={arXiv:2502.18688}, publisher={arXiv}, author={Taylor, Angelique and Tanjim, Tauhid and Sack, Michael Joseph and Hirsch, Maia and Cheng, Kexin and Ching, Kevin and George, Jonathan St and Roumen, Thijs and Jung, Malte F. and Lee, Hee Rin}, year={2025}, month=feb }

@article{Kharat_Dudhani_Kouser_Subramanian_Bhattacharjee_Jhamb_2024, title={Exploring the Impact of 3D Printing Technology on Patient-Specific Prosthodontic Rehabilitation: A Comparative Study}, volume={16}, url={https://pmc.ncbi.nlm.nih.gov/articles/PMC11001056/}, DOI={10.4103/jpbs.jpbs_643_23}, number={Suppl 1}, journal={Journal of Pharmacy \& Bioallied Sciences}, author={Kharat, Swati and Dudhani, Soumya Iranna and Kouser, Afreen and Subramanian, Ponniah and Bhattacharjee, Debarshi and Jhamb, Vikram}, year={2024}, month=feb, pages={S423}, language={en} }

@article{Mamo_Adamiak_Kunwar_2023, title={3D printed biomedical devices and their applications: A review on state-of-the-art technologies, existing challenges, and future perspectives}, volume={143}, ISSN={1751-6161}, url={https://www.sciencedirect.com/science/article/pii/S1751616123002837}, DOI={10.1016/j.jmbbm.2023.105930}, journal={Journal of the Mechanical Behavior of Biomedical Materials}, author={Mamo, Hana Beyene and Adamiak, Marcin and Kunwar, Anil}, year={2023}, month=jul, pages={105930} }

@book{Mongeon_2016, address={Burlington, MA Abingdon, Oxon}, title={3D technology in fine art and craft: exploration of 3D printing, scanning, sculpting and milling}, ISBN={9781138844339}, publisher={Focal Press}, author={Mongeon, Bridgette}, year={2016}, language={eng} }

@book{Hoskins_2018, address={New York}, edition={Second edition}, title={3D printing for artists, designers and makers}, ISBN={9781474248723}, callNumber={TS171.95}, publisher={Bloomsbury Visual Arts, An imprint of Bloomsbury Publishing Plc}, author={Hoskins, Stephen}, year={2018} }

@article{Fraile-Narváez_Chidean_2025, title={Between machines and art: The impact of cnc technology on artistic creation}, ISSN={2590-0056}, url={https://www.sciencedirect.com/science/article/pii/S2590005625000931}, DOI={10.1016/j.array.2025.100466}, journal={Array}, author={Fraile-Narváez, Marcelo and Chidean, Mihaela I.}, year={2025}, month=jul, pages={100466} }

@book{Vygotskiĭ_Cole_1978, address={Cambridge}, title={Mind in society: the development of higher psychological processes}, ISBN={9780674576285}, callNumber={BF311 .V93 1978}, publisher={Harvard University Press}, author = {Vygotsky, L. S. and Cole, Michael},
 year={1978} }

@article{Wood_Bruner_Ross_1976, title={THE ROLE OF TUTORING IN PROBLEM SOLVING*}, volume={17}, rights={http://onlinelibrary.wiley.com/termsAndConditions#vor}, ISSN={0021-9630, 1469-7610}, url={https://acamh.onlinelibrary.wiley.com/doi/10.1111/j.1469-7610.1976.tb00381.x}, DOI={10.1111/j.1469-7610.1976.tb00381.x}, number={2}, journal={Journal of Child Psychology and Psychiatry}, author={Wood, David and Bruner, Jerome S. and Ross, Gail}, year={1976}, month=apr, pages={89–100}, language={en} }

@book{Sweller_Ayres_Kalyuga_2011, address={New York Dordrecht Heidelberg London}, series={Explorations in the learning sciences, instructional systems and performance technologies}, title={Cognitive load theory}, ISBN={9781441981257}, abstractNote={Over the last 25 years, cognitive load theory has become one of the world’s leading theories of instructional design. It is heavily researched by many educational and psychological researchers and is familiar to most practicing instructional designers, especially designers using computer and related technologies. The theory can be divided into two aspects that closely inter-relate and influence each other: human cognitive architecture and the instructional designs and prescriptions that flow from that architecture. The cognitive architecture is based on biological evolution. The resulting description of human cognitive architecture is novel and accordingly, the instructional designs that flow from the architecture also are novel. All instructional procedures are routinely tested using randomized, controlled experiments. Roughly 1/3 of the book will be devoted to cognitive architecture and its evolutionary base with 2/3 devoted to the instructional implications that follow, including technology-based instruction. Researchers, teachers and instructional designers need the book because of the explosion of interest in cognitive load theory over the last few years.The theory is represented in countless journal articles but a detailed, modern overview presenting the theory and its implications in one location is not available}, publisher={Springer}, author={Sweller, John and Ayres, Paul and Kalyuga, Slava}, year={2011}, collection={Explorations in the learning sciences, instructional systems and performance technologies}, language={eng} }

@misc{Tabula_rasa_2025, rights={Creative Commons Attribution-ShareAlike License}, url={https://en.wikipedia.org/w/index.php?title=Tabula_rasa&oldid=1301231875}, abstractNote={Tabula rasa (; Latin for “blank slate”) is the idea of individuals being born empty of any built-in mental content, so that all knowledge comes from later perceptions or sensory experiences. Proponents typically form the extreme “nurture” side of the nature versus nurture debate, arguing that humans are born without any “natural” psychological traits and that all aspects of one’s personality, social and emotional behaviour, knowledge, or sapience are later imprinted by one’s environment onto the mind as one would onto a wax tablet. This idea is the central view posited in the theory of knowledge known as empiricism. Empiricists disagree with the doctrines of innatism or rationalism, which hold that the mind is born already in possession of specific knowledge or rational capacity.}, note={Page Version ID: 1301231875}, journal={Wikipedia}, year={2025}, month=jul, language={en} }

@article{Burnett_Stumpf_Macbeth_Makri_Beckwith_Kwan_Peters_Jernigan_2016, title={GenderMag: A Method for Evaluating Software’s Gender Inclusiveness}, volume={28}, ISSN={0953-5438, 1873-7951}, url={https://academic.oup.com/iwc/article-lookup/doi/10.1093/iwc/iwv046}, DOI={10.1093/iwc/iwv046}, number={6}, journal={Interacting with Computers}, author={Burnett, Margaret and Stumpf, Simone and Macbeth, Jamie and Makri, Stephann and Beckwith, Laura and Kwan, Irwin and Peters, Anicia and Jernigan, William}, year={2016}, month=nov, pages={760–787}, language={en} }

@inproceedings{fossdal2023VespidaeProgrammingFrameworkb,
  title = {Vespidae: {{A Programming Framework}} for {{Developing Digital Fabrication Workflows}}},
  shorttitle = {Vespidae},
  booktitle = {Proceedings of the 2023 {{ACM Designing Interactive Systems Conference}}},
  author = {Fossdal, Frikk H and Nguyen, Vinh and Heldal, Rogardt and Cobb, Corie L. and Peek, Nadya},
  year = {2023},
  month = jul,
  pages = {2034--2049},
  publisher = {ACM},
  address = {Pittsburgh PA USA},
}

@inproceedings{tran_oleary_tandem_2024,
  address = {New York, NY, USA},
  series = {{CHI} '24},
  title = {Tandem: {Reproducible} {Digital} {Fabrication} {Workflows} as {Multimodal} {Programs}},
  isbn = {979-8-4007-0330-0/24/05},
  shorttitle = {Tandem},
  url = {https://dl.acm.org/doi/10.1145/3613904.3642751},
  doi = {10.1145/3613904.3642751},
  urldate = {2023-08-23},
  booktitle = {Proceedings of the 2024 {CHI} {Conference} on {Human} {Factors} in {Computing} {Systems}},
  publisher = {Association for Computing Machinery},
  author = {Tran O'Leary, Jasper and Ramesh, Thrisha and Zhang, Octi and Peek, Nadya},
  month = may,
  year = {2024},
  pages = {1--15},
}

@inproceedings{Beyer_Gurevich_Mueller_Chen_Baudisch_2015, address={Seoul Republic of Korea}, title={Platener: Low-Fidelity Fabrication of 3D Objects by Substituting 3D Print with Laser-Cut Plates}, ISBN={9781450331456}, url={https://dl.acm.org/doi/10.1145/2702123.2702225}, DOI={10.1145/2702123.2702225}, booktitle={Proceedings of the 33rd Annual ACM Conference on Human Factors in Computing Systems}, publisher={ACM}, author={Beyer, Dustin and Gurevich, Serafima and Mueller, Stefanie and Chen, Hsiang-Ting and Baudisch, Patrick}, year={2015}, month=apr, pages={1799–1806}, language={en} }

@inproceedings{Mueller_Mohr_Guenther_Frohnhofen_Baudisch_2014, address={Toronto Ontario Canada}, title={faBrickation: fast 3D printing of functional objects by integrating construction kit building blocks}, ISBN={9781450324731}, url={https://dl.acm.org/doi/10.1145/2556288.2557005}, DOI={10.1145/2556288.2557005}, booktitle={Proceedings of the SIGCHI Conference on Human Factors in Computing Systems}, publisher={ACM}, author={Mueller, Stefanie and Mohr, Tobias and Guenther, Kerstin and Frohnhofen, Johannes and Baudisch, Patrick}, year={2014}, month=apr, pages={3827–3834}, language={en} }

@inproceedings{Teibrich_Patching_Physical_Objects, address={Charlotte NC USA}, title={Patching Physical Objects}, ISBN={9781450337793}, url={https://dl.acm.org/doi/10.1145/2807442.2807467}, DOI={10.1145/2807442.2807467}, booktitle={Proceedings of the 28th Annual ACM Symposium on User Interface Software \& Technology}, publisher={ACM}, author={Teibrich, Alexander and Mueller, Stefanie and Guimbretière, François and Kovacs, Robert and Neubert, Stefan and Baudisch, Patrick}, year={2015}, month=nov, pages={83–91}, language={en} }

@inproceedings{Nisser_Liao_Chai_Adhikari_Hodges_Mueller_2021, address={Yokohama Japan}, title={LaserFactory: A Laser Cutter-based Electromechanical Assembly and Fabrication Platform to Make Functional Devices \& Robots}, ISBN={9781450380966}, url={https://dl.acm.org/doi/10.1145/3411764.3445692}, DOI={10.1145/3411764.3445692}, booktitle={Proceedings of the 2021 CHI Conference on Human Factors in Computing Systems}, publisher={ACM}, author={Nisser, Martin and Liao, Christina Chen and Chai, Yuchen and Adhikari, Aradhana and Hodges, Steve and Mueller, Stefanie}, year={2021}, month=may, pages={1–15}, language={en} }

@inbook{Bjork_1994, title={Memory and Metamemory Considerations in the Training of Human Beings}, ISBN={9780262279697}, url={https://direct.mit.edu/books/book/3931/chapter/164557/Memory-and-Metamemory-Considerations-in-the}, DOI={10.7551/mitpress/4561.003.0011}, booktitle={Metacognition}, publisher={The MIT Press}, author={Bjork, Robert A.}, editor={Metcalfe, Janet and Shimamura, Arthur P.}, year={1994}, month=apr, pages={185–206}, language={en} }

@article{Kapur_2014, title={Productive Failure in Learning Math}, volume={38}, ISSN={0364-0213, 1551-6709}, url={https://onlinelibrary.wiley.com/doi/10.1111/cogs.12107}, DOI={10.1111/cogs.12107}, author={Kapur, Manu}, year={2014}, month=jun, pages={1008–1022}, language={en} }

@article{Markovits_Sowder_1994, title={Developing Number Sense: An Intervention Study in Grade 7}, volume={25}, ISSN={00218251}, url={https://www.jstor.org/stable/749290?origin=crossref}, DOI={10.2307/749290}, number={1}, journal={Journal for Research in Mathematics Education}, author={Markovits, Zvia and Sowder, Judith}, year={1994}, month=jan, pages={4} }

@article{McNeil_Alibali_2005, title={Why Won’t You Change Your Mind? Knowledge of Operational Patterns Hinders Learning and Performance on Equations}, volume={76}, ISSN={0009-3920, 1467-8624}, url={https://srcd.onlinelibrary.wiley.com/doi/10.1111/j.1467-8624.2005.00884.x}, DOI={10.1111/j.1467-8624.2005.00884.x},  number={4}, journal={Child Development}, author={McNeil, Nicole M. and Alibali, Martha W.}, year={2005}, month=jul, pages={883–899}, language={en} }

@article{Langer_2000, title={Mindful Learning}, volume={9}, rights={https://journals.sagepub.com/page/policies/text-and-data-mining-license}, ISSN={0963-7214, 1467-8721}, url={https://journals.sagepub.com/doi/10.1111/1467-8721.00099}, DOI={10.1111/1467-8721.00099},  number={6}, journal={Current Directions in Psychological Science}, author={Langer, Ellen J.}, year={2000}, month=dec, pages={220–223}, language={en} }

@article{DeCaro_2016, title={Inducing mental set constrains procedural flexibility and conceptual understanding in mathematics}, volume={44}, ISSN={0090-502X, 1532-5946}, url={http://link.springer.com/10.3758/s13421-016-0614-y}, DOI={10.3758/s13421-016-0614-y}, number={7}, journal={Memory \& Cognition}, author={DeCaro, Marci S.}, year={2016}, month=oct, pages={1138–1148}, language={en} }

@inproceedings{Tohidi_Buxton_Baecker_Sellen_2006, address={Montréal Québec Canada}, title={Getting the right design and the design right}, ISBN={9781595933720}, url={https://dl.acm.org/doi/10.1145/1124772.1124960}, DOI={10.1145/1124772.1124960}, booktitle={Proceedings of the SIGCHI Conference on Human Factors in Computing Systems}, publisher={ACM}, author={Tohidi, Maryam and Buxton, William and Baecker, Ronald and Sellen, Abigail}, year={2006}, month=apr, pages={1243–1252}, language={en} }

@inbook{Kulkarni_Dow_Klemmer_2014, address={Cham}, title={Early and Repeated Exposure to Examples Improves Creative Work}, ISBN={9783319013022}, url={https://link.springer.com/10.1007/978-3-319-01303-9_4}, DOI={10.1007/978-3-319-01303-9_4}, booktitle={Design Thinking Research}, publisher={Springer International Publishing}, author={Kulkarni, Chinmay and Dow, Steven P. and Klemmer, Scott R}, editor={Leifer, Larry and Plattner, Hasso and Meinel, Christoph}, year={2014}, pages={49–62}, language={en} }

@inproceedings{qiu_curriculum_2013,
    address = {New York, NY, USA},
    series = {{IDC} '13},
    title = {A curriculum for teaching computer science through computational textiles},
    isbn = {978-1-4503-1918-8},
    url = {https://dl.acm.org/doi/10.1145/2485760.2485787},
    doi = {10.1145/2485760.2485787},
    urldate = {2025-08-31},
    booktitle = {Proceedings of the 12th {International} {Conference} on {Interaction} {Design} and {Children}},
    publisher = {Association for Computing Machinery},
    author = {Qiu, Kanjun and Buechley, Leah and Baafi, Edward and Dubow, Wendy},
    month = jun,
    year = {2013},
    pages = {20--27},
}

@inproceedings{buechley_computational_2022,
    address = {New York, NY, USA},
    series = {{SCF} '22},
    title = {A {Computational} {Fabrication} {Course}: {Exploring} {Philosophical} {Reflection}, {Real}-{World} {Use}, {Personal} {Expression}, and {Social} {Connection}},
    isbn = {978-1-4503-9872-5},
    shorttitle = {A {Computational} {Fabrication} {Course}},
    url = {https://dl.acm.org/doi/10.1145/3559400.3562006},
    doi = {10.1145/3559400.3562006},
    urldate = {2025-08-31},
    booktitle = {Proceedings of the 7th {Annual} {ACM} {Symposium} on {Computational} {Fabrication}},
    publisher = {Association for Computing Machinery},
    author = {Buechley, Leah and Bustos, Alyshia and Edreva, Eleonora and Fenske, Tyler and Fresquez, Reuben and Hafer, Samuel and Handey, Aislinn and Louie, Michelle and Ng, John and Shen, Alan and Smith, Randi and Sustaita, Amber and Truong, Michael and Vallon, Kai and Weiss, Kage Micaiah},
    month = oct,
    year = {2022},
    pages = {1--13},
}

@inproceedings{fernandez_toward_2021,
    address = {New York, NY, USA},
    series = {{FabLearn} '20},
    title = {Toward a sustainable model for maker education in public education: {Teachers} as co-designers in an implementation of educational makerspaces},
    isbn = {978-1-4503-7543-6},
    shorttitle = {Toward a sustainable model for maker education in public education},
    url = {https://dl.acm.org/doi/10.1145/3386201.3386218},
    doi = {10.1145/3386201.3386218},
    urldate = {2025-08-31},
    booktitle = {Proceedings of the {FabLearn} 2020 - 9th {Annual} {Conference} on {Maker} {Education}},
    publisher = {Association for Computing Machinery},
    author = {Fernandez, Cassia and Hochgreb-Haegele, Tatiana and Blikstein, Paulo},
    month = nov,
    year = {2021},
    pages = {46--53},
}

@inproceedings{higgins_towards_2023,
    address = {New York, NY, USA},
    series = {{ASSETS} '23},
    title = {Towards a {Social} {Justice} {Aligned} {Makerspace}: {Co}-designing {Custom} {Assistive} {Technology} within a {University} {Ecosystem}},
    isbn = {9798400702204},
    shorttitle = {Towards a {Social} {Justice} {Aligned} {Makerspace}},
    url = {https://dl.acm.org/doi/10.1145/3597638.3608393},
    doi = {10.1145/3597638.3608393},
    urldate = {2025-08-31},
    booktitle = {Proceedings of the 25th {International} {ACM} {SIGACCESS} {Conference} on {Computers} and {Accessibility}},
    publisher = {Association for Computing Machinery},
    author = {Higgins, Erin and Oliver, Zaria and Hamidi, Foad},
    month = oct,
    year = {2023},
    pages = {1--13},
}

@misc{Comparing_Novice_and_Expert_Designers’_Approaches_to_Design_Thinking_and_Decision_Making, 
author={Alsagour,Mahdi},
year={2020},
title={Comparing Novice and Expert Designers’ Approaches to Design Thinking and Decision Making},
journal={ProQuest Dissertations and Theses},
pages={140},
note={Copyright - Database copyright ProQuest LLC; ProQuest does not claim copyright in the individual underlying works; Last updated - 2023-11-20},
isbn={9798557046817},
language={English},
url={https://www.proquest.com/dissertations-theses/comparing-novice-expert-designers-approaches/docview/2488270158/se-2},
}

@article{Cheung_Kulasegaram_Woods_Brydges_2019, title={Why Content and Cognition Matter: Integrating Conceptual Knowledge to Support Simulation-Based Procedural Skills Transfer}, volume={34}, ISSN={1525-1497}, url={https://doi.org/10.1007/s11606-019-04959-y}, DOI={10.1007/s11606-019-04959-y}, abstractNote={Curricular constraints require being selective about the type of content trainees practice in their formal training. Teaching trainees procedural knowledge about “how” to perform steps of a skill along with conceptual knowledge about “why” each step is performed can support skill retention and transfer (i.e., the ability to adapt knowledge to novel problems). However, how best to organize how and why content for procedural skills training is unknown.}, number={6}, journal={Journal of General Internal Medicine}, author={Cheung, Jeffrey J. H. and Kulasegaram, Kulamakan M. and Woods, Nicole N. and Brydges, Ryan}, year={2019}, month=jun, pages={969–977}, language={en} }

@article{Kulkarni_Magda_2025, title={Deploying computation-based Making projects in authentic public school classrooms at scale: Lessons learned}, volume={45}, ISSN={22128689}, url={https://linkinghub.elsevier.com/retrieve/pii/S2212868925000182}, DOI={10.1016/j.ijcci.2025.100738}, journal={International Journal of Child-Computer Interaction}, author={Kulkarni, Abhishek and Magda, David and Ward, Rebecca and Jimenez, Yerika and Hernandez, Monica and Liu, Ting and Gardner-McCune, Christina and Quek, Francis and Schlegel, Rebecca and Chu, Sharon Lynn}, year={2025}, month=sep, pages={100738}, language={en} }

@article{Gentner_Markman_1997, title={Structure mapping in analogy and similarity.}, volume={52}, ISSN={1935-990X, 0003-066X}, url={https://doi.apa.org/doi/10.1037/0003-066X.52.1.45}, DOI={10.1037/0003-066X.52.1.45}, number={1}, journal={American Psychologist}, author={Gentner, Dedre and Markman, Arthur B.}, year={1997}, month=jan, pages={45–56}, language={en} }

@article{Loewenstein_Thompson_Gentner_1999, title={Analogical encoding facilitates knowledge transfer in negotiation}, volume={6}, rights={http://www.springer.com/tdm}, ISSN={1069-9384, 1531-5320}, url={http://link.springer.com/10.3758/BF03212967}, DOI={10.3758/BF03212967}, number={4}, journal={Psychonomic Bulletin \& Review}, author={Loewenstein, Jeffrey and Thompson, Leigh and Gentner, Dedre}, year={1999}, month=dec, pages={586–597}, language={en} }

@article{Rittle-Johnson_Schneider_Star_2015, title={Not a One-Way Street: Bidirectional Relations Between Procedural and Conceptual Knowledge of Mathematics}, volume={27}, ISSN={1040-726X, 1573-336X}, url={http://link.springer.com/10.1007/s10648-015-9302-x}, DOI={10.1007/s10648-015-9302-x}, number={4}, journal={Educational Psychology Review}, author={Rittle-Johnson, Bethany and Schneider, Michael and Star, Jon R.}, year={2015}, month=dec, pages={587–597}, language={en} }

@article{Kapur_2016, title={Examining Productive Failure, Productive Success, Unproductive Failure, and Unproductive Success in Learning}, volume={51}, ISSN={0046-1520, 1532-6985}, url={http://www.tandfonline.com/doi/full/10.1080/00461520.2016.1155457}, DOI={10.1080/00461520.2016.1155457}, number={2}, journal={Educational Psychologist}, author={Kapur, Manu}, year={2016}, month=apr, pages={289–299}, language={en} }

@inproceedings{Bishop_Verleger_2013, address={Atlanta, Georgia}, title={The Flipped Classroom: A Survey of the Research}, url={http://peer.asee.org/22585}, DOI={10.18260/1-2--22585}, booktitle={2013 ASEE Annual Conference \& Exposition Proceedings}, publisher={ASEE Conferences}, author={Bishop, Jacob and Verleger, Matthew}, year={2013}, month=jun, pages={23.1200.1-23.1200.18} }

@article{Lo_Hew_2017, title={A critical review of flipped classroom challenges in K-12 education: possible solutions and recommendations for future research}, volume={12}, ISSN={1793-7078}, url={http://telrp.springeropen.com/articles/10.1186/s41039-016-0044-2}, DOI={10.1186/s41039-016-0044-2}, number={1}, journal={Research and Practice in Technology Enhanced Learning}, author={Lo, Chung Kwan and Hew, Khe Foon}, year={2017}, month=dec, pages={4}, language={en} }

@inproceedings{Roumen_2020, address={Virtual Event USA}, title={Portable Laser Cutting}, rights={https://www.acm.org/publications/policies/copyright_policy#Background}, ISBN={9781450375153}, url={https://dl.acm.org/doi/10.1145/3379350.3415802}, DOI={10.1145/3379350.3415802}, booktitle={Adjunct Publication of the 33rd Annual ACM Symposium on User Interface Software and Technology}, publisher={ACM}, author={Roumen, Thijs}, year={2020}, month=oct, pages={157–161}, language={en} }

@article{Tan_Otto_Wood_2017, title={A comparison of design decisions made early and late in development}, ISSN={2220-4342}, url={https://www.designsociety.org/publication/39558/a_comparison_of_design_decisions_made_early_and_late_in_development}, abstractNote={The occurrence rates and cost impact of design changes made early and later in the design process were studied, to test and quantify the 80-20 rule of design cost impacts, that early design decisions account for the majority of costs in a development program. Cost and schedule impact of decisions made throughout the development process was carried out at a large aerospace firm on two programs covering 7 years of development with 275 person-years effort. The underlying data used was the rate and cost of design changes made. We found no significant difference in the rate of occurrence of design change decisions made, but we found a significant difference in the cost impact of the design changes. Overall, early design change decisions cost 5 times more than later design change decisions. This difference is primarily due to the inability to determine if an early design decision is correct until later in development during testing.}, journal={DS 87-2 Proceedings of the 21st International Conference on Engineering Design (ICED 17) Vol 2: Design Processes, Design Organisation and Management, Vancouver, Canada, 21-25.08.2017}, author={Tan, James and Otto, Kevin and Wood, Kristin}, year={2017}, pages={041–050}, language={en} }

@inproceedings{Guo_Jin_Sun_Li_Li_Shi_Cao_2021, address={Yokohama Japan}, title={Vinci: An Intelligent Graphic Design System for Generating Advertising Posters}, ISBN={9781450380966}, url={https://dl.acm.org/doi/10.1145/3411764.3445117}, DOI={10.1145/3411764.3445117}, booktitle={Proceedings of the 2021 CHI Conference on Human Factors in Computing Systems}, publisher={ACM}, author={Guo, Shunan and Jin, Zhuochen and Sun, Fuling and Li, Jingwen and Li, Zhaorui and Shi, Yang and Cao, Nan}, year={2021}, month=may, pages={1–17}, language={en} }

@inproceedings{Suh_Zhao_Law_2022, address={Bend OR USA}, title={CodeToon: Story Ideation, Auto Comic Generation, and Structure Mapping for Code-Driven Storytelling}, ISBN={9781450393201}, url={https://dl.acm.org/doi/10.1145/3526113.3545617}, DOI={10.1145/3526113.3545617}, booktitle={Proceedings of the 35th Annual ACM Symposium on User Interface Software and Technology}, publisher={ACM}, author={Suh, Sangho and Zhao, Jian and Law, Edith}, year={2022}, month=oct, pages={1–16}, language={en} }

@inproceedings{Deiner_Fraser_2024, address={Lisbon Portugal}, title={NuzzleBug: Debugging Block-Based Programs in Scratch}, ISBN={9798400702174}, url={https://dl.acm.org/doi/10.1145/3597503.3623331}, DOI={10.1145/3597503.3623331}, booktitle={Proceedings of the IEEE/ACM 46th International Conference on Software Engineering}, publisher={ACM}, author={Deiner, Adina and Fraser, Gordon}, year={2024}, month=feb, pages={1–13}, language={en} }

@inproceedings{Li__2024, address={Pittsburgh PA USA}, title={E-Joint: Fabrication of Large-Scale Interactive Objects Assembled by 3D Printed Conductive Parts with Copper Plated Joints}, ISBN={9798400706288}, url={https://dl.acm.org/doi/10.1145/3654777.3676398}, DOI={10.1145/3654777.3676398}, booktitle={Proceedings of the 37th Annual ACM Symposium on User Interface Software and Technology}, publisher={ACM}, author={Li, Xiaolong and Yao, Cheng and Shi, Shang and Feng, Shuyue and Zhou, Yujie and Dong, Haoye and Huang, Shichao and Cai, Xueyan and Jin, Kecheng and Ying, Fangtian and Wang, Guanyun}, year={2024}, month=oct, pages={1–18}, language={en} }

@inproceedings{Magrisso_Mizrahi_Zoran_2018, address={Montreal QC Canada}, title={Digital Joinery For Hybrid Carpentry}, ISBN={9781450356206}, url={https://dl.acm.org/doi/10.1145/3173574.3173741}, DOI={10.1145/3173574.3173741}, booktitle={Proceedings of the 2018 CHI Conference on Human Factors in Computing Systems}, publisher={ACM}, author={Magrisso, Shiran and Mizrahi, Moran and Zoran, Amit}, year={2018}, month=apr, pages={1–11}, language={en} }

@inproceedings{Rosner_Ryokai_2009, address={Berkeley California USA}, title={Reflections on craft: probing the creative process of everyday knitters}, ISBN={9781605588650}, url={https://dl.acm.org/doi/10.1145/1640233.1640264}, DOI={10.1145/1640233.1640264}, booktitle={Proceedings of the seventh ACM conference on Creativity and cognition}, publisher={ACM}, author={Rosner, Daniela K. and Ryokai, Kimiko}, year={2009}, month=oct, pages={195–204}, language={en} }

@inproceedings{Resnick_Silverman_2005, address={Boulder Colorado}, title={Some reflections on designing construction kits for kids}, ISBN={9781595930965}, url={https://dl.acm.org/doi/10.1145/1109540.1109556}, DOI={10.1145/1109540.1109556}, booktitle={Proceedings of the 2005 conference on Interaction design and children}, publisher={ACM}, author={Resnick, Mitchel and Silverman, Brian}, year={2005}, month=june, pages={117–122}, language={en} }

@article{Song_2024, title={Unmaking and HCI: Techniques, Technologies, Materials, and Philosophies beyond Making}, volume={31}, ISSN={1073-0516, 1557-7325}, url={https://dl.acm.org/doi/10.1145/3689047}, DOI={10.1145/3689047}, number={6}, journal={ACM Transactions on Computer-Human Interaction}, author={Song, Katherine W. and Sabie, Samar and Jackson, Steven J. and Lindström, Kristina and Paulos, Eric and Ståhl, Åsa and Wakkary, Ron}, year={2024}, month=dec, pages={1–6}, language={en} }


%% file: thijs-papers.bib
@inproceedings{higgins_creating_2022,
	address = {New York, NY, USA},
	series = {{ASSETS} '22},
	title = {Creating {3D} {Printed} {Assistive} {Technology} {Through} {Design} {Shortcuts}: {Leveraging} {Digital} {Fabrication} {Services} to {Incorporate} {3D} {Printing} into the {Physical} {Therapy} {Classroom}: {Leveraging} {Digital} {Fabrication} {Services} to {Incorporate} {3D} {Printing} into the {Physical} {Therapy} {Classroom}},
	isbn = {978-1-4503-9258-7},
	shorttitle = {Creating {3D} {Printed} {Assistive} {Technology} {Through} {Design} {Shortcuts}},
	url = {https://dl.acm.org/doi/10.1145/3517428.3544816},
	doi = {10.1145/3517428.3544816},
	urldate = {2024-08-27},
	booktitle = {Proceedings of the 24th {International} {ACM} {SIGACCESS} {Conference} on {Computers} and {Accessibility}},
	publisher = {Association for Computing Machinery},
	author = {Higgins, Erin and Easley, William Berkley and Gordes, Karen L. and Hurst, Amy and Hamidi, Foad},
	month = oct,
	year = {2022},
	pages = {1--16},
}

@article{hartikainen_making_2024,
	title = {Making {The} {Future} {School}: {An} {Analysis} of {Teens}' {Collaborative} {Digital} {Fabrication} {Project}},
	volume = {8},
	shorttitle = {Making {The} {Future} {School}},
	url = {https://dl.acm.org/doi/10.1145/3637368},
	doi = {10.1145/3637368},
	number = {CSCW1},
	urldate = {2024-08-27},
	journal = {Proc. ACM Hum.-Comput. Interact.},
	author = {Hartikainen, Heidi and Ventä-Olkkonen, Leena and Cortés Orduña, Marta and Sanchez Milara, Ivan and Käsmä, Marjukka and Kuure, Leena},
	month = apr,
	year = {2024},
	pages = {91:1--91:37},
}

@inproceedings{parry-hill_understanding_2017,
	address = {New York, NY, USA},
	series = {{CHI} '17},
	title = {Understanding {Volunteer} {AT} {Fabricators}: {Opportunities} and {Challenges} in {DIY}-{AT} for {Others} in e-{NABLE}},
	isbn = {978-1-4503-4655-9},
	url = {https://doi.org/10.1145/3025453.3026045},
	doi = {10.1145/3025453.3026045},
	booktitle = {Proceedings of the 2017 {CHI} {Conference} on {Human} {Factors} in {Computing} {Systems}},
	publisher = {Association for Computing Machinery},
	author = {Parry-Hill, Jeremiah and Shih, Patrick C. and Mankoff, Jennifer and Ashbrook, Daniel},
	year = {2017},
	note = {event-place: Denver, Colorado, USA},
	pages = {6184--6194},
}

@inproceedings{roumen_structure-preserving_2022,
	address = {Seattle WA USA},
	title = {Structure-{Preserving} {Editing} of {Plates} and {Volumes} for {Laser} {Cutting}},
	copyright = {All rights reserved},
	isbn = {978-1-4503-9872-5},
	url = {https://dl.acm.org/doi/10.1145/3559400.3561996},
	doi = {10.1145/3559400.3561996},
	language = {en},
	urldate = {2024-01-20},
	booktitle = {Proceedings of the 7th {Annual} {ACM} {Symposium} on {Computational} {Fabrication}},
	publisher = {ACM},
	author = {Roumen, Thijs and Apel, Ingo and Kern, Thomas and Taraz, Martin and Sharma, Ritesh and Schlueter, Ole and Johnson, Jeffrey and Meier, Dominik and Lempert, Conrad and Baudisch, Patrick},
	month = oct,
	year = {2022},
	pages = {1--12},
}

@inproceedings{bermano_state_2017,
	title = {State of the art in methods and representations for fabrication-aware design},
	volume = {36},
	booktitle = {Computer {Graphics} {Forum}},
	publisher = {Wiley Online Library},
	author = {Bermano, Amit H and Funkhouser, Thomas and Rusinkiewicz, Szymon},
	year = {2017},
	note = {Issue: 2},
	pages = {509--535},
}

@article{baudisch_personal_2017,
	title = {Personal fabrication},
	volume = {10},
	number = {3–4},
	journal = {Foundations and Trends® in Human–Computer Interaction},
	author = {Baudisch, Patrick and Mueller, Stefanie and {others}},
	year = {2017},
	note = {Publisher: Now Publishers, Inc.},
	pages = {165--293},
}

@inproceedings{zisimatos_open-source_2014,
	title = {Open-source, affordable, modular, light-weight, underactuated robot hands},
	booktitle = {2014 {IEEE}/{RSJ} {International} {Conference} on {Intelligent} {Robots} and {Systems}},
	publisher = {IEEE},
	author = {Zisimatos, Agisilaos G and Liarokapis, Minas V and Mavrogiannis, Christoforos I and Kyriakopoulos, Kostas J},
	year = {2014},
	pages = {3207--3212},
}

@inproceedings{baudisch_kyub_2019,
	address = {Glasgow Scotland Uk},
	title = {Kyub: {A} {3D} {Editor} for {Modeling} {Sturdy} {Laser}-{Cut} {Objects}},
	copyright = {All rights reserved},
	isbn = {978-1-4503-5970-2},
	shorttitle = {Kyub},
	url = {https://dl.acm.org/doi/10.1145/3290605.3300796},
	doi = {10.1145/3290605.3300796},
	language = {en},
	urldate = {2024-01-20},
	booktitle = {Proceedings of the 2019 {CHI} {Conference} on {Human} {Factors} in {Computing} {Systems}},
	publisher = {ACM},
	author = {Baudisch, Patrick and Silber, Arthur and Kommana, Yannis and Gruner, Milan and Wall, Ludwig and Reuss, Kevin and Heilman, Lukas and Kovacs, Robert and Rechlitz, Daniel and Roumen, Thijs},
	month = may,
	year = {2019},
	pages = {1--12},
}

@inproceedings{roumen_grafter_2018,
	address = {Montreal QC Canada},
	title = {Grafter: {Remixing} {3D}-{Printed} {Machines}},
	copyright = {All rights reserved},
	isbn = {978-1-4503-5620-6},
	shorttitle = {Grafter},
	url = {https://dl.acm.org/doi/10.1145/3173574.3173637},
	doi = {10.1145/3173574.3173637},
	language = {en},
	urldate = {2024-01-20},
	booktitle = {Proceedings of the 2018 {CHI} {Conference} on {Human} {Factors} in {Computing} {Systems}},
	publisher = {ACM},
	author = {Roumen, Thijs Jan and Müller, Willi and Baudisch, Patrick},
	month = apr,
	year = {2018},
	pages = {1--12},
}

@inproceedings{park_foolproofjoint_2022,
	address = {New Orleans LA USA},
	title = {{FoolProofJoint}: {Reducing} {Assembly} {Errors} of {Laser} {Cut} {3D} {Models} by {Means} of {Custom} {Joint} {Patterns}},
	copyright = {All rights reserved},
	isbn = {978-1-4503-9157-3},
	shorttitle = {{FoolProofJoint}},
	url = {https://dl.acm.org/doi/10.1145/3491102.3501919},
	doi = {10.1145/3491102.3501919},
	language = {en},
	urldate = {2024-01-20},
	booktitle = {{CHI} {Conference} on {Human} {Factors} in {Computing} {Systems}},
	publisher = {ACM},
	author = {Park, Keunwoo and Lempert, Conrad and Abdullah, Muhammad and Katakura, Shohei and Shigeyama, Jotaro and Roumen, Thijs and Baudisch, Patrick},
	month = apr,
	year = {2022},
	pages = {1--12},
}

@inproceedings{mccrae_flatfitfab_2014,
	address = {New York, NY, USA},
	title = {{FlatFitFab}: interactive modeling with planar sections},
	booktitle = {Proceedings of the 27th annual {ACM} symposium on {User} interface software and technology},
	publisher = {ACM},
	author = {McCrae, James and Umetani, Nobuyuki and Singh, Karan},
	year = {2014},
	pages = {13--22},
}

@inproceedings{abdullah_fastforce_2021,
	address = {Yokohama Japan},
	title = {{FastForce}: {Real}-{Time} {Reinforcement} of {Laser}-{Cut} {Structures}},
	copyright = {All rights reserved},
	isbn = {978-1-4503-8096-6},
	shorttitle = {{FastForce}},
	url = {https://dl.acm.org/doi/10.1145/3411764.3445466},
	doi = {10.1145/3411764.3445466},
	language = {en},
	urldate = {2024-01-20},
	booktitle = {Proceedings of the 2021 {CHI} {Conference} on {Human} {Factors} in {Computing} {Systems}},
	publisher = {ACM},
	author = {Abdullah, Muhammad and Taraz, Martin and Kommana, Yannis and Katakura, Shohei and Kovacs, Robert and Shigeyama, Jotaro and Roumen, Thijs and Baudisch, Patrick},
	month = may,
	year = {2021},
	pages = {1--12},
}

@inproceedings{hurst_empowering_2011,
	title = {Empowering individuals with do-it-yourself assistive technology},
	booktitle = {The proceedings of the 13th international {ACM} {SIGACCESS} conference on {Computers} and accessibility},
	author = {Hurst, Amy and Tobias, Jasmine},
	year = {2011},
	pages = {11--18},
}

@inproceedings{roumen_autoassembler_2021,
	address = {Virtual Event USA},
	title = {{autoAssembler}: {Automatic} {Reconstruction} of {Laser}-{Cut} {3D} {Models}},
	copyright = {All rights reserved},
	isbn = {978-1-4503-8635-7},
	shorttitle = {{autoAssembler}},
	url = {https://dl.acm.org/doi/10.1145/3472749.3474776},
	doi = {10.1145/3472749.3474776},
	language = {en},
	urldate = {2024-01-20},
	booktitle = {The 34th {Annual} {ACM} {Symposium} on {User} {Interface} {Software} and {Technology}},
	publisher = {ACM},
	author = {Roumen, Thijs and Lempert, Conrad and Apel, Ingo and Brendel, Erik and Brand, Markus and Seidel, Laurenz and Rambold, Lukas and Baudisch, Patrick},
	month = oct,
	year = {2021},
	pages = {652--662},
}

@inproceedings{roumen_assembler3_2021,
	address = {Yokohama Japan},
	title = {Assembler3: {3D} {Reconstruction} of {Laser}-{Cut} {Models}},
	copyright = {All rights reserved},
	isbn = {978-1-4503-8096-6},
	shorttitle = {Assembler3},
	url = {https://dl.acm.org/doi/10.1145/3411764.3445453},
	doi = {10.1145/3411764.3445453},
	language = {en},
	urldate = {2024-01-20},
	booktitle = {Proceedings of the 2021 {CHI} {Conference} on {Human} {Factors} in {Computing} {Systems}},
	publisher = {ACM},
	author = {Roumen, Thijs and Kommana, Yannis and Apel, Ingo and Lempert, Conrad and Brand, Markus and Brendel, Erik and Seidel, Laurenz and Rambold, Lukas and Goedecken, Carl and Crenzin, Pascal and Hurdelhey, Ben and Abdullah, Muhammad and Baudisch, Patrick},
	month = may,
	year = {2021},
	pages = {1--11},
}
